\documentclass[12pt,reqno]{amsart}
\usepackage{graphicx} 
\usepackage{amssymb}
\usepackage{amsmath}
\usepackage{cite}

\newtheorem{proposition}{Proposition}[section]
\newtheorem{corollary}{Corollary}[section]
\newtheorem{lemma}{Lemma}[section]
\newtheorem{remark}{Remark}[section]
\newcommand{\fun}[3]{{#1}^{({#2})}_{{#3}}}
\newcommand{\tu}[1]{\tau^{(#1)}}

\begin{document}

\title[Solving bi-directional soliton equations]{Solving bi-directional
  soliton equations in the KP hierarchy by gauge transformation}

\author{Jingsong He\dag\ddag, Yi Cheng\dag and Rudolf A.\
R\"{o}mer\ddag}
\dedicatory { \dag\ Department of Mathematics,
University of Science and
  Technology of China, Hefei, 230026 Anhui, P.R.\ China\\
\ddag\ Department of Physics and Centre for Scientific Computing,
  University of Warwick, Coventry CV4 7AL, United Kingdom}

\begin{abstract}
We present a systematic way to construct solutions of the
$(n=5)$-reduction of the BKP and CKP hierarchies from the general
$\tau$ function $\tu{n+k}$ of the KP hierarchy. We obtain the
one-soliton, two-soliton, and periodic solution for the
bi-directional Sawada-Kotera (bSK), the bi-directional
Kaup-Kupershmidt (bKK) and also the bi-directional Satsuma-Hirota
(bSH) equation. Different solutions such as left- and right-going
solitons are classified according to the symmetries of the $5$th
roots of $e^{i\varepsilon}$.
Furthermore, we show that the soliton solutions of the
$n$-reduction of the BKP and CKP hierarchies with $n= 2 j +1$,
$j=1, 2, 3, \ldots$, can propagate along $j$ directions in the
$1+1$ space-time domain. Each such direction corresponds to one
symmetric distribution of the $n$th roots of $e^{i\varepsilon}$.
Based on this classification, we detail the existence of two-peak
solitons of the $n$-reduction from the Grammian $\tau$ function of
the sub-hierarchies BKP and CKP. If $n$ is even, we again find
two-peak solitons. Last, we obtain the "stationary" soliton for
the higher-order KP hierarchy.
\end{abstract}

\date{$Revision: 1.2 $, compiled \today}
 \maketitle



\section{Introduction}

The Kadomtsev-Petviashvili (KP) hierarchy is of central interest for
integrable systems and includes several well-known partial differential
equations such as the Korteweg-de Vries (KdV) and the KP equation. With
pseudo-differential Lax operator $L$ given as \cite{dkjm,ost,dl1}
\begin{equation}
    L=\partial+u_2 \partial^{-1}+
      u_3\partial^{-2}+\cdots ,
\label{1,1}
\end{equation}
the corresponding generalized Lax equation
\begin{equation}\label{1.2}
 \frac{\partial L}{\partial {t_n}}=[B_n, L], \quad n=1, 2, 3, \cdots,
\end{equation}
gives rise to the infinite number of partial differential equations
(PDEs) of the KP hierarchy with dynamical variables $\left\{ u_i(t_1,
  t_2, t_3, \cdots) \right\}$ with $i= 2, 3, 4, \cdots$.
Here $B_n=\sum\limits_{i=0}^n b_{n,i}\partial^i \equiv (L^n)_+$
denotes the {\em differential} part of $L^n$ and in following we
will use $L^n_- \equiv L^n-B_n$ to denote the {\em integral} part.

The simplest nontrivial PDE constructed from (\ref{1.2}) is the KP
equation given as
\begin{equation}
\frac{\partial }{\partial x}\left(4 \frac{\partial u_2}{\partial t_3}
-12u_2 \frac{\partial u_2}{\partial x}- \frac{\partial^3 u_2}{\partial
x^3} \right)-3 \frac{\partial^2 u_2}{\partial t_2}=0 \quad .
\end{equation}
In Table \ref{tab-KPexamples} we show the Lax operator and corresponding
$(1+1)$-dimensional examples of sub-hierarchies of the KP hierarchy.
An alternative way to express the KP hierarchy is given by the
Zakharov-Shabat (ZS) equation \cite{zs},
\begin{equation}\label{1.8}
\frac{\partial B_n}{\partial t_m}- \frac{\partial B_m}{\partial
t_n}+[B_n,B_m]=0,\quad m, n=2, 3, 4, \cdots \quad .
\end{equation}
The {\em eigenfunction} $\phi$ and the {\em adjoint eigenfunction}
$\psi$ of the KP hierarchy associated with equation (\ref{1.8}) are defined by
\begin{equation}\label{zslax1}
\frac{\partial \phi}{\partial t_n}=B_n\phi, \quad
\frac{\partial \psi}{\partial t_n}=-B_n^{*}\psi, 
\end{equation}
where $\phi=\phi(\lambda;\bar{t})$ and $\psi=\psi(\lambda; \bar{t} )$
and $\bar{t}=(t_1, t_2, \cdots)$.

The $n$-reduction of the KP hierarchy corresponds to the situation
$L^n_{-}=0$ such that $L^n=B_n=\partial^{n}
+v_{n-2}\partial^{n-2}+\cdots+v_1\partial +v_0$. Then the $v_i$,\
$i=0, 1, \cdots, n-2$,\  are independent of $( t_n, t_{2n},
t_{3n},\cdots)$. In this way the Lax pair of the
$(1+1)$-dimensional integrable system can be found. Well-known
examples of such $n$-reductions include the $4$-reduction of the
KP hierarchy \cite{sh} with Lax pair
\begin{equation}
(\partial_x^4 +4 u \partial_x^2 +4u_x \partial_x + 2u_{xx} +4
u^2+v)\phi=\lambda \phi,  \label{bSHlax1}
\end{equation}
\begin{equation}
\partial_t\phi=(\partial^3_x +3u\partial_x
+\frac{3}{2}u_x)\phi,\quad t_1=x,\quad t_3=t, \label{bSHlax2}
\end{equation}
corresponding to the Satsuma-Hirota (SH) equation \cite{sh}
\begin{equation}\label{bSH0}
-4u_t+ 12uu_{x}+u_{xxx}+3v_x=0,\quad 2v_t +6uv_x+v_{xxx}=0.
\end{equation}
Furthermore, eliminating $v$ in the above equations, we can obtain
a $6$th order equation ($u=z_x$)
\begin{equation}\label{bSH}
-8z_{tt}+z_{xxxxxx}-2z_{xxxt}+18z_xz_{xxxx}+36z_{xx}z_{xxx}+72z_x^2z_{xx}=0,
\end{equation}
which has been called bi-directional Satsuma-Hirota (bSH) equation
\cite{vm1}. Naturally, there also exist $n$-reductions of the BKP
and CKP hierarchies. For example, the $5$-reduction of the BKP
hierarchy with $u=u_2$ is given as
\begin{equation}
\left[\partial_x^5 +5 u \partial_x^3 +5 u_x \partial_x^2+(
5u^2+\frac{10}{3}u_{xx}+\frac{5}{3}z_t )\partial_x
\right]\phi=\lambda \phi, \label{bSKlax1}
\end{equation}
\begin{equation}
\partial_t\phi=(\partial^3_x +3u\partial_x )\phi,\quad
u=z_x,t_3=t,t_1=x, \label{bSKlax2}
\end{equation}
which is the Lax pair corresponding to bi-directional
Sawada-Kotera (bSK) equation \cite{dp1,dp2}
\begin{equation}
\left(z_{xxxxx}+ 15z_xz_{xxx}+15 z_x^3-15z_xz_t-5z_{xxt}
\right)_x-5z_{tt}=0. \label{bSK}
\end{equation}

The 5-reduction of the CKP hierarchy ($u=u_2$) with Lax pair \cite{dp1,dp2}
\begin{equation}
\left[\partial_x^5 +5 u \partial_x^3 +\frac{15}{2}u_x
\partial_x^2+(
5u^2+\frac{35}{6}u_{xx}+\frac{5}{3}z_t )\partial_x
+5uu_x+\frac{5}{3}u_{xxx}+\frac{5}{6}u_t \right]\phi=\lambda \phi,
\label{bKKlax1}
\end{equation}
\begin{equation}
\partial_t\phi=(\partial^3_x +3u\partial_x
+\frac{3}{2}u_x)\phi,\quad u=z_x,t_3=t,t_1=x, \label{bKKlax2}
\end{equation}
gives the bi-directional Kaup-Kupershmidt (bKK) equation
\begin{equation}
\left(z_{xxxxx}+
15z_xz_{xxx}+15z_x^3-15z_xz_t-5z_{xxt}+\frac{45}{4}z_{xx}^2
\right)_x-5z_{tt}=0.\label{bKK}
\end{equation}

An essential characteristic of the KP hierarchy is the existence
of the $\tau$-function and all dynamical variables \{$u_i$\},
$i=2, 3, \ldots $, \  can be constructed from it \cite{dkjm,ost},
e.g.,
\begin{align}
&u_2=\frac{\partial^2}{\partial x^2}\log \tau,\\
&u_3=\frac{1}{2}\left(-\frac{\partial^3}{\partial
x^3}+\frac{\partial^2}{\partial x \partial t_2} \right)\log \tau, \\
&\quad  \vdots \notag
\end{align}
So it is a central task to construct the $\tau$-function in order to
solve the nonlinear PDEs associated with the KP hierarchy. In the
following, we will show that $\phi$ and $\psi$ play a key role in this
construction.

Gauge transformations \cite{zm,ms} offer an efficient route towards the
construction of the $\tau$ function of the KP hierarchy. In Ref.\
\cite{csy2} two kinds of such a gauge transformation have been proposed,
namely,
\begin{equation}
T_D(\phi)=\phi \partial \phi^{-1}, 
\quad
T_I(\psi)=\psi^{-1} \partial^{-1}\psi \label{gti}
\end{equation}
resulting in a very general and universal $\tau$ function (see equation
(3.17) of \cite{csy2} and also $IW_{k,n}$ in \cite{hlc1}).  The {\em
  determinant representation} of the gauge transformation operators with
$(n+k)$ steps is given in Ref.\ \cite{hlc1}.  In particular, the
Grammian $\tau$ function \cite{nakamura} of the KP hierarchy can be
generated by an iteration of the transformation \cite{csy2,os,njj}. This
is straightforwardly understood from Chau's $\tau$ function and the
determinant representation \cite{hlc1} if we impose a restriction on the
generating functions of the gauge transformation. Grammian $\tau$
function have also been used to solve the reduction of the constrained
BKP and CKP hierarchies \cite{il1,lw2,ow,hlc2}.

There are two issues that arise when one wants to study the
solutions of the $(1+1)$-dimensional solitons equations given by
the $n$-reduction of the BKP and CKP hierarchies. The first is how
it retain the restrictions, i.e.\ $L^{*}=-\partial L
\partial^{-1}$for BKP and $L^{*}=-L$ for CKP, for the transformed Lax operators
$L^{(1)}=TLT^{-1}$. In other words, the problem is how to obtain
the $\tau$-functions $\tau^{(n+k)}_{\rm BKP}$and
$\tau^{(n+k)}_{\rm CKP}$ from the general $\tau$-function
$\tau^{(n+k)}=IW_{k,n}\tau^{(0)}$ with the gauge transformation
$T_{n+k}$ of the KP hierarchy. Here $\tau^{(0)}$ is the initial
value of the $\tau$-function of the KP hierarchy. Also, the
generating functions $\phi_i$, $\psi_i$ of the gauge
transformation will be complex-valued and related to the $n$-th
roots of $e^{i\varepsilon}$. The second issue therefore is how to
choose generating functions $\phi_i=\phi(\lambda_i; x,t)$ and
$\psi_i=\psi(\mu_i; x,t)$) such that $\tau_{\rm BKP}^{(n+k)}$ and
$\tau^{(n+k)}_{\rm CKP}$ correspond to a {\em physical}
$\tau$-function $\hat{\tau}^{(n+k)}_{Eq}$, which is real and
positive on the full $(x,t)$ plane.

\begin{table}[bp]\label{tab-KPexamples}
\caption{{\small
  Examples of sub-hierarchies of the KP hierarchy, Lax operators used to
  construct them and resulting equations. The symbol $*$ indicates the
  conjugation, for example, $\partial^*=-\partial$.
  There are some abbreviations used in Table:
   Sawada-Kotera (SK), bi-directional Sawada-Kotera (bSK),
   Kaup-Kupershmidt (KK), bi-directional Kaup-Kupershmidt (bKK),
   Satsuma-Hirota (SH), Yajima-Oikawa (YO), Modified KdV (MKdV),
   Non-linear Schr\"{o}dinger (NLS),constrained KP(cKP).} }
\begin{center}
\begin{tabular}{@{}lll}
\cline{1-3}\\
sub-hierarchy & Lax operator & example equation  \\[0.5ex] \cline{1-3}\\[1ex]

BKP \cite{dkjm,jm}
& $L^*=-\partial L\partial^{-1}$
& SK\cite{sk,cdg}, bSK\cite{dp1,dp2}\\

CKP \cite{jm}
& $L^*=-L$
& KK\cite{kaup,kupershmidt}, bKK\cite{dp1,dp2}\\

$n$-thKdV \cite{gd}
& $L^n_{-}=0$
& KdV\cite{kdv}, Boussinesq-type\cite{ost}, SH\cite{sh} \\

&
&$n=2, 3, 4$ \\

cKP \cite{kss,cy2} & $L=\partial+\phi\partial^{-1} \psi$ &
YO\cite{yo},
MKdV\cite{wadati}, NLS\cite{zs2}\\[0.5ex]
\cline{1-3}
\end{tabular}

\end{center}
\end{table}

In fact, the bKK and bSK equations have been introduced recently
by Dye and Parker \cite{dp1,dp2} when looking for the
bidirectional soliton analogues of the Sawada-Kotera (SK)
\cite{sk,cdg} and Kaup-Kupershmidt (KK) \cite{kaup,kupershmidt}
equations. The Lax pairs of bKK and bSK related similarly as the
Lax pairs of KdV and Boussineq equation, thus ensuring their
integrability. Both bKK and bSK equation have a {\em
bidirectional} soliton solution \cite{dp1,dp2} which have been
obtained by the Hirota bilinear method \cite{hirota1}. The profile
of the bKK solitons depend on their direction of propagation. The
right-going solitons of bKK are standard one-peak solitons, but
the left-going solitons have two peaks.  Very recently, Verhoeven
and Musette \cite{vm1} have plotted the bi-directional solitons
for the bKK and bSH equation based on the Grammian $\tau$
function.

In this paper, we want to study why the $5$-reduction of the BKP
and CKP hierarchies have bidirectional soliton solutions, whereas
their $3$-reduction does not. As a first step, we will therefore
exhibit the relationship between the periodic, left-going and
right-going solitons of the $5$-reduction and the $5$-th roots of
$e^{i\varepsilon}$. In order to do so, we derive the $\tau$
functions of the BKP and CKP hierarchies in sections 2-4. The
explicit formulas of the corresponding $\tau$-functions for
solitons as well as for the periodic solutions of bSK and bKK are
given and the two-peak soliton is discussed in detail. In section
5, we will prove that no two-peak solitons exist for the bSH
equation. The one-peak soliton has bi-directional motion and we
also obtain the periodic and two-soliton solutions. In section 6,
we will discuss the lower and higher-order reductions of BKP and
CKP hierarchies and also the $n=$even-reductions of the KP
hierarchy. We will show that the soliton of the $(2j+1)$-reduction
of BKP and CKP hierarchies can move along $j$ directions ($j=1, 2,
\dots$), investigate the relationship with the symmetric
distribution of the $(2j+1)$-th roots of $e^{i\varepsilon}$. In
particular, we will obtain the "stationary" soliton for the higher
reduction of the KP hierarchy. For the higher-order equation and
even-reduction of KP hierarchy, we can again find a two-peak
soliton.

\section{$\tau$ functions for BKP and CKP hierarchies}
\label{sec-tau-BKP-CKP}
Let us first define the generalized Wronskian determinant
\begin{align}
IW_{k,n} &\equiv IW_{k,n}(\fun{g}{0}{k}, \fun{g}{0}{k-1}, \cdots,
\fun{g}{0}{1};
      \fun{f}{0}{1}, \fun{f}{0}{2},\cdots , \fun{f}{0}{n} ) \notag
      \\[1ex]
&=\left|
\begin{array}{ccccc}
\int \fun{g}{0}{k}\cdot\fun{f}{0}{1} & \int
\fun{g}{0}{k}\cdot\fun{f}{0}{2} &\int
\fun{g}{0}{k}\cdot\fun{f}{0}{3} &\cdots
&\int \fun{g}{0}{k}\cdot\fun{f}{0}{n}\\  
\int \fun{g}{0}{k-1}\cdot\fun{f}{0}{1} & \int
\fun{g}{0}{k-1}\cdot\fun{f}{0}{2} &\int
\fun{g}{0}{k-1}\cdot\fun{f}{0}{3} &\cdots
&\int \fun{g}{0}{k-1}\cdot\fun{f}{0}{n}\\  
\vdots&\vdots&\vdots&\cdots&\vdots\\        
\int \fun{g}{0}{1}\cdot\fun{f}{0}{1} & \int
\fun{g}{0}{1}\cdot\fun{f}{0}{2} &\int
\fun{g}{0}{1}\cdot\fun{f}{0}{3} &\cdots
&\int \fun{g}{0}{1}\cdot\fun{f}{0}{n}\\  
\fun{f}{0}{1}&  \fun{f}{0}{2}&\fun{f}{0}{3}
& \cdots&\fun{f}{0}{n}\\
\fun{f}{0}{1,x}&\fun{f}{0}{2,x}&\fun{f}{0}{3,x}
&\cdots&\fun{f}{0}{n,x}\\
\vdots&\vdots&\vdots&\cdots&\vdots\\        
(\fun{f}{0}{1})^{(n-k-1)}&(\fun{f}{0}{2})^{(n-k-1)}
&(\fun{f}{0}{3})^{(n-k-1)}&\cdots&(\fun{f}{0}{n})^{(n-k-1)}
\end{array}
\right|
\end{align}
In particular, $IW_{0,n}= W_n \left( f_1^{(0)}, f_2^{(0)}, \cdots,
  f_n^{(0)} \right)$ with
$\left(\fun{f}{0}{i}\right)^{(k)}=\frac{\partial^k \fun{f}{0}{i}
} {\partial x^k}$ is the usual Wronskian determinant of functions
$\{\fun{f}{0}{1},\fun{f}{0}{2},\cdots, \fun{f}{0}{n}\}$. We shall
also use the abbreviation $\int f =\int f dx$ with integration
constant equal to zero.
\begin{lemma}[\cite{csy2,hlc1}] \label{lemtaukp}
  The $\tau$ function of the KP hierarchy generated by the gauge
  transformation $T_{n+k}$ is given as
\begin{equation}\tu{n+k}
=IW_{k,n}(\fun{\psi}{0}{k}, \fun{\psi}{0}{k-1}, \cdots,
\fun{\psi}{0}{1};
      \fun{\phi}{0}{1}, \fun{\phi}{0}{2},\cdots , \fun{\phi}{0}{n} )\tu{0}\label{taukp},
\end{equation}
where
$\left(\fun{\phi}{0}{i},\fun{\psi}{0}{j}\right)=\left(\phi(\lambda_i;
  \stackrel{\_}{t}), \psi(\mu_j; \stackrel{\_}{t})\right)$ are solutions
of equation (\ref{zslax1}) with initial value $\tu{0}$ for the
$\tau$-function and the initial values of the $\{u_i\}$ are
$\{\fun{u}{0}{i}\}$. Here $\{\fun{\phi}{0}{i},\fun{\psi}{0}{j}\}$
are called generating functions(GFs) of gauge
  transformation.
\end{lemma}

Let us now discuss how to reduce the $\tu{n+k}$ in (\ref{taukp})
to the $\tau$ function of the BKP hierarchy.  The key problem is
how to keep the restriction $( L^{(n+k)})^*=-\partial
L^{(n+k)}\partial^{-1}$ under the gauge transformation $T_{n+k}$
\cite{hlc1}. It should be noted that $\bar{t}=\left( t_1, t_3,
t_5, \cdots \right)$ in BKP hierarchy.
\begin{proposition}[see Refs.\ \cite{njj,am}] \label{proptauBKP}
  \begin{enumerate}
  \item The Lax operator transforms as
    $L^{(n+k)}=T_{n+k}LT_{n+k}^{-1}$ under the gauge transformation
    $T_{n+k}$ with $n=k$ and generating functions
    $\fun{\psi}{0}{i}=\fun{\phi}{0}{i,x}$ for $i=1, 2, \cdots, n$.

  \item The $\tau$ function $\fun{\tau}{n+n}{\rm BKP}$ of the {\rm BKP} hierarchy
    is
\begin{eqnarray}
\fun{\tau}{n+n}{\rm BKP} &= & IW_{n,n}(\fun{\phi}{0}{n,x},
\fun{\phi}{0}{n-1,x}, \cdots, \fun{\phi}{0}{1,x};
      \fun{\phi}{0}{1}, \fun{\phi}{0}{2},\cdots , \fun{\phi}{0}{n} )
\end{eqnarray}
\begin{equation}
= \left|
\begin{array}{lccccc}
\int \fun{\phi}{0}{n,x}\cdot\fun{\phi}{0}{1} &
\int\fun{\phi}{0}{n,x}\cdot\fun{\phi}{0}{2} &
\int\fun{\phi}{0}{n,x}\cdot\fun{\phi}{0}{3} &\cdots
&\int \fun{\phi}{0}{n,x}\cdot\fun{\phi}{0}{n-1} &\frac{1}{2}(\fun{\phi}{0}{n})^2\\  
\int \fun{\phi}{0}{n-1,x}\cdot\fun{\phi}{0}{1} & \int
\fun{\phi}{0}{n-1,x}\cdot\fun{\phi}{0}{2} &\int
\fun{\phi}{0}{n-1,x}\cdot\fun{\phi}{0}{3} &\cdots
& \frac{1}{2}(\fun{\phi}{0}{n-1})^2 &\int \fun{\phi}{0}{n-1,x}\cdot\fun{\phi}{0}{n}\\  
 \mbox{}\hspace{0.5cm}\vdots&\vdots&\vdots&\cdots&\vdots &\vdots \\        
 \int \fun{\phi}{0}{2,x}\cdot\fun{\phi}{0}{1} &
  \frac{1}{2}(\fun{\phi}{0}{2})^{2}&\int
\fun{\phi}{0}{2,x}\cdot\fun{\phi}{0}{3} &\cdots
 &\int\fun{\phi}{0}{2,x}\cdot\fun{\phi}{0}{n-1} &\int
\fun{\phi}{0}{2,x}\cdot\fun{\phi}{0}{n}\\ 
\frac{1}{2}(\fun{\phi}{0}{1})^2 & \int
\fun{\phi}{0}{1,x}\cdot\fun{\phi}{0}{2} &\int
\fun{\phi}{0}{1,x}\cdot\fun{\phi}{0}{3} &\cdots
 &\int\fun{\phi}{0}{1,x}\cdot\fun{\phi}{0}{n-1} &\int
\fun{\phi}{0}{1,x}\cdot\fun{\phi}{0}{n}
\end{array} \right|\tu{0}_{\rm BKP}
\end{equation}
  \end{enumerate}
\end{proposition}
\begin{proof}
  \begin{enumerate}
  \item It is clear that a single step of the gauge transformations
    $T_D$ or $T_I$ can not keep the restriction. So we use
\begin{equation}
T \equiv T_{1+1}=
 T_I\left(\fun{\psi}{1}{1}\right) \cdot
 T_D\left(\fun{\phi}{0}{1}\right)
\end{equation}
such that the lax operator is $L^{(2)} = TLT^{-1}$. Let us check whether
it satisfies the required restriction
\begin{equation}
\left(L^{(2)}\right)^*=-\partial L^{(2)}\partial^{-1} \label{BKPL2}
\end{equation}
which means in terms of $T$ that
\begin{equation}
T_D\left(\fun{\psi}{1}{1}\right)T_I\left(\fun{\phi}{0}{1}\right)\partial
=\partial
T_I\left(\fun{\psi}{1}{1}\right)T_D\left(\fun{\phi}{0}{1}\right).\label{BKPTDTI}
\end{equation}
Based on the determinant representation of $T$ \cite{hlc1} we see
from (\ref{BKPTDTI}) that
\begin{eqnarray}
\mbox{r.h.s} &= &\partial -\left(\frac{\fun{\phi}{0}{1}}{\int
\fun{\phi}{0}{1}\fun{\psi}{0}{1}}\right)_x\partial^{-1}\fun{\psi}{0}{1}-
\frac{\fun{\phi}{0}{1}\fun{\psi}{0}{1}  }{\int
\fun{\phi}{0}{1}\fun{\psi}{0}{1}},  \\
\mbox{l.h.s} &= &\partial + \left(\frac{\fun{\psi}{0}{1}}{\int
\fun{\phi}{0}{1}\fun{\psi}{0}{1}}\right)\partial^{-1}\fun{\phi}{0}{1,x}-
\frac{\fun{\phi}{0}{1}\fun{\psi}{0}{1}  }{\int
\fun{\phi}{0}{1}\fun{\psi}{0}{1}}.
\end{eqnarray}
This implies $\fun{\psi}{0}{1}=\fun{\phi}{0}{1,x}$. So we have
seen that in order to keep the restriction of the Lax operator, we
have to regard $T=T_{1+1}$ as basic building block in iteration of
the gauge transformations $T_{n+k}$. In particular,
\begin{equation}
\begin{array}{l}
T_{2+2}=T_I\left(\fun{\phi}{3}{2,x}\right)T_D\left(\fun{\phi}{2}{2}\right)
T_I\left(\fun{\phi}{1}{1,x}\right)T_D\left(\fun{\phi}{0}{1}\right),\\[1ex]
T_{3+3}=T_I\left(\fun{\phi}{5}{3,x}\right)T_D\left(\fun{\phi}{4}{3}\right)
T_I\left(\fun{\phi}{3}{2,x}\right)T_D\left(\fun{\phi}{2}{2}\right)
T_I\left(\fun{\phi}{1}{1,x}\right)T_D\left(\fun{\phi}{0}{1}\right),
\end{array}
\end{equation}
and so on such that $k=n$ and
$\fun{\psi}{0}{i}=\fun{\phi}{0}{i,x}$ for $i=1, 2, \ldots, n$.
\item According to the determinant of $T_{n+k}$\cite{hlc1} and
  $\tu{n+k}$\cite{csy2,hlc1} with $k=n$ and
  $\fun{\psi}{0}{i}=\fun{\phi}{0}{i,x}$, $i=1, 2, \cdots, n$,
  $\tu{n+n}_{\rm BKP}$ can be obtained directly from $\tu{n+k}$ as in
  Lemma \ref{lemtaukp}.
  \end{enumerate}
\end{proof}

For the CKP hierarchy, we have again $\bar{t}=( t_1, t_3, t_5, \ldots)$ and the restriction is $( L^{(n+k)})^*=- L^{(n+k)}$.
\begin{proposition}[see Refs.\ \cite{njj,am}]\label{proptauCKP}
  \begin{enumerate}
  \item The appropriate gauge transformation $T_{n+k}$ is given by $n=k$
    and generating functions
    $\fun{\psi}{0}{i}=\fun{\phi}{0}{i}$ for $i=1,2,\ldots,n$.
  \item The $\tau$ function $\fun{\tau}{n+n}{\rm CKP}$ of the {\rm CKP} hierarchy
    has the form
\begin{eqnarray}
\fun{\tau}{n+n}{\rm CKP} &= &IW_{n,n}(\fun{\phi}{0}{n},
\fun{\phi}{0}{n-1}, \cdots, \fun{\phi}{0}{1};
      \fun{\phi}{0}{1}, \fun{\phi}{0}{2},\cdots , \fun{\phi}{0}{n} )
\end{eqnarray}
\begin{equation}
= \left|
\begin{array}{lccccc}
\int \fun{\phi}{0}{n}\cdot\fun{\phi}{0}{1} &
\int\fun{\phi}{0}{n}\cdot\fun{\phi}{0}{2} &
\int\fun{\phi}{0}{n}\cdot\fun{\phi}{0}{3} &\cdots
&\int \fun{\phi}{0}{n}\cdot\fun{\phi}{0}{n-1} &\int\fun{\phi}{0}{n}\cdot\fun{\phi}{0}{n}\\  
\int \fun{\phi}{0}{n-1}\cdot\fun{\phi}{0}{1} & \int
\fun{\phi}{0}{n-1}\cdot\fun{\phi}{0}{2} &\int
\fun{\phi}{0}{n-1}\cdot\fun{\phi}{0}{3} &\cdots
& \int\fun{\phi}{0}{n-1}\cdot\fun{\phi}{0}{n-1} &\int \fun{\phi}{0}{n-1}\cdot\fun{\phi}{0}{n}\\  
\vdots&\vdots&\vdots&\cdots&\vdots\\        
 \int \fun{\phi}{0}{2}\cdot\fun{\phi}{0}{1} &
  \int\fun{\phi}{0}{2}\cdot\fun{\phi}{0}{2}&\int
\fun{\phi}{0}{2}\cdot\fun{\phi}{0}{3} &\cdots
 &\int\fun{\phi}{0}{2}\cdot\fun{\phi}{0}{n-1} &\int
\fun{\phi}{0}{2}\cdot\fun{\phi}{0}{n}\\ 
\int\fun{\phi}{0}{1}\cdot\fun{\phi}{0}{1} & \int
\fun{\phi}{0}{1}\cdot\fun{\phi}{0}{2} &\int
\fun{\phi}{0}{1}\cdot\fun{\phi}{0}{3} &\cdots
 &\int\fun{\phi}{0}{1}\cdot\fun{\phi}{0}{n-1} &\int
\fun{\phi}{0}{1}\cdot\fun{\phi}{0}{n}
\end{array} \right|\tu{0}_{\rm CKP}
\end{equation}
  \end{enumerate}
\end{proposition}
\begin{proof}
  \begin{enumerate}
  \item Similar to the BKP hierarchy we have to try the two-step gauge
    transformation
\begin{equation}
T\equiv T_{1+1}=
T_I\left(\fun{\psi}{1}{1}\right)T_D\left(\fun{\phi}{0}{1}\right).
\end{equation}
With $L^{(2)}=TLT^{-1}$, the restriction
$\left(L^{(2)}\right)^*=-L^{(2)}$ then implies
\begin{equation}
T_D\left(\fun{\psi}{1}{1}\right)T_I\left(\fun{\phi}{0}{1}\right)=
T_I\left(\fun{\psi}{1}{1}\right)T_D\left(\fun{\phi}{0}{1}\right).\label{CKPTDTI}
\end{equation}
Based on the determinant representation of $T$ \cite{hlc1}, we
find from (\ref{CKPTDTI}) that
\begin{eqnarray}
\mbox{r.h.s} &= &\partial -\frac{\fun{\phi}{0}{1}}{\int
\fun{\phi}{0}{1}\fun{\psi}{0}{1}}\partial^{-1}\fun{\psi}{0}{1},  \\
\mbox{l.h.s} &= &\partial - \frac{\fun{\psi}{0}{1}}{\int
\fun{\phi}{0}{1}\fun{\psi}{0}{1}}\partial^{-1}\fun{\phi}{0}{1}.
\end{eqnarray}
Then $\fun{\psi}{0}{1}=\fun{\phi}{0}{1}$. Again, we have to regard
$T=T_I\left(\fun{\phi}{1}{1}\right)T_D\left(\fun{\phi}{0}{1}\right)$
as basic building block such that
\begin{equation}
\begin{array}{l}
T_{2+2}=T_I\left(\fun{\phi}{3}{2}\right)T_D\left(\fun{\phi}{2}{2}\right)
T_I\left(\fun{\phi}{1}{1}\right)T_D\left(\fun{\phi}{0}{1}\right),\\[1ex]
T_{3+3}=T_I\left(\fun{\phi}{5}{3}\right)T_D\left(\fun{\phi}{4}{3}\right)
T_I\left(\fun{\phi}{3}{2}\right)
T_D\left(\fun{\phi}{2}{2}\right)T_I\left(\fun{\phi}{1}{1}\right)T_D\left(\fun{\phi}{0}{1}\right),
\end{array}
\end{equation}
so $k=n$ and $\fun{\psi}{0}{i}=\fun{\phi}{0}{i}$ for
$i=1,2,\ldots,n$. \item According to the determinant of $T_{n+k}$
\cite{hlc1} and
  $\tu{n+k}$ \cite{csy2,hlc1} with $k=n$ and
  $\fun{\psi}{0}{i}=\fun{\phi}{0}{i}$, $i=1, 2, \cdots, n$,
  $\tu{n+n}_{\rm CKP}$ is obtained directly from $\tu{n+k}$ in Lemma
  \ref{lemtaukp}.
  \end{enumerate}
\end{proof}
In fact, we can let $\fun{\psi}{0}{i}=c_i \fun{\phi}{0}{i,x}$ (or
  $\fun{\psi}{0}{i}=c_i \fun{\phi}{0}{i}$) with constants $c_i$. However, the new
  $\tu{n+n}_{\rm BKP}$ (or $\tu{n+n}_{\rm CKP}$ ) associated with
  $\fun{\psi}{0}{i}=c_i \fun{\phi}{0}{i,x}$ (or $\fun{\psi}{0}{i}=c_i
  \fun{\phi}{0}{i}$) is equivalent to the ones in Proposition
  \ref{proptauBKP} (or Proposition \ref{proptauCKP}).
  Although Refs.\ \cite{njj,am} have results similar to our Propositions
  \ref{proptauBKP} and \ref{proptauCKP}, our approach is more direct and
  simpler for the construction $\tu{n+n}_{\rm BKP}$ and $\tu{n+n}_{\rm CKP}$.
If the initial values of dynamical variables $\{u_i\}$ of BKP(CKP)
hierarchy are zero, then the equations in (\ref{zslax1})  of
$\left(\fun{\phi}{0}{i},\fun{\psi}{0}{j}\right)=\left(\phi(\lambda_i;
\stackrel{\_}{t}), \psi(\mu_j; \stackrel{\_}{t})\right)$ become
more simpler as
\begin{align}
& \frac{\partial \phi(\lambda;\stackrel{\_}{t})}{\partial t_n}
=\left[\partial_x^n\phi(\lambda;\stackrel{\_}{t})\right],
\quad \stackrel{\_}{t}=\left(t_1,t_3,t_5,\cdots \right),  \label{zerozslax1}\\
& \frac{\partial \psi(\mu;\stackrel{\_}{t})}{\partial
t_n}=(-1)^{n+1}\left[\partial_x^{n}\psi(\mu;\stackrel{\_}{t})\right],
\quad \stackrel{\_}{t}=\left(t_1,t_3,t_5,\cdots \right),
\label{zerozslax2}
\end{align}
and $\tu{0}_{\rm BKP}=1(\tu{0}_{\rm CKP}=1)$. Last, we note that
for the generalized KP (gKP) hierarchy with Lax operator
$\hat{L}=L^n$, $n=2, 4, 6, 8, \ldots$, and $\hat{L}^*=\hat{L}$,
the $\tau$ function $\tu{n+k}_{\rm gKP}$ generated by gauge
transformations $T_{n+k}$ has the same form as for the CKP
hierarchy. This result will afford a simple way to construct the
$\tau$ function of bSH equation in Section \ref{sectbSH}.

\section{Soliton solutions of the bSK equation}

As pointed out in the introduction, there are two steps en route
from a $\tau$ function $\tu{n+k}$ generated by the gauge
transformations $T_{n+k}$ of the KP hierarchy to the $\tau$
function of equations as the $n$-reduction of BKP or CKP
hierarchies.  The second step is to build physical $\tau$
functions from the complex-valued $\tu{n+n}_{\rm BKP}$ and
$\tu{n+n}_{\rm CKP}$ constructed in the last section. In the
following Sections, we will illustrate our approach by computing
the $\tau$ function for the $5$-reduction of BKP and CKP, i.e.,
for the bSK and bKK equations.

The $5$-reduction of the BKP hierarchy is the bSK equation (\ref{bSK}).
Assume for the initial value $u = 0$ in equations (\ref{bSKlax1}) and
(\ref{bSKlax2}), then $\fun{\phi}{0}{i}=\phi(\lambda_i;x,t)$ are
solutions of
\begin{equation} \label{zerobSKlax1}
\partial_x^5\phi(\lambda_i;x,t)=\lambda_i
\phi(\lambda_i;x,t),\quad \frac{\partial
\phi(\lambda_i;x,t)}{\partial
t}=\left(\partial_x^{3}\phi(\lambda_i;x,t)\right). 
\end{equation}
So proposition \ref{proptauBKP} with $\tu{0}_{\rm BKP}=1$ implies
that the $\tau$ function of bSK is given as follows.
\begin{proposition}\label{propzerotaubSK}
  The $\tau$ function of the {\rm bSK} equation generated by $T_{n+n}$ from
  initial value $1$ is
\begin{equation}\label{zeronntaubSK}
\fun{\tau}{n+n}{\rm bSK}=IW_{n,n}\left(\fun{\phi}{0}{n,x},
\fun{\phi}{0}{n-1,x}, \cdots,
  \fun{\phi}{0}{1,x};
  \fun{\phi}{0}{1}, \fun{\phi}{0}{2},\cdots , \fun{\phi}{0}{n} \right)
\end{equation}
and the solution of the {\rm bSK} equation generated by $T_{n+n}$
from initial value $0$ is
\begin{equation}
u=\partial_x^2\left(\log \fun{\tau}{n+n}{\rm bSK}\right)
\label{zeroubSK}
\end{equation}
Here $\fun{\phi}{0}{i}=\phi(\lambda_i;x,t)$ are solutions of equation
(\ref{zerobSKlax1}).
\end{proposition}
In general, this $\tau$ function $\fun{\tau}{n+n}{\rm bSK}$ for
bSK is complex and related to $5$-th roots of $e^{i\varepsilon}$.
We have to find the real and non-zero $\tau$ function from it such
that $u$ in equation (\ref{zeroubSK}) is a real and smooth solution of
bSK. This is main task of this section. We start by analysing the
solution $\phi(\lambda;x,t)$ of equation (\ref{zerobSKlax1})and make
the universal ansatz
\begin{equation}
\phi(\lambda;x,t)=\sum\limits_{j=1}^5 A_j e^{xp_j+tp_j^3},
\quad\mbox{with}\quad
p_j^5= \lambda.
\end{equation}
Here $p_j=k\exp\left({\frac{\varepsilon+2 \pi j }{5}i}\right)$,
$k^5=|\lambda|,k\in \mathbb{R}$, $0\leq \varepsilon < 2\pi$ and
$j=0,1,2,3,4$. There are two important ingredients which we can
use to find the desired solution.  The first is that the $5$-th
roots $\varepsilon_j= \exp\left({\frac{\varepsilon+2 \pi
j}{5}i}\right) $ of $e^{i\varepsilon}$ are distributed uniformly
on the unit circle in $\mathbb{C}$. So for a suitable value of
$\varepsilon$ there exist combinations of $p_j$'s which are
symmetric upon reflection on the $x$-axes; similarly for the
$y$-axes for other values of $\varepsilon$. The second ingredient
is that $\tau_{\rm bSK}$ and $\exp\left({\alpha x+\beta
t}\right)\tau_{\rm bSK}$ will imply the same solution $u$ since
$u=\partial^2_x \log \tau_{\rm bSK}$. Here, $\alpha$ and $\beta$
are arbitrary, complex constants. Therefore we can obtain the
desired real and smooth solutions of the bSK if $\tau_{\rm bSK}$
can be expressed as $\tau_{\rm bSK}=e^{\alpha x +\beta t}
\hat{\tau}_{\rm bSK}\approxeq \hat{\tau}_{\rm bSK}$, in which
$\hat{\tau}_{\rm bSK}$ is a real and nonzero function although
$\tau_{\rm bSK}$ is complex. We call $\hat{\tau}_{\rm bSK}$ the
{\em physical} $\tau$ function for the bSK equation. Based on the
above arguments, let us make the refined ansatz
\begin{align}
&\phi(\lambda_1;x,t)=A_1 e^{p_1 x+ p_1^3 t}+B_1 e^{q_1 x+ q_1^3
t},\quad p_1=k_1
e^{i \varepsilon_1}, q_1=-k_1 e^{-i \varepsilon_1}, k_1^5=|\lambda_1|,k_1\in \mathbb{R}, \label{zerophi1a} \\
\intertext{or}
 &\phi(\lambda_1;x,t)=A_1 e^{p_1 x+ p_1^3 t}+B_1
e^{q_1 x+ q_1^3 t},\quad p_1=k_1 e^{i \varepsilon_1}, q_1= k_1
e^{-i \varepsilon_1},k_1^5=|\lambda_1|,k_1\in \mathbb{R},
\label{zerophi1b}
\end{align}
and in the next step we need to fix the ratio $\frac{B_1}{A_1}$.
We stress that the above analysis is also true for the derivation
of the bKK equation.
\begin{proposition}\label{proponesolitonbSK}
Define $\xi_1= xk_1\cos \varepsilon_1+tk_1^3\cos 3\varepsilon_1$.
Then the physical $\tau$ function of {\rm bSK} generated by
$T_{1+1}$ is
\begin{equation}
\hat{\tau}^{(1+1)}_{\rm bSK}= e^{2
\xi_1}+\left(\frac{B_1}{A_1}\right)^2 e^{-2
\xi_1}+2\left(\frac{B_1}{A_1}\right)
\end{equation}
and the corresponding one soliton $u =\partial^2_x
\log\hat{\tau}^{(1+1)}_{\rm bSK}$ is given as
\begin{equation}
u=\frac{16\left(\frac{B_1}{A_1}\right)^2 k_1^2\cos^2\varepsilon_1}
{\left[ e^{-2 \xi_1}+\left(\frac{B_1}{A_1}\right)^2 e^{-2 \xi_1}+2
\left(\frac{B_1}{A_1}\right)\right]^2}\ .  \label{onesolitonbSK}
\end{equation}
Here $\frac{B_1}{A_1}>0$. The velocity of the moving soliton is
$v=-k_1^2\frac{\cos 3\varepsilon_1 }{\cos \varepsilon_1}$ and can
be both positive and negative depending on the choice of
$\varepsilon_1$. Specifically, we have
$v_{\_}=v|_{\varepsilon_1=\frac{\pi}{10}} < 0$ and
$v_{+}=v|_{\varepsilon_1=\frac{3\pi}{10}} > 0$.
\end{proposition}
\begin{proof}
\begin{equation}
\tu{1+1}_{\rm bSK}=\left(\fun{\phi}{0}{1}\right)^2= A_1^2
e^{2i(xk_1\sin \varepsilon_1+tk_1^3\sin 3\varepsilon_1)}\left[
e^{2 \xi_1}+\left(\frac{B_1}{A_1}\right)^2 e^{-2
\xi_1}+2\left(\frac{B_1}{A_1}\right) \right]
\end{equation}
and $\fun{\phi}{0}{1}=\phi(\lambda_1;x,t)$ defined by equation
(\ref{zerophi1a}).
\end{proof}
  Let us point out a relation between the distribution of the $5$-th
  roots of $e^{i\varepsilon}$ and the direction of movement for the soliton.
  \begin{enumerate}
  \item
    $\left(e^{i\varepsilon_1}|_{\varepsilon_1=\pi/10},
    -e^{-i\varepsilon_1}|_{\varepsilon_1=\pi/10}\right)$ $\longrightarrow$
    one distribution of 5-th roots of $e^{i\varepsilon}$ $\longrightarrow$
    $(p_1=k_1e^{i\varepsilon_1}|_{\varepsilon_1=\pi/10},\\ q_1=-k_1e^{-i\varepsilon_1}|_{\varepsilon_1=\pi/10})$
    in equation (\ref{zerophi1a})$\longrightarrow$ $v|_{\varepsilon_1=\pi/10} < 0$, left-going soliton $u$ in equation (\ref{onesolitonbSK}) ;
  \item
    $\left(e^{i\varepsilon_1}|_{\varepsilon_1=3\pi/10},
    -e^{-i\varepsilon_1}|_{\varepsilon_1=3\pi/10}\right)$
    $\longrightarrow$
    another distribution of 5-th roots of $e^{i\varepsilon}$ $\longrightarrow$
    $(p_1=k_1e^{i\varepsilon_1}|_{\varepsilon_1=3\pi/10},\\ q_1=-k_1e^{-i\varepsilon_1}|_{\varepsilon_1=3\pi/10})$
    in equation (\ref{zerophi1a}) $\longrightarrow $
$v|_{\varepsilon_1=3\pi/10} > 0$, right-going soliton $u$ in
equation (\ref{onesolitonbSK}) .
  \end{enumerate}

We can see from equation (\ref{onesolitonbSK}) that the one-soliton of
bSK has only one peak in its profile.  The process of generating a
two-soliton by $T_{2+2}$ is more complicated.
\begin{lemma}\label{lemtau2+2bSK}
  With $\phi(\lambda_1;x,t)$ and $\phi(\lambda_2;x,t)$ as in equation
  (\ref{zerophi1a}), and using Proposition \ref{propzerotaubSK},
  $\tu{2+2}_{\rm bSK}$ is given by
\begin{align}\label{taubSK(2+2)}
\tu{2+2}_{\rm bSK}&=A^2_1A^2_2
e^{2i\left[x(k_1\sin\varepsilon_1+k_2\sin\varepsilon_2) +t(k_1^3\sin
3\varepsilon_1+k_2^3\sin 3\varepsilon_2)\right]} \times
\nonumber \\ 
&\Big\{ \frac{(4 z_1-f_1) e^{2(\xi_1+\xi_2)}
}{4\left[k^2_1+k^2_2+2k_1k_2\cos(\varepsilon_1-\varepsilon_2)
\right]^2} + \frac{(4z^*_1 -f_1)e^{-2(\xi_1+\xi_2)}
}{4\left[k^2_1+k^2_2+2k_1k_2\cos(\varepsilon_1-\varepsilon_2)
\right]^2}\left(\frac{B_1}{A_1}\right)^2\left(\frac{B_2}{A_2}\right)^2
\nonumber \\ 
&+ \frac{-(4z_3+f_3) e^{2(\xi_1-\xi_2)}
}{4\left[k^2_1+k^2_2-2k_1k_2\cos(\varepsilon_1+\varepsilon_2)
\right]^2}\left( \frac{B_2}{A_2}\right)^2 + \frac{-(4z^*_3+f_3)
e^{-(2\xi_1-\xi_2)}
}{4\left[k^2_1+k^2_2-2k_1k_2\cos(\varepsilon_1+\varepsilon_2)
\right]^2}\left( \frac{B_1}{A_1}\right)^2
\nonumber \\  
&+\frac{(2iz_5-f_5)e^{2\xi_1}}
 {2\left[k^2_1+k^2_2+2k_1k_2\cos(\varepsilon_1-\varepsilon_2)
\right]\left[k^2_1+k^2_2-2k_1k_2\cos(\varepsilon_1+\varepsilon_2)
\right]}\left(\frac{B_2}{A_2}\right)
\nonumber \\ 
&+\frac{(-2iz^*_5-f_5)e^{-2\xi_1}}
 {2\left[k^2_1+k^2_2+2k_1k_2\cos(\varepsilon_1-\varepsilon_2)
\right]\left[k^2_1+k^2_2-2k_1k_2\cos(\varepsilon_1+\varepsilon_2)
\right]}\left(\frac{B_1}{A_1}\right)^2\left(\frac{B_2}{A_2}\right)
\nonumber  \\  
&+\frac{(-2iz_7-f_5)e^{2\xi_2}}
 {2\left[k^2_1+k^2_2+2k_1k_2\cos(\varepsilon_1-\varepsilon_2)
\right]\left[k^2_1+k^2_2-2k_1k_2\cos(\varepsilon_1+\varepsilon_2)
\right]}\left(\frac{B_1}{A_1}\right)
\nonumber  \\  
&+\frac{(2iz^*_7-f_5)e^{-2\xi_2}}
 {2\left[k^2_1+k^2_2+2k_1k_2\cos(\varepsilon_1-\varepsilon_2)
\right]\left[k^2_1+k^2_2-2k_1k_2\cos(\varepsilon_1+\varepsilon_2)
\right]}\left(\frac{B_1}{A_1}\right)\left(\frac{B_2}{A_2}\right)^2
\nonumber  \\  
&+\frac{-(k^2_1+k^2_2)^2}
 {2\left[k^2_1+k^2_2+2k_1k_2\cos(\varepsilon_1-\varepsilon_2)
\right]\left[k^2_1+k^2_2-2k_1k_2\cos(\varepsilon_1+\varepsilon_2)
\right]}\left(\frac{B_1}{A_1}\right)\left(\frac{B_2}{A_2}\right)        
\Big\}
\end{align}
Here the $z_i$, $i= 1,3,5,7$ are given in \ref{appbSK} and
$f_1=\left[k^2_1+k^2_2+2k_1k_2\cos(\varepsilon_1-\varepsilon_2)
\right]^2$,\\
$f_3=\left[k^2_1+k^2_2-2k_1k_2\cos(\varepsilon_1+\varepsilon_2)
\right]^2$ and $f_5=\sqrt{f_1f_3}$, as well as $\xi_i=x k_i\cos
\varepsilon_i+tk_i^3\cos 3\varepsilon_i$ for $i=1$ and $2$.
\end{lemma}

We now need to find a suitable solution of $\frac{B_i}{A_i}, i=1, 2$
such that the summation of terms in the $\{ \}$ bracket of equation
(\ref{taubSK(2+2)}) is a positive function on the whole (x,t) plane. The
following two lemmas is useful. Let $z_1^{\prime}=4z_1-f_1,
z_3^{\prime}=4z_3+f_3, z_5^{\prime}=2iz_5-f_5, z_7^{\prime}=-2iz_7-f_5$,
and $z_i, i=1, 3, 5, 7$ given in \ref{appbSK}.
\begin{lemma}For $z_i^{\prime}, i=1, 3, 5, 7$ there exist relations
\begin{equation}
-z_1^{\prime}z_3^{\prime}=\left(z_5^{\prime}\right)^2;\quad
-z_1^{\prime}\left(z_3^{\prime}\right)^*=\left(z_7^{\prime}\right)^2.
\end{equation}
\end{lemma}
\begin{lemma}\label{lemA1B1A2B2bSK}
Let $\frac{B_1}{A_1}=\frac{z_1^{\prime}}{z_7^{\prime}},
\frac{B_2}{A_2}=\frac{z_1^{\prime}}{z_5^{\prime}}$, and
$g_2=\frac{|z_1^{\prime}|^2}{|z_3^{\prime}|^2},\
g_6=g_8=\frac{|z_1^{\prime}|^2}{|z_5^{\prime}|^2},\
 g_9=\frac{1}{|z_3^{\prime}|}$ then
\begin{align}
&-\frac{z_3^{\prime}}{z_1^{\prime}}\left(\frac{B_2}{A_2}
\right)^2=1,
&-\frac{(z_3^{\prime})^*}{z_1^{\prime}}\left(\frac{B_1}{A_1}\right)^2=1,    \\
&\frac{(z_1^{\prime})^*}{z_1^{\prime}}\left(\frac{B_1}{A_1}\right)^2\left(\frac{B_2}{A_2}\right)^2=g_2,
&\frac{-1}{z_1^{\prime}}\left(\frac{B_1}{A_1}
\right)\left(\frac{B_2}{A_2} \right)=g_9,   \\ 
&\frac{(z_5^{\prime})^*}{z_1^{\prime}}\left(\frac{B_1}{A_1}
\right)^2\left(\frac{B_2}{A_2} \right)=g_6,
&\frac{(z_7^{\prime})^*}{z_1^{\prime}}\left(\frac{B_1}{A_1}\right)
\left(\frac{B_2}{A_2}\right)^2=g_8,    
\end{align}
hold.
\end{lemma}
The $\frac{B_i}{A_i}, i=1, 2$ in Lemma \ref{lemA1B1A2B2bSK} are
what we are looking for, and then the {\em physical} $\tau$
function $\hat{\tau}^{(2+2)}_{\rm bSK}$ is give by following
proposition.

\begin{figure}[htbp]
  \centerline{
  \includegraphics[width=0.5\textwidth]{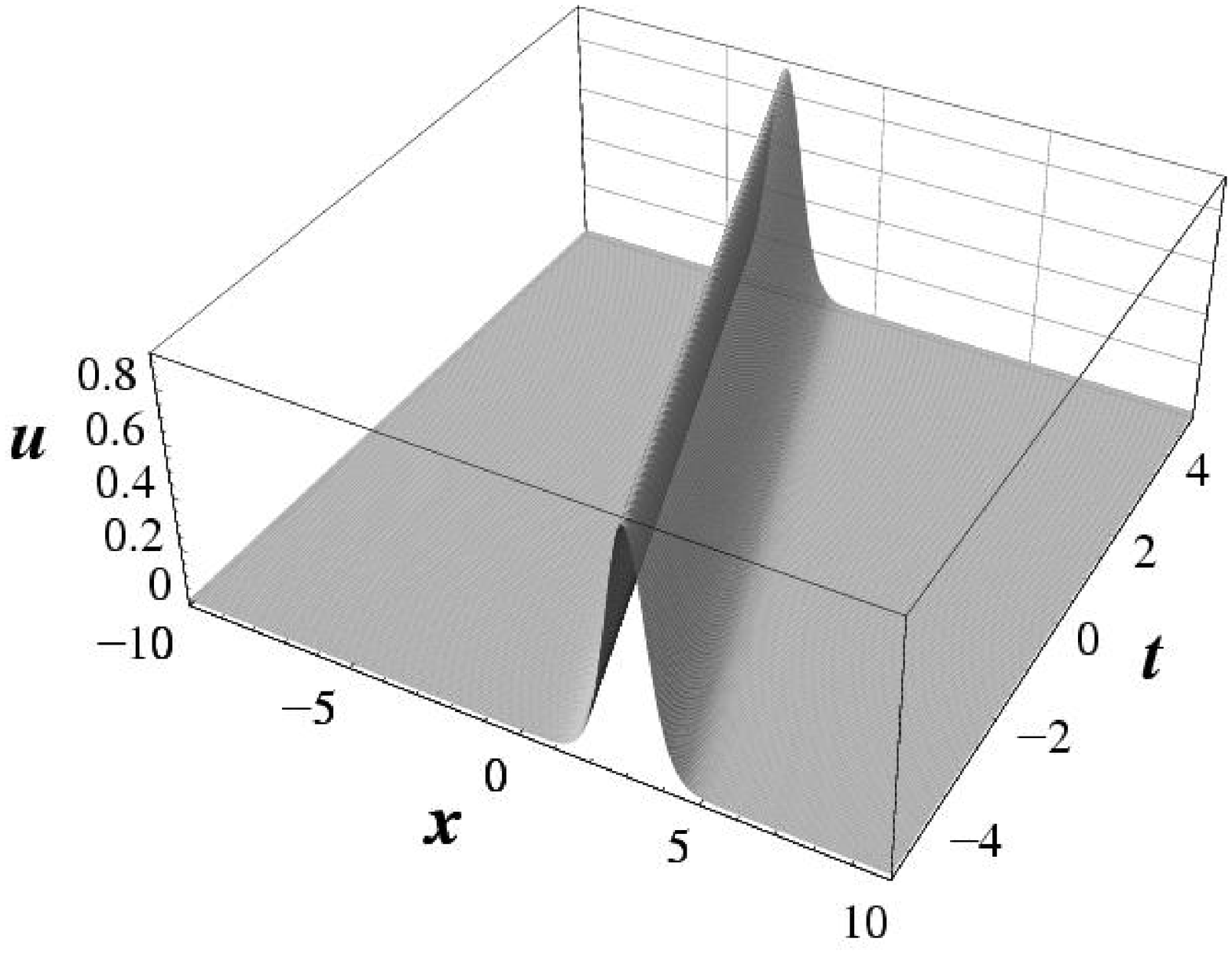}
  \includegraphics[width=0.5\textwidth]{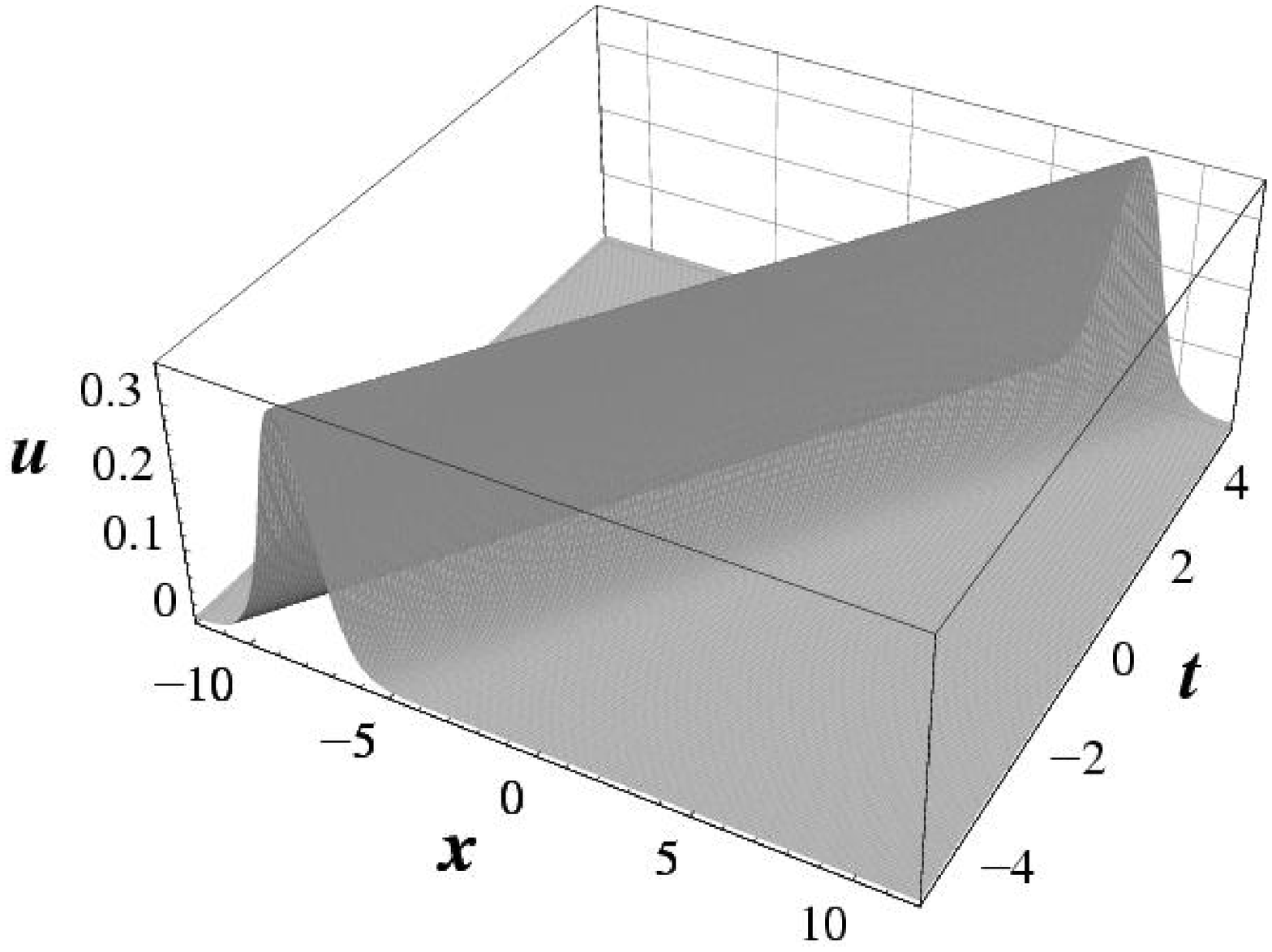}
}
  \caption{\label{fig:bsk1soliton}
    Single left- and right-going solitons for the bSK equation (\ref{bSK})
    :\ $\varepsilon_1=\frac{\pi}{10}$ (left), $\varepsilon_1=\frac{3\pi}{10}$ (right). }
\end{figure}

\begin{figure}[htbp]
  \centerline{
    \includegraphics[width=0.5\textwidth]{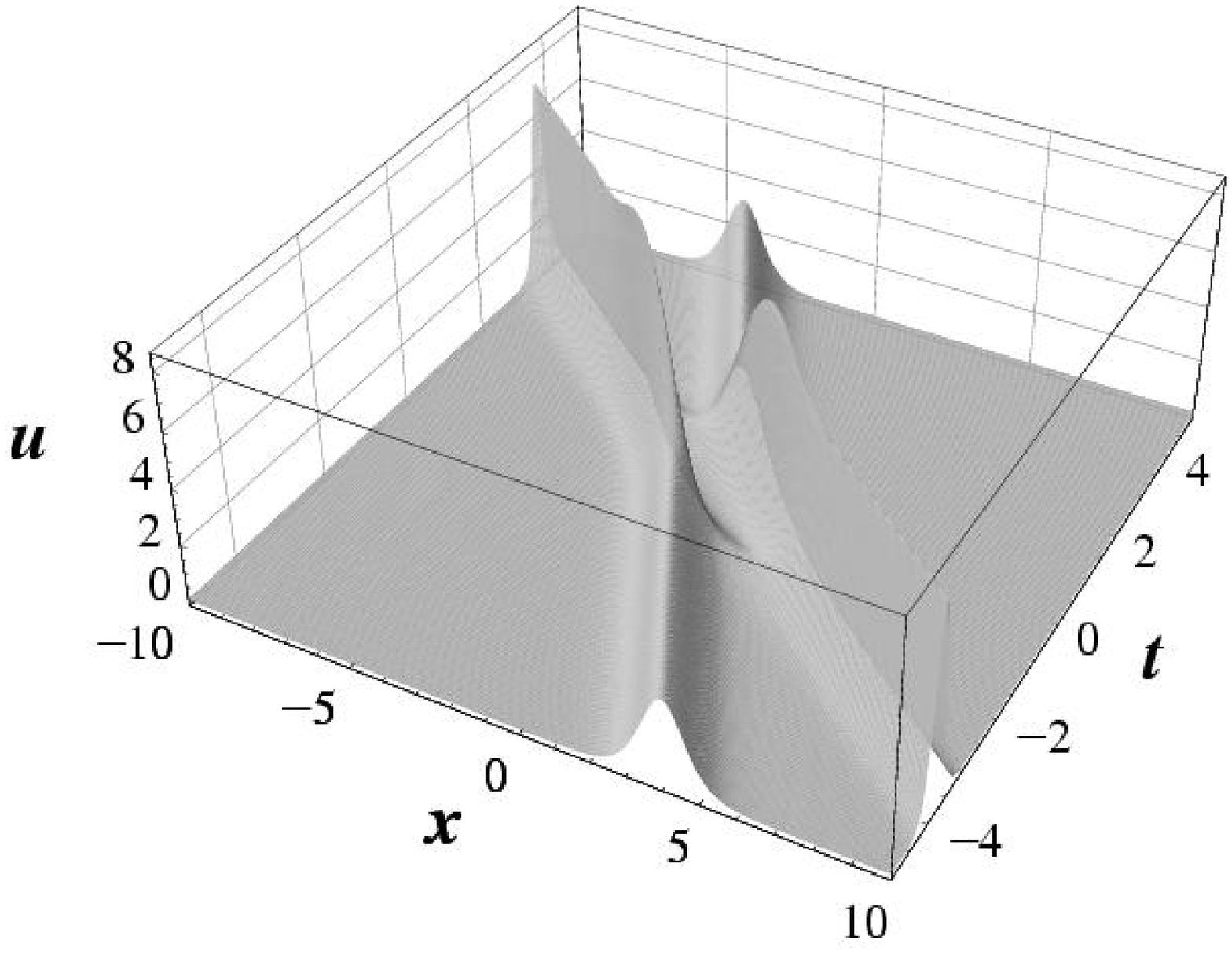}
    \includegraphics[width=0.5\textwidth]{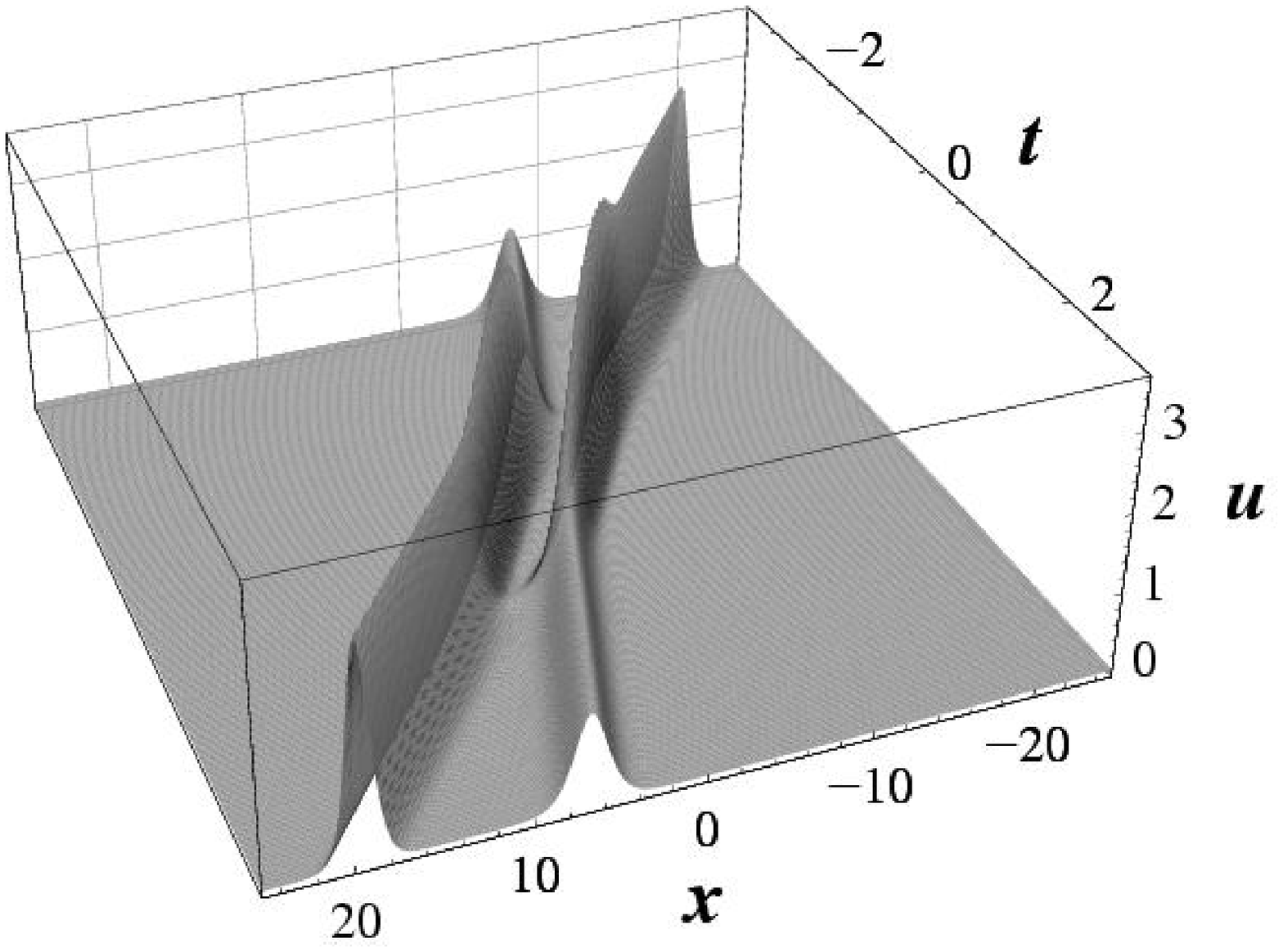}
  }
  \centerline{
    \includegraphics[width=0.5\textwidth]{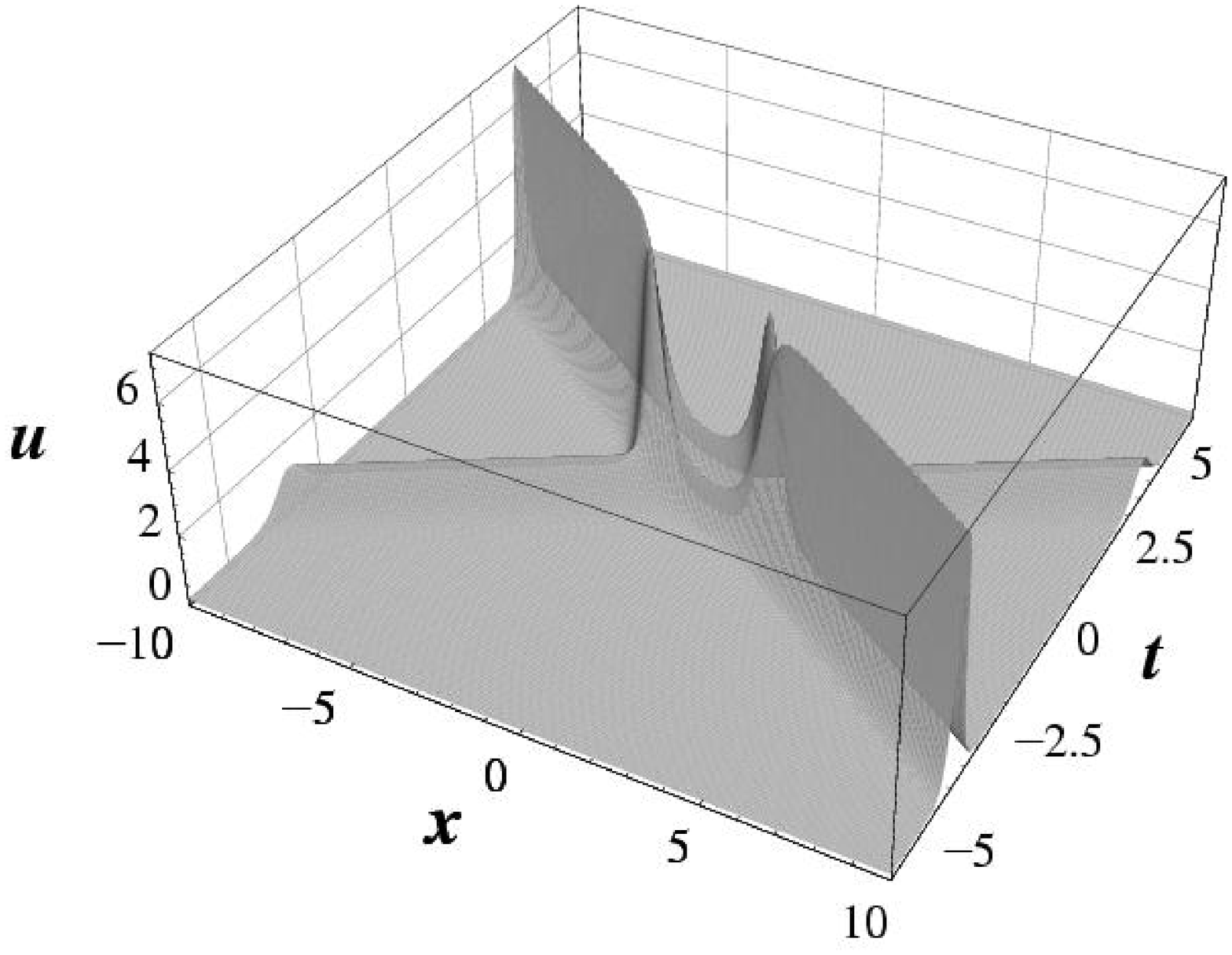}
  }
  \caption{\label{fig:bsk2solitons}
    Two left- and right-going as well as head-on colliding solitons for
    the bSK equation (\ref{bSK}):\
    $\varepsilon_1=\varepsilon_2=\frac{\pi}{10}$ (left),
    $\varepsilon_1=\varepsilon_2=\frac{3\pi}{10}$ (right) and
    $\varepsilon_1=\frac{\pi}{10},\varepsilon_2=\frac{3\pi}{10}$
    (collision).}

\end{figure}

\begin{proposition}\label{proptwosolitonbSK}
\begin{align}
&\hat{\tau}^{(2+2)}_{\rm bSK}=   \nonumber \\ 
&\Big\{ \frac{ e^{2(\xi_1+\xi_2)}
}{4\big(k^2_1+k^2_2+2k_1k_2\cos(\varepsilon_1-\varepsilon_2)
\big)^2} + \frac{g_2 e^{-2(\xi_1+\xi_2)}
}{4\big(k^2_1+k^2_2+2k_1k_2\cos(\varepsilon_1-\varepsilon_2)
\big)^2} \nonumber \\ 
&+ \frac{ e^{2(\xi_1-\xi_2)}
}{4\big(k^2_1+k^2_2-2k_1k_2\cos(\varepsilon_1+\varepsilon_2)
\big)^2} + \frac{ e^{-(2\xi_1-\xi_2)}
}{4\big(k^2_1+k^2_2-2k_1k_2\cos(\varepsilon_1+\varepsilon_2)
\big)^2}                   \nonumber \\  
&+\frac{e^{2\xi_1}}
 {2\big(k^2_1+k^2_2+2k_1k_2\cos(\varepsilon_1-\varepsilon_2)
\big)\big(k^2_1+k^2_2-2k_1k_2\cos(\varepsilon_1+\varepsilon_2)
\big)}                        \nonumber \\ 
&+\frac{g_6e^{-2\xi_1}}
 {2\big(k^2_1+k^2_2+2k_1k_2\cos(\varepsilon_1-\varepsilon_2)
\big)\big(k^2_1+k^2_2-2k_1k_2\cos(\varepsilon_1+\varepsilon_2)
\big)}     \nonumber  \\  
&+\frac{e^{2\xi_2}}
 {2\big(k^2_1+k^2_2+2k_1k_2\cos(\varepsilon_1-\varepsilon_2)
\big)\big(k^2_1+k^2_2-2k_1k_2\cos(\varepsilon_1+\varepsilon_2)
\big)}     \nonumber  \\  
&+\frac{g_8e^{-2\xi_2}}
 {2\big(k^2_1+k^2_2+2k_1k_2\cos(\varepsilon_1-\varepsilon_2)
\big)\big(k^2_1+k^2_2-2k_1k_2\cos(\varepsilon_1+\varepsilon_2)
\big)}     \nonumber  \\  
&+\frac{(k^2_1+k^2_2)^2g_9}
 {2\big(k^2_1+k^2_2+2k_1k_2\cos(\varepsilon_1-\varepsilon_2)
\big)\big(k^2_1+k^2_2-2k_1k_2\cos(\varepsilon_1+\varepsilon_2)
\big)}       
\Big\},
\end{align}
the two soliton solution is $u=(\partial_x^2\log
\hat{\tau}^{(2+2)}_{\rm bSK})$. In particular,
$\varepsilon_1=\varepsilon_2=\frac{\pi}{10}$ results in two
overtaking  solitons and moving in negative direction;
$\varepsilon_1=\varepsilon_2=\frac{3\pi}{10}$ produces two
overtaking  solitons and moving in positive direction;
$\varepsilon_1=\frac{\pi}{10},\varepsilon_2=\frac{3\pi}{10}$
results in two head-on  solitons.
\end{proposition}
We have plotted the one solitons in figure \ref{fig:bsk1soliton}
associated with parameters $A_1=B_1=1, k_1=1$ of equation
(\ref{onesolitonbSK}). The two solitons in Proposition
\ref{proptwosolitonbSK} are shown  in figure \ref{fig:bsk2solitons}
associated with parameters $k_1=2$ and $k_2=1.3$.

\section{Periodic and soliton solutions of  bKK equation}
The 5-reduction of the CKP hierarchy yields the bKK equation equation
(\ref{bKK}). Let the initial value be $u = 0$ in Eqs.\
(\ref{bKKlax1}) and (\ref{bKKlax2}), then
$\fun{\phi}{0}{i}=\phi(\lambda_i;x,t)$ are solutions of
\begin{equation} \label{zerobKKlax1}
 \partial_x^5\phi(\lambda_i;x,t)=\lambda_i
 \phi(\lambda_i;x,t),\quad
  \frac{\partial \phi(\lambda_i;x,t)}{\partial
t}=(\partial_x^{3}\phi(\lambda_i;x,t)). 
\end{equation}
So the Proposition \ref{proptauCKP} implies the $\tau$ function of
bKK equation.

\begin{proposition}\label{propzerotaubKK}
The $\tau$ function of the {\rm bKK} equation  generated by
$T_{n+n}$ from initial value $1$ is
\begin{align}
&\fun{\tau}{n+n}{\rm bKK} = IW_{n,n}(\fun{\phi}{0}{n},
\fun{\phi}{0}{n-1}, \cdots, \fun{\phi}{0}{1};
      \fun{\phi}{0}{1}, \fun{\phi}{0}{2},\cdots , \fun{\phi}{0}{n} ) \notag \\
&=\left|
\begin{array}{lccccc}
\int \fun{\phi}{0}{n}\cdot\fun{\phi}{0}{1} &
\int\fun{\phi}{0}{n}\cdot\fun{\phi}{0}{2} &
\int\fun{\phi}{0}{n}\cdot\fun{\phi}{0}{3} &\cdots
&\int \fun{\phi}{0}{n}\cdot\fun{\phi}{0}{n-1} &\int\fun{\phi}{0}{n}\cdot\fun{\phi}{0}{n}\\  
\int \fun{\phi}{0}{n-1}\cdot\fun{\phi}{0}{1} & \int
\fun{\phi}{0}{n-1}\cdot\fun{\phi}{0}{2} &\int
\fun{\phi}{0}{n-1}\cdot\fun{\phi}{0}{3} &\cdots
& \int\fun{\phi}{0}{n-1}\cdot\fun{\phi}{0}{n-1} &\int \fun{\phi}{0}{n-1}\cdot\fun{\phi}{0}{n}\\  
\vdots&\vdots&\vdots&\cdots&\vdots\\        
 \int \fun{\phi}{0}{2}\cdot\fun{\phi}{0}{1} &
  \int\fun{\phi}{0}{2}\cdot\fun{\phi}{0}{2}&\int
\fun{\phi}{0}{2}\cdot\fun{\phi}{0}{3} &\cdots
 &\int\fun{\phi}{0}{2}\cdot\fun{\phi}{0}{n-1} &\int
\fun{\phi}{0}{2}\cdot\fun{\phi}{0}{n}\\ 
\int\fun{\phi}{0}{1}\cdot\fun{\phi}{0}{1} & \int
\fun{\phi}{0}{1}\cdot\fun{\phi}{0}{2} &\int
\fun{\phi}{0}{1}\cdot\fun{\phi}{0}{3} &\cdots
 &\int\fun{\phi}{0}{1}\cdot\fun{\phi}{0}{n-1} &\int
\fun{\phi}{0}{1}\cdot\fun{\phi}{0}{n}
\end{array} \right|
\end{align}
and the solution $u$ of the {\rm bKK} from initial value $zero$ is
\begin{equation}
u=\left(\partial_x^2 \log \fun{\tau}{n+n}{\rm bKK}\right).
\label{zeroubKK}
\end{equation}
Here $\fun{\phi}{0}{i}=\phi(\lambda_i;x,t)$ are solutions of
 equation (\ref{zerobKKlax1}).
\end{proposition}
As before, $\fun{\tau}{n+n}{\rm bKK}$ is complex and related to
the 5-th roots of $e^{i\varepsilon}$ and again we have to find a
{\em physical} $\tau$\ function $\hat{\tau}^{(n+n)}_{\rm bKK}$
such that $u$ in equation (\ref{zeroubKK}) is real and smooth solution
includes solitons and periodic solutions. The case of $n=1$ and
$n=2$ will be discussed in detail. Similar to the bSK equation, we
should assume the solutions of equation (\ref{zerobKKlax1}) as
\begin{align}
&\phi(\lambda_1;x,t)=A_1 e^{p_1 x+ p_1^3 t}+B_1 e^{q_1 x+ q_1^3
t},\quad p_1=k_1
e^{i \varepsilon_1}, q_1=-k_1 e^{-i \varepsilon_1}, k_1^5=|\lambda_1|,k_1\in \mathbb{R},
\label{bKKzerophi1a} \\
\intertext{or}
 &\phi(\lambda_1;x,t)=A_1 e^{p_1 x+ p_1^3 t}+B_1
e^{q_1 x+ q_1^3 t},\quad p_1=k_1 e^{i \varepsilon_1}, q_1= k_1
e^{-i \varepsilon_1},k_1^5=|\lambda_1|,k_1\in \mathbb{R},
\label{bKKzerophi1b}
\end{align}
to extract {\em physical}\ $\tau$ function
$\hat{\tau}^{(n+n)}_{\rm bKK}$ from $\tu{n+n}_{\rm bKK}$.  At
first, we would like to give the two simple cases which are
generated by the gauge transformations $T_{1+1}$.

\begin{proposition}\label{proponesolitonbKK}
Let $\xi_1= xk_1\cos \varepsilon_1+tk_1^3\cos 3\varepsilon_1$,
 $\frac{B_1}{A_1}=ie^{-i\varepsilon_1}$ and $\fun{\phi}{0}{1}=\phi(\lambda_1;x,t)$ defined by
equation (\ref{bKKzerophi1a}), then the physical $\tau$ function of
{\rm bKK} extracted from $\left.\tau^{(n+n)}_{\rm
bKK}\right|_{n=1}$ is
\begin{align}
&\hat{\tau}^{(1+1)}_{\rm bKK}= e^{2 \xi_1}+ e^{-2
\xi_1}+ \frac{2}{\sin\varepsilon_1} \label{efficienttauonesolitonbKK} \\
\intertext{and the corresponding one soliton $u
=\left(\partial^2_x \log\hat{\tau}^{(1+1)}_{\rm bKK}\right)$ is }
&u=\frac{4 k_1^2 \left(\cos\varepsilon_1\right)^2\left(1+
\frac{\cosh 2\xi_1}{\sin\varepsilon_1} \right) }
 { \left(\cosh2\xi_1+\frac{1}{\sin\varepsilon_1} \right)^2}  \label{onesolitonbKK}
\end{align}
with $\varepsilon_1=\frac{\pi}{10}$ or $\frac{3\pi}{10}$. The
velocity of the soliton is $v=-k_1^2\frac{\cos 3\varepsilon_1
}{\cos \varepsilon_1}$. In particular, the left-going soliton have
two peaks in its profile and the negative speed
$v_{\_}=\left.v\right|_{\varepsilon_1=\frac{\pi}{10}}$; the
right-going soliton have only one peak and positive speed
$v_{+}=\left.v\right|_{\varepsilon_1=\frac{3\pi}{10}}$.
\end{proposition}
\begin{proof}
Taking  $\fun{\phi}{0}{1}=\phi(\lambda_1;x,t)$ of equation
(\ref{bKKzerophi1a}) and $n=1$ back into Proposition
\ref{propzerotaubKK}, the straightforward calculation leads to
\begin{equation}
\tu{1+1}_{\rm bKK}=\int
\left(\fun{\phi}{0}{1}\right)^2=\frac{A^2_1e^{2i(xk_1\sin
\varepsilon_1+tk_1^3\sin 3\varepsilon_1)}}{2p_1}\left[
e^{2\xi_1}+e^{-2\xi_1} + \frac{2}{\sin\varepsilon_1}\right].
\end{equation}
Here $\xi_1= xk_1\cos \varepsilon_1+tk_1^3\cos 3\varepsilon_1$.
\end{proof}
If $\fun{\phi}{0}{1}=\phi(\lambda_1;x,t)$ defined by equation
(\ref{bKKzerophi1b}), then we can get a periodic solution as
following proposition.

\begin{proposition}\label{proponeparameterbKK}
Let $\eta_1= xk_1\sin \varepsilon_1+tk_1^3\sin 3\varepsilon_1$,
 and $\fun{\phi}{0}{1}=\phi(\lambda_1;x,t)$ defined by
equation (\ref{bKKzerophi1b}), $A_1=B_1=1$ in $\fun{\phi}{0}{1}$, then
the physical $\tau$ function of {\rm bKK} extracted from
$\left.\tau^{(n+n)}_{\rm bKK}\right|_{n=1}$ is
\begin{align}
&\hat{\tau}^{(1+1)}_{\rm bKK}= \frac{1}{\cos\varepsilon_1}+ \cos(2\eta_1-\varepsilon_1)\\
\intertext{and the corresponding  solution}
& u=\left(\partial^2_x
\log\hat{\tau}^{(1+1)}_{\rm bKK}\right)
=\frac{-4k_1^2\sin^2\varepsilon_1\left(\frac{\cos(2\eta_1-2\varepsilon_1)}{\cos\varepsilon_1}
+1 \right) }
 { \left(\frac{1}{\cos\varepsilon_1}+ \cos(2\eta_1-\varepsilon_1)\right)^2}  \label{oneparameterbKK}
\end{align}
is periodic. Here $ \varepsilon_1=\frac{2\pi}{10}$ or
$\frac{4\pi}{10}$. The velocity for  the solution is
$v=-k_1^2\frac{\sin 3\varepsilon_1 }{\sin \varepsilon_1}$. If
$\varepsilon_1=\frac{2\pi}{10}$, $u$ in equation
(\ref{oneparameterbKK}) is a left-going periodic wave. If
$\varepsilon_1=\frac{4\pi}{10}$, $u$ in equation
(\ref{oneparameterbKK}) is a right-going periodic wave.
\end{proposition}
\begin{proof}
\begin{equation}
\tu{1+1}_{\rm bKK}=\int
\left(\fun{\phi}{0}{1}\right)^2=\frac{e^{2(xk_1\cos
\varepsilon_1+tk_1^3\cos 3\varepsilon_1)}}{4k_1}\left[
\frac{1}{\cos\varepsilon_1}+\cos(2\eta_1-\varepsilon_1) \right].
\end{equation}
Here $\eta_1= xk_1\sin \varepsilon_1+tk_1^3\sin 3\varepsilon_1$,
 and $\fun{\phi}{0}{1}=\phi(\lambda_1;x,t)$ is defined by
equation (\ref{bKKzerophi1b}).
\end{proof}

Based on Propositions \ref{proponesolitonbKK} and
\ref{proponeparameterbKK}, we can find following corresponding
relationship between symmetrical distributions of 5-th roots of
$e^{i\varepsilon}$ and moving direction of solutions.
\begin{enumerate}
  \item
$(e^{i\varepsilon_1},-e^{-i\varepsilon_1})|_{\varepsilon_1=\frac{\pi}{10}}$ $\longrightarrow$ the first distribution of 5-th roots of
$e^{i\varepsilon}$ $\longrightarrow$
$(p_1=k_1e^{i\varepsilon_1},q_1=-k_1e^{-i\varepsilon_1})|_{\varepsilon_1={\pi\over
10}}$ in equation (\ref{bKKzerophi1a}) $\longrightarrow$ left-going
two-peak soliton in equation (\ref{onesolitonbKK});
  \item
$(e^{i\varepsilon_1},-e^{-i\varepsilon_1})|_{\varepsilon_1={3\pi\over
10}}$ $\longrightarrow$ the second distribution of 5-th roots of
$e^{i\varepsilon}$ $\longrightarrow$
$(p_1=k_1e^{i\varepsilon_1},q_1=-k_1e^{-i\varepsilon_1})|_{\varepsilon_1={3\pi\over
10}}$ in equation (\ref{bKKzerophi1a}) $\longrightarrow$ right-going
one-peak soliton in equation (\ref{onesolitonbKK});
  \item $(e^{i\varepsilon_1},e^{-i\varepsilon_1})|_{\varepsilon_1={2\pi\over
10}}$ $\longrightarrow$ the third distribution of 5-th roots of
$e^{i\varepsilon}$ $\longrightarrow$
$(p_1=k_1e^{i\varepsilon_1},q_1=k_1e^{-i\varepsilon_1})|_{\varepsilon_1={2\pi\over
10}}$ in equation (\ref{bKKzerophi1b}) $\longrightarrow$ left-going
periodic wave in equation (\ref{oneparameterbKK});
  \item
$(e^{i\varepsilon_1},e^{-i\varepsilon_1})|_{\varepsilon_1={4\pi\over
10}}$ $\longrightarrow$ the fourth distribution of 5-th roots of
$e^{i\varepsilon}$ $\longrightarrow$
$(p_1=k_1e^{i\varepsilon_1},q_1=k_1e^{-i\varepsilon_1})|_{\varepsilon_1={4\pi\over
10}}$ in equation (\ref{bKKzerophi1b}) $\longrightarrow$ right-going
periodic wave in equation (\ref{oneparameterbKK}).
  \end{enumerate}
There are only four distributions of 5-th roots of
$e^{i\varepsilon}$, which are symmetric respect with x-axes or
y-axes. However, there exist several other pairs of roots in the
above four distributions which will result in divergent solutions
of bKK through the above procedure. For example, $p_1= k_1
e^{i\frac{13\pi}{10}},q_1=-k_1 e^{-i\frac{13\pi}{10}}$ or $p_1=
k_1 e^{i\frac{8\pi}{10}},q_1=k_1 e^{-i\frac{8\pi}{10}}$.

\par
Let us now concentrate on the two-peak soliton solution in equation
(\ref{onesolitonbKK}).
\begin{lemma}\label{lembKKpoint}
Let $x\geq 1$, constant $a>0$ and function
\begin{equation}
 y=y(x)=\frac{1+\frac{x}{a}}{\left(x+\frac{1}{a}\right)^2},
\end{equation}
then
\begin{enumerate}
\item if\ $a>1/2$, $\frac{\partial y}{\partial x} < 0$;
\item if\ $a=1/2$, then $\frac{\partial y}{\partial x}|_{x=1}=0 $;
\item if\ $\frac{1}{2}> a>0$, then there exists one point $x_1>1$ such  that
$\left.\frac{\partial y}{\partial x}\right|_{x=x_1}=0 $, $x_1$ is
one extreme maximum point of $y$, and $\left.\frac{\partial
y}{\partial x}\right|_{x=1}>0 $.
\end{enumerate}
\end{lemma}
\begin{proof} We have
\begin{equation}
y_x=\frac{\partial y}{\partial x}=\frac{\frac{1}{a}\big(
\frac{1}{a}-2a -x \big)}{\big(x+ \frac{1}{a}\big)^3}.
\end{equation}
Firstly, $y_x<0$ if $a>1/2$. Secondly, if $a=1/2$, $y_x=0$ when
$x=1$. At last, if $ 1/2 >a>0 $, there exist $x_1>1$ such that
$y_x=0$. Note that $y_x >0$ if $x \in (1,x_1)$, $y_x < 0$ if $x >
x_1 $. So $x_1$ is one extreme maximum point of $y$.
\end{proof}

\begin{proposition}\label{propdiscusstwopeakbKK}
Let  $a, b,k$ be positive constants, $\xi=kx+ct, c\in \mathbb{R}$,
for following kind of solution
\begin{equation}\label{modeltwopeaksoliton}
u=\frac{b\big(1+ \frac{\cosh 2\xi}{a} \big) }{
\big(\cosh2\xi+\frac{1}{a} \big)^2},
\end{equation}
\begin{enumerate}
\item if $a\geq 1/2$, $u$ has one peak in its profile defined by
  $\xi=0$;
\item if $0<a <\frac{1}{2}$, then there exist two peaks in profile;
\item There exist no more than two peaks in a soliton give by equation
(\ref{modeltwopeaksoliton}).
\end{enumerate}
\end{proposition}
\begin{proof} By calculation, we have
\begin{equation}
u_x=\frac{2kb\sinh 2\xi\left(\frac{1}{a} -2a-\cosh 2\xi
\right)}{a\left(\cosh2\xi+\frac{1}{a} \right)^3}.
\end{equation}
According to the Lemma \ref{lembKKpoint}, we have
\begin{enumerate}
\item $a>1/2$, there exist $\xi=0$ such that
$u_x=0$ because $\left.\sinh 2\xi \right|_{\xi=0}=0$. Note
$\left(\frac{1}{a} -2a-\cosh 2\xi \right)$ $<0$.
\item $a=1/2$, there exist $\xi=0$ such that $u_x=0$ because $\sinh
  2\xi|_{\xi=0}=0$ and $\left.\left(\frac{1}{a} -2a-\cosh 2\xi
    \right)\right|_{\xi=0}$ $=0$. However, let $|\xi|$ be sufficiently
  small, we have $u_x< 0 $ if $\xi>0$ and $u_x> 0 $ if $\xi<0$. So
  $\xi=kx+ct=0$ defines one extreme maximum line of $u(x,t)$ on (x,t)
  plane.
\item $1/2 >a>0$, there exist $\xi=0$ and $\xi_1>0$ and $\xi_2=-\xi_1<0$
  such that $u_x =0$. But $u_x>0$ if $\xi<-\xi_1$; $u_x<0$ if $\xi\in
  (-\xi_1,0)$; $u_x>0$ if $\xi \in (0,\xi_1)$; $u_x<0$ if $\xi>\xi_1$.
  So $\xi=kx+ct=0$ defines one extreme minimal line on (x,t) plane;
  $0<\xi_1=kx+ct$ and $0>-\xi_1=-(kx+ct)$ define two extreme maximum
  lines on the (x,t) plane.
\end{enumerate}
Using $u\rightarrow 0$ if $|\xi| \rightarrow \infty $,
conclusions are proven.
\end{proof}
Comparing equation (\ref{modeltwopeaksoliton}) with equation
(\ref{onesolitonbKK}) we get $a=\sin\varepsilon_1$, and then can
understand why $\sin\varepsilon_1|_{\varepsilon_1=\pi/10} $ will
lead to two peaks in one soliton of bKK but
$\sin\varepsilon_1|_{\varepsilon_1=3\pi/10} $ will lead only to
one peak in one soliton of bKK.  On the other hand, one soliton
solution of $u$ in equation (\ref{onesolitonbKK}) have one peak or two
peaks(maximum case) in its profile. According to analysis above,
we can claim from the point of view of reduction in KP hierarchy
that the existence of two peaks in the soliton is traced to three
facts:
\begin{enumerate}
  \item  The Grammian $\tau $ function in
Proposition \ref{propzerotaubKK} which determines the form of
soliton in equation (\ref{onesolitonbKK});
 \item  The order of  n-reduction, i.e. $n \geq 5 $  can produce two
peaks soliton in KP hierarchy;
 \item The phase $\varepsilon_1$ of
n-th root($n \geq 5$) of $e^{i\varepsilon}$, such that
$0<a=\sin\varepsilon_1 < 1/2$.
\end{enumerate}

\par
Now we turn to the  more complicated $\tu{2+2}_{\rm bKK}$ from
Proposition \ref{propzerotaubKK}, which generates the two soliton
and periodic solution with two spectral parameters of bKK
equation. The first case is the two soliton solution.

\begin{lemma}\label{lemtau2+2bKK}
Let $\fun{\phi}{0}{i}=\phi(\lambda_i;x,t), i=1,2$,  defined by
Eq.({\ref{bKKzerophi1a}}), $\xi_i=xk_i\cos\varepsilon_i
+tk^3_i\cos3\varepsilon_i, \eta_i=xk_i\sin\varepsilon_i
+tk^3_i\sin3\varepsilon_i, i=1, 2$, then $\tu{n+n}_{\rm
bKK}|_{n=2}$ gives out
\begin{align}
&\tu{2+2}_{\rm bKK}=A^2_1A^2_2e^{2i(\eta_1+\eta_2)}\times  \nonumber \\ 
&\Big\{ \frac{z_1^* e^{2(\xi_1+\xi_2)}
}{4\big(k^2_1+k^2_2+2k_1k_2\cos(\varepsilon_1-\varepsilon_2)
\big)^2} + \frac{z_1e^{-2(\xi_1+\xi_2)}
}{4\big(k^2_1+k^2_2+2k_1k_2\cos(\varepsilon_1-\varepsilon_2)
\big)^2}\left(\frac{B_1}{A_1}\right)^2\left(\frac{B_2}{A_2}\right)^2 \nonumber \\ 
&+ \frac{-z^*_3 e^{2(\xi_1-\xi_2)}
}{4ik_1k_2\big(k^2_1+k^2_2-2k_1k_2\cos(\varepsilon_1+\varepsilon_2)
\big)^2}\left( \frac{B_2}{A_2}\right)^2 + \frac{-z_3
e^{-(2\xi_1-\xi_2)}
}{4ik_1k_2\big(k^2_1+k^2_2-2k_1k_2\cos(\varepsilon_1+\varepsilon_2)
\big)^2}\left( \frac{B_1}{A_1}\right)^2                    \nonumber \\  
&+\frac{z^*_2e^{2\xi_1}}
 {2ik_1k_2\sin\varepsilon_2\big(k^2_1+k^2_2+2k_1k_2\cos(\varepsilon_1-\varepsilon_2)
\big)\big(k^2_1+k^2_2-2k_1k_2\cos(\varepsilon_1+\varepsilon_2)
\big)}\left(\frac{B_2}{A_2}\right)                        \nonumber \\ 
&+\frac{-z_2e^{-2\xi_1}}
 {2ik_1k_2\sin\varepsilon_2\big(k^2_1+k^2_2+2k_1k_2\cos(\varepsilon_1-\varepsilon_2)
\big)\big(k^2_1+k^2_2-2k_1k_2\cos(\varepsilon_1+\varepsilon_2)
\big)}\left(\frac{B_1}{A_1}\right)^2\left(\frac{B_2}{A_2}\right)     \nonumber  \\  
&+\frac{z^*_4e^{2\xi_2}}
 {2ik_1k_2\sin\varepsilon_1\big(k^2_1+k^2_2+2k_1k_2\cos(\varepsilon_1-\varepsilon_2)
\big)\big(k^2_1+k^2_2-2k_1k_2\cos(\varepsilon_1+\varepsilon_2)
\big)}\left(\frac{B_1}{A_1}\right)     \nonumber  \\  
&+\frac{-z_4e^{-2\xi_2}}
 {2ik_1k_2\sin\varepsilon_1\big(k^2_1+k^2_2+2k_1k_2\cos(\varepsilon_1-\varepsilon_2)
\big)\big(k^2_1+k^2_2-2k_1k_2\cos(\varepsilon_1+\varepsilon_2)
\big)}\left(\frac{B_1}{A_1}\right)\left(\frac{B_2}{A_2}\right)^2     \nonumber  \\  
&+\frac{-\big((k^2_1+k^2_2)^2-4k^2_1k^2_2(\cos^2\varepsilon_1\cos^2\varepsilon_2+
\sin^2\varepsilon_1\sin^2\varepsilon_2) \big)}
 {2k_1k_2\sin\varepsilon_1\sin\varepsilon_2\big(k^2_1+k^2_2+2k_1k_2\cos(\varepsilon_1-\varepsilon_2)
\big)\big(k^2_1+k^2_2-2k_1k_2\cos(\varepsilon_1+\varepsilon_2)
\big)}\left(\frac{B_1}{A_1}\right)\left(\frac{B_2}{A_2}\right)        
\Big\}
\end{align}
Here $z_i, i=1, 2, 3, 4$, are given in  \ref{appbKK2S}.
$z_i^*$ means the complex conjugation of $z_i$.
\end{lemma}
In order to  extract physical $\tau$ function
$\hat{\tau}^{(2+2)}_{\rm bKK}$, we need following two Lemmas for
suitable $\frac{B_i}{A_i}, i=1, 2$.
\begin{lemma}For $z_i, i=1, 2, 3, 4$, as given in  \ref{appbKK2S},
  the following identities
\begin{equation}
z_2^2=z_1z_3, \quad z_4^2=z_1z_3^*
\end{equation}
hold.
\end{lemma}

\begin{lemma}\label{lemA1B1A2B2bKK}
Let $\frac{B_1}{A_1}=i\frac{z_1^{*}}{z_4^{*}},
\frac{B_2}{A_2}=i\frac{z_1^{*}}{z^*_2}$, and
$g_5=\frac{1}{|z_3|}$, $g_6=g_8=\frac{|z_1|^2}{|z_2|^2}$,
$g_9=\frac{|z_1|^4}{|z_2|^4}$, then
\begin{align}
&-\frac{z_3^*}{z_1^{*}}\left(\frac{B_2}{A_2} \right)^2=1,
&-\frac{z_3}{z_1^{*}}\left(\frac{B_1}{A_1}\right)^2=1,    \\
&\frac{z_1}{z_1^*}\left(\frac{B_1}{A_1}\right)^2\left(\frac{B_2}{A_2}\right)^2=g_9,
&\frac{-1}{z_1^{*}}\left(\frac{B_1}{A_1}
\right)\left(\frac{B_2}{A_2} \right)=g_5,  \\ 
&\frac{z_2}{-iz_1^{*}}\left(\frac{B_1}{A_1}
\right)^2\left(\frac{B_2}{A_2} \right)=g_8,
&\frac{(z_4)^*}{-iz_1^{*}}\left(\frac{B_1}{A_1}\right)\left(\frac{B_2}{A_2}\right)^2=g_6    
\end{align}
hold.
\end{lemma}
Taking $\frac{B_i}{A_i}, i=1, 2$,  and relations in Lemma
\ref{lemA1B1A2B2bKK} back into Lemma \ref{lemtau2+2bKK}, the
physical $\tau$ function $\hat{\tau}^{(2+2)}_{\rm bKK}$ is
obtained.

\begin{figure}[htbp]
  \centerline{
  \includegraphics[width=0.5\textwidth]{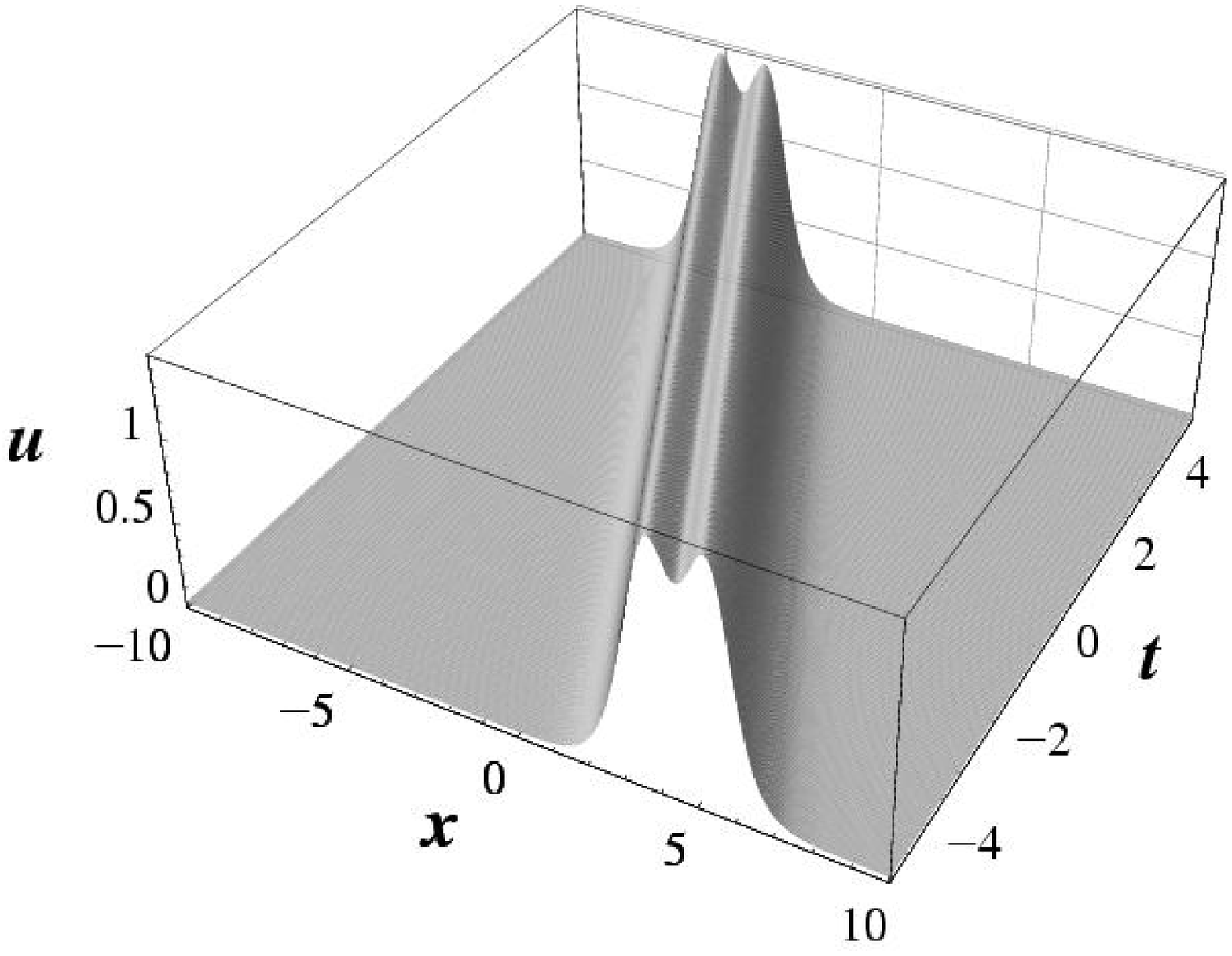}
  \includegraphics[width=0.5\textwidth]{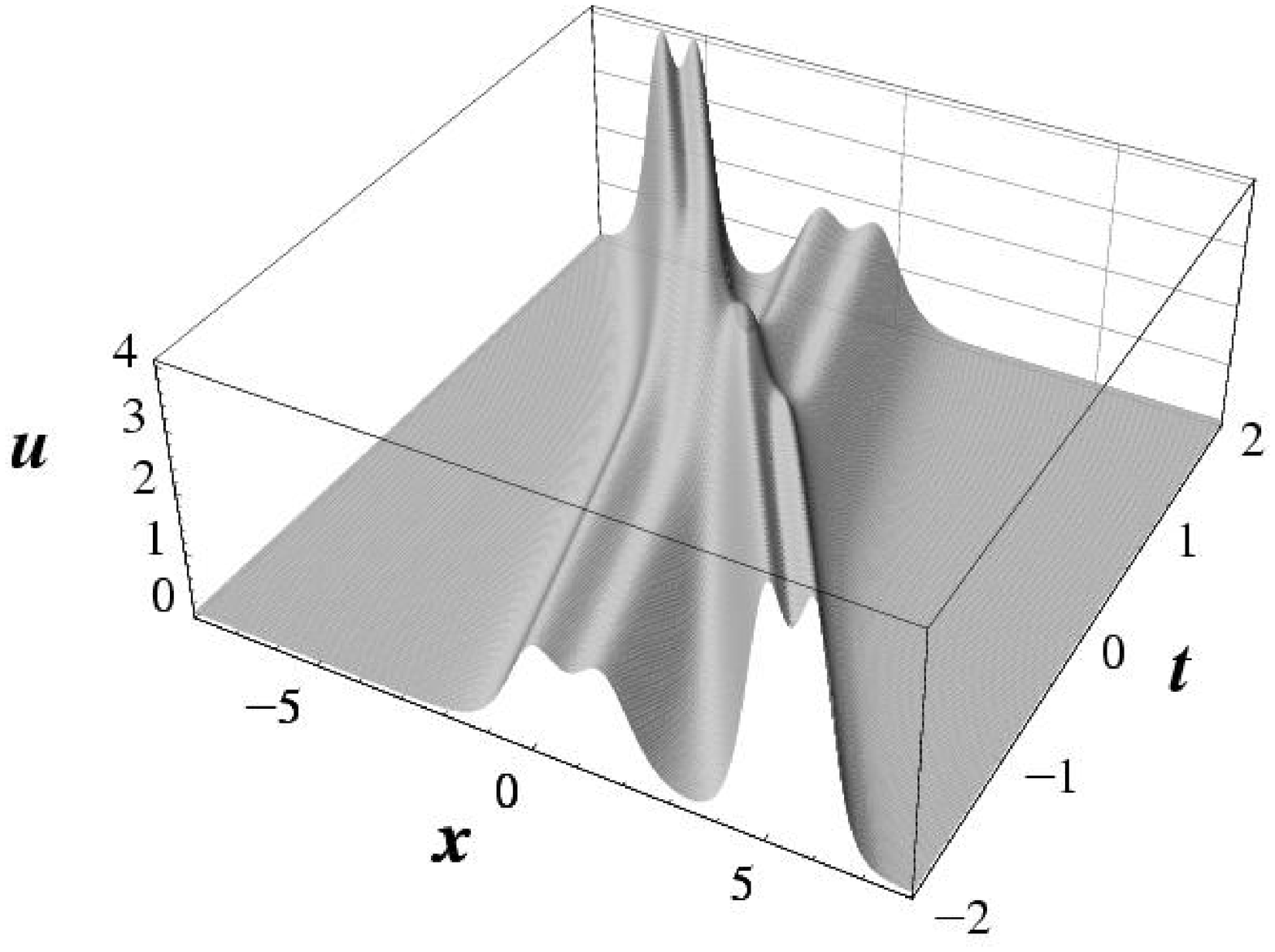}
} \centerline{
  \includegraphics[width=0.5\textwidth]{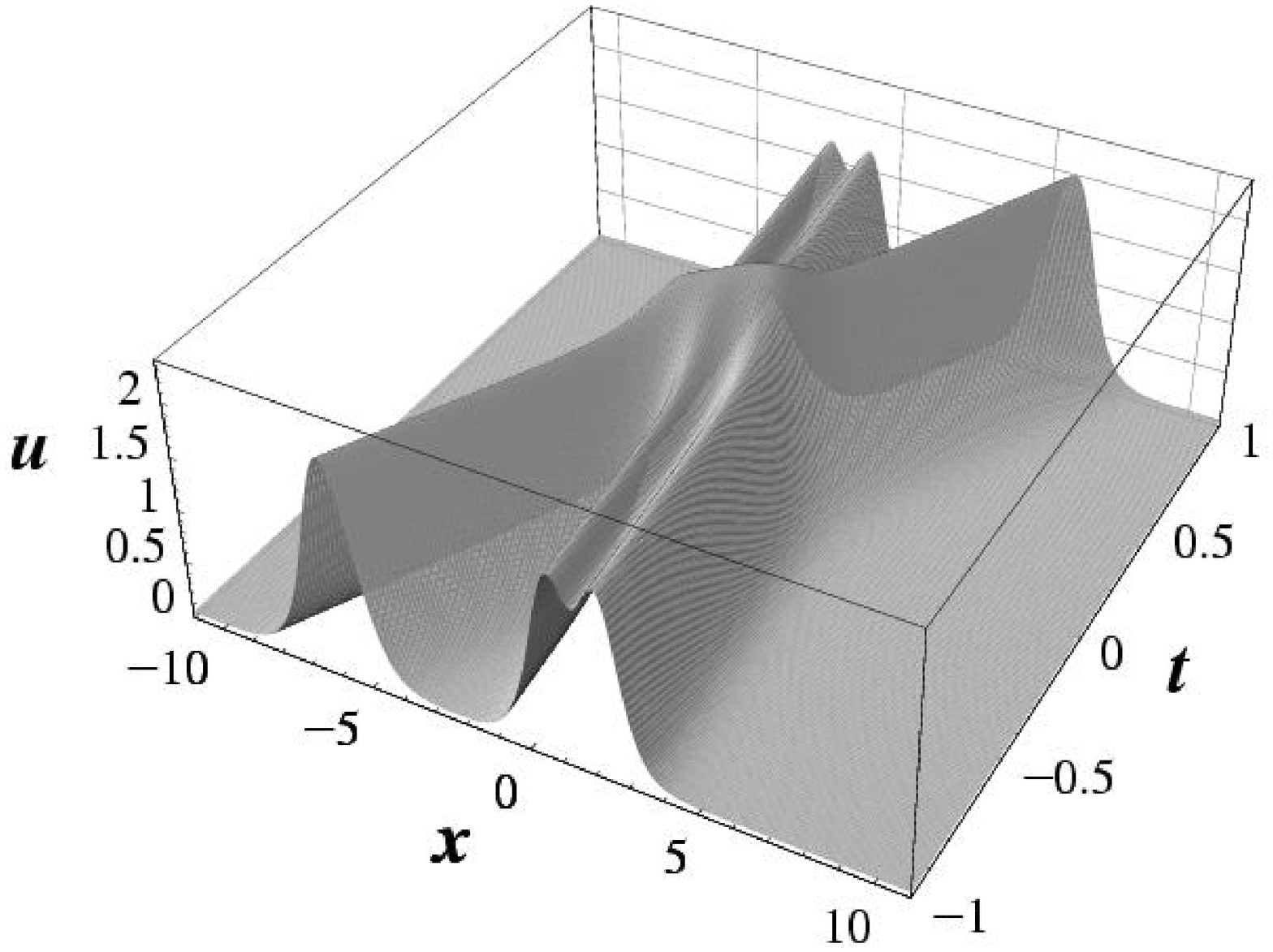}}
  \caption{\label{fig:bKKonePara}
Soliton solutions of the bKK equation (\ref{bKK}). Top left: one
left-going (two-peak) soliton when $\varepsilon_1={\pi \over 10}$
and $k_1=1.2$. Top right: two left-going soliton when $k_1=2,
k_2=1.3, \varepsilon_1=\varepsilon_2={\pi \over 10}$. Bottom:
Head-on collision of left- and right-going solitons when $k_1=1.8,
k_2=1.3, \varepsilon_1={3\pi \over 10},\varepsilon_2={\pi \over
10}$.
  }

\end{figure}

\begin{figure}[htbp]
\centerline{
  \includegraphics[width=0.5\textwidth]{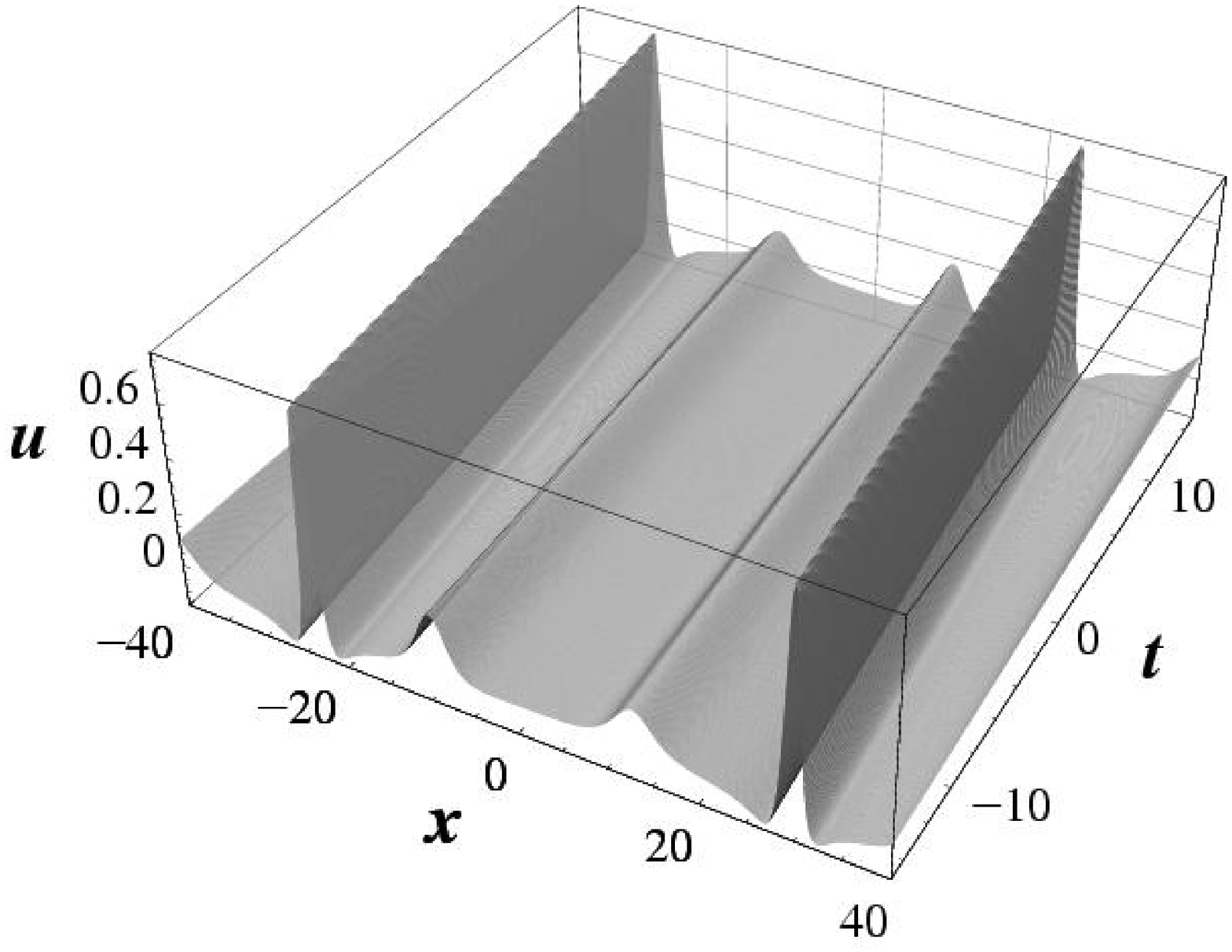}
  \includegraphics[width=0.5\textwidth]{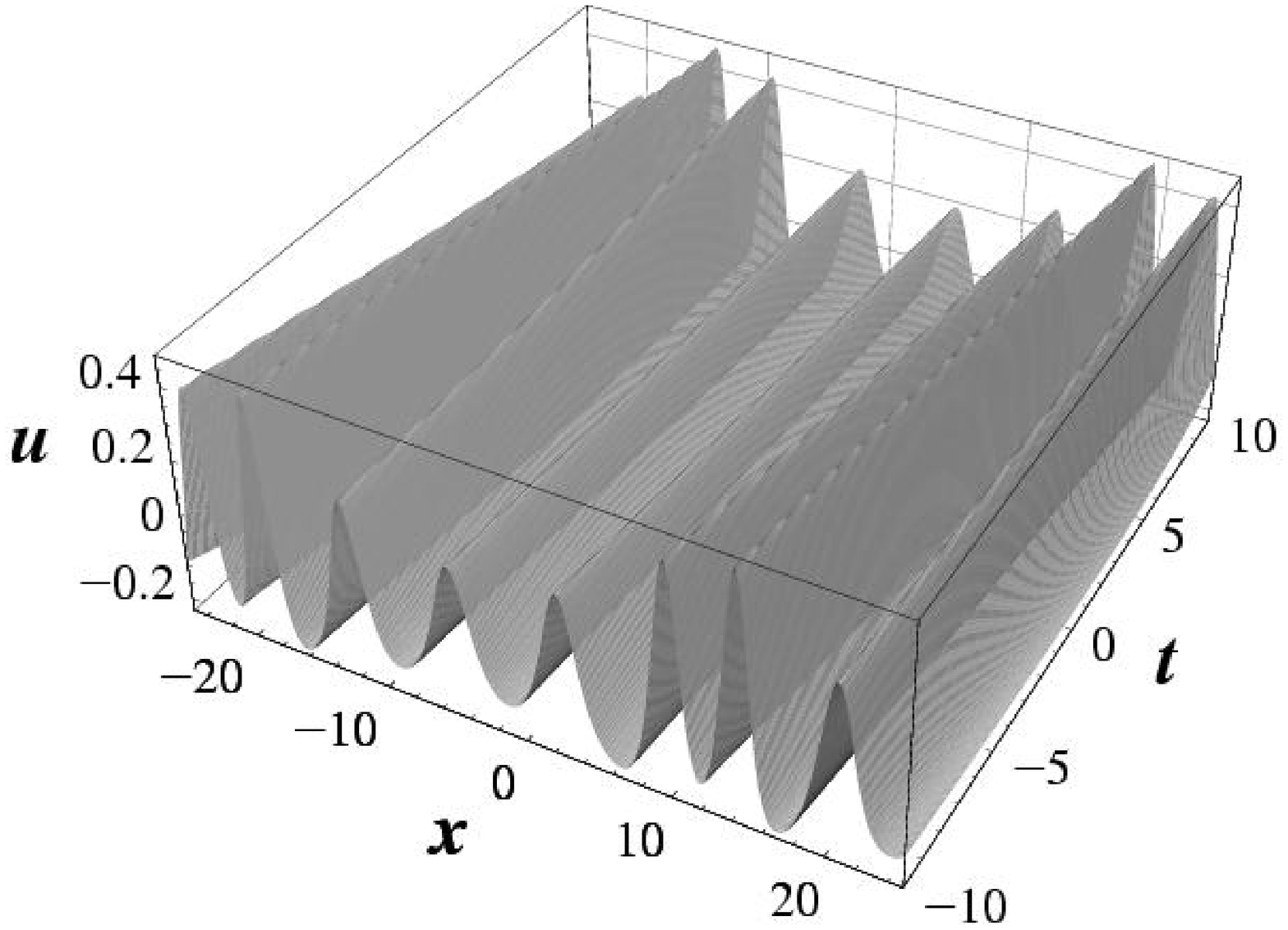}
}
\centerline{
  \includegraphics[width=0.5\textwidth]{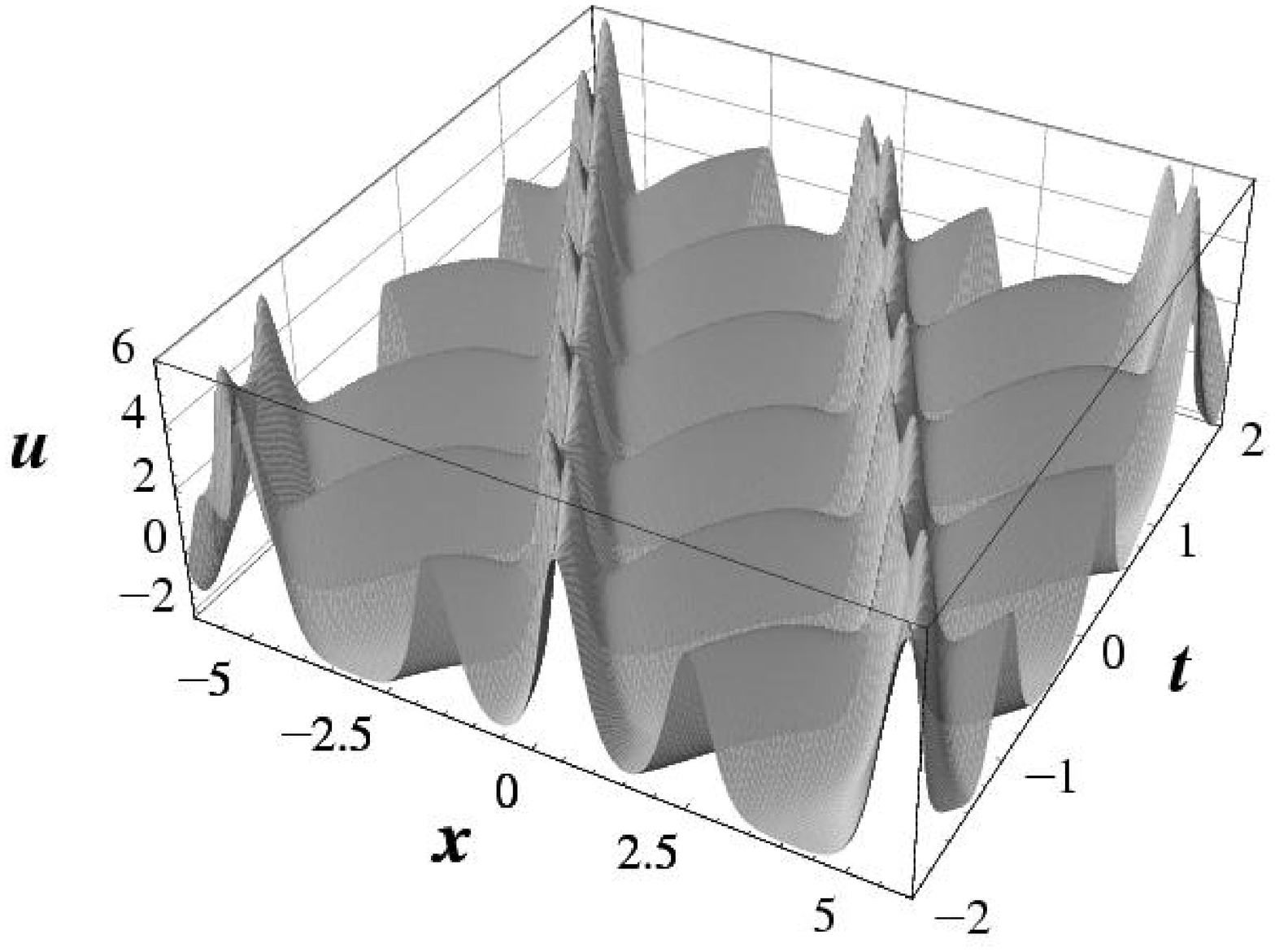}
}

  \caption{\label{fig:bKKtwoPara}
    Periodic solutions with two spectral parameters of  bKK equation
    (\ref{bKK}).
    Top left: left-going periodic solution with $k_1=0.2,
k_2=0.3, \varepsilon_1=\varepsilon_2={2\pi \over 10}$. Top right:
right-going periodic solution when $k_1=0.4, k_2=0.5,
\varepsilon_1=\varepsilon_2={4\pi \over 10}$.
    Bottom: Collision of left- and right-going
    periodic solution when $k_1=1,
k_2=1.5, \varepsilon_1={2\pi \over 10}, \varepsilon_2={4\pi \over
10}$.}
\end{figure}

\begin{proposition}\label{proptwosolitonbKK}
\begin{align}
&\hat{\tau}^{(2+2)}_{\rm bKK}=   \nonumber \\ 
&\Big\{ \frac{ e^{2(\xi_1+\xi_2)}
}{4k_1k_2\big(k^2_1+k^2_2+2k_1k_2\cos(\varepsilon_1-\varepsilon_2)
\big)^2} + \frac{g_9 e^{-2(\xi_1+\xi_2)}
}{4k_1k_2\big(k^2_1+k^2_2+2k_1k_2\cos(\varepsilon_1-\varepsilon_2)
\big)^2} \nonumber \\ 
&+ \frac{ e^{2(\xi_1-\xi_2)}
}{4k_1k_2\big(k^2_1+k^2_2-2k_1k_2\cos(\varepsilon_1+\varepsilon_2)
\big)^2} + \frac{ e^{-(2\xi_1-\xi_2)}
}{4k_1k_2\big(k^2_1+k^2_2-2k_1k_2\cos(\varepsilon_1+\varepsilon_2)
\big)^2}                   \nonumber \\  
&+\frac{e^{2\xi_1}}
 {2k_1k_2\sin\varepsilon_2\big(k^2_1+k^2_2+2k_1k_2\cos(\varepsilon_1-\varepsilon_2)
\big)\big(k^2_1+k^2_2-2k_1k_2\cos(\varepsilon_1+\varepsilon_2)
\big)}                        \nonumber \\ 
&+\frac{g_8e^{-2\xi_1}}
 {2k_1k_2\sin\varepsilon_2\big(k^2_1+k^2_2+2k_1k_2\cos(\varepsilon_1-\varepsilon_2)
\big)\big(k^2_1+k^2_2-2k_1k_2\cos(\varepsilon_1+\varepsilon_2)
\big)}     \nonumber  \\  
&+\frac{e^{2\xi_2}}
 {2k_1k_2\sin\varepsilon_1\big(k^2_1+k^2_2+2k_1k_2\cos(\varepsilon_1-\varepsilon_2)
\big)\big(k^2_1+k^2_2-2k_1k_2\cos(\varepsilon_1+\varepsilon_2)
\big)}     \nonumber  \\  
&+\frac{g_6e^{-2\xi_2}}
 {2k_1k_2\sin\varepsilon_2\big(k^2_1+k^2_2+2k_1k_2\cos(\varepsilon_1-\varepsilon_2)
\big)\big(k^2_1+k^2_2-2k_1k_2\cos(\varepsilon_1+\varepsilon_2)
\big)}     \nonumber  \\  
&+\frac{g_5\big((k^2_1+k^2_2)^2-4k^2_1k^2_2(\cos^2\varepsilon_1\cos^2\varepsilon_2+
\sin^2\varepsilon_1\sin^2\varepsilon_2) \big)}
 {k_1k_2\sin\varepsilon_1\sin\varepsilon_2\big(k^2_1+k^2_2+2k_1k_2\cos(\varepsilon_1-\varepsilon_2)
\big)\big(k^2_1+k^2_2-2k_1k_2\cos(\varepsilon_1+\varepsilon_2)
\big)}       
\Big\}
\end{align}
The two solitons solution is $u=(\partial_x^2\log
\hat{\tau}^{(2+2)}_{\rm bKK})$. In particular,
$\varepsilon_1=\varepsilon_2=\frac{\pi}{10}$ results in two
overtaking  soltions moving in negative direction;
$\varepsilon_1=\varepsilon_2=\frac{3\pi}{10}$ results in two
overtaking soltions moving in positive direction;
$\varepsilon_1=\frac{\pi}{10},\varepsilon_2=\frac{3\pi}{10}$
results in head-on colliding two soltions.
\end{proposition}
The second  case is a periodic solution with two spectral
parameters of bKK equation from Proposition \ref{propzerotaubKK}.
\begin{lemma}\label{lemperiodictau2+2bKK}
Let $\fun{\phi}{0}{i}=\phi(\lambda_i;x,t), i=1, 2$,  defined by
Eq.( \ref{bKKzerophi1b}), $\xi_i=xk_i\cos\varepsilon_i
+tk^3_i\cos3\varepsilon_i, \eta_i=xk_i\sin\varepsilon_i
+tk^3_i\sin3\varepsilon_i(i=1,2)$, then $\left.\tu{n+n}_{\rm
bKK}\right|_{n=2}$ gives
\begin{align}
&\tu{2+2}_{\rm bKK}=A^2_1A^2_2e^{2i(\xi_1+\xi_2)}\times  \nonumber \\ 
&\Big\{ \frac{z_1^* e^{2i(\eta_1+\eta_2)}
}{4\big(k^2_1+k^2_2+2k_1k_2\cos(\varepsilon_1-\varepsilon_2)
\big)^2} + \frac{z_1e^{-2i(\eta_1+\eta_2)}
}{4\big(k^2_1+k^2_2+2k_1k_2\cos(\varepsilon_1-\varepsilon_2)
\big)^2}\left(\frac{B_1}{A_1}\right)^2\left(\frac{B_2}{A_2}\right)^2 \nonumber \\ 
&+ \frac{z^*_3 e^{2i(\eta_1-\eta_2)}
}{4k_1k_2\big(k^2_1+k^2_2+2k_1k_2\cos(\varepsilon_1+\varepsilon_2)
\big)^2}\left( \frac{B_2}{A_2}\right)^2 + \frac{z_3
e^{-i(2\eta_1-\eta_2)}
}{4k_1k_2\big(k^2_1+k^2_2+2k_1k_2\cos(\varepsilon_1+\varepsilon_2)
\big)^2}\left( \frac{B_1}{A_1}\right)^2                    \nonumber \\  
&+\frac{z^*_2e^{2i\eta_1}}
 {2k_1k_2\cos\varepsilon_2\big(k^2_1+k^2_2+2k_1k_2\cos(\varepsilon_1-\varepsilon_2)
\big)\big(k^2_1+k^2_2+2k_1k_2\cos(\varepsilon_1+\varepsilon_2)
\big)}\left(\frac{B_2}{A_2}\right)                        \nonumber \\ 
&+\frac{z_2e^{-2i\eta_1}}
 {2k_1k_2\cos\varepsilon_2\big(k^2_1+k^2_2+2k_1k_2\cos(\varepsilon_1-\varepsilon_2)
\big)\big(k^2_1+k^2_2+2k_1k_2\cos(\varepsilon_1+\varepsilon_2)
\big)}\left(\frac{B_1}{A_1}\right)^2\left(\frac{B_2}{A_2}\right)     \nonumber  \\  
&+\frac{z^*_4e^{2i\eta_2}}
 {2k_1k_2\cos\varepsilon_1\big(k^2_1+k^2_2+2k_1k_2\cos(\varepsilon_1-\varepsilon_2)
\big)\big(k^2_1+k^2_2+2k_1k_2\cos(\varepsilon_1+\varepsilon_2)
\big)}\left(\frac{B_1}{A_1}\right)     \nonumber  \\  
&+\frac{z_4e^{-2i\eta_2}}
 {2k_1k_2\cos\varepsilon_1\big(k^2_1+k^2_2+2k_1k_2\cos(\varepsilon_1-\varepsilon_2)
\big)\big(k^2_1+k^2_2+2k_1k_2\cos(\varepsilon_1+\varepsilon_2)
\big)}\left(\frac{B_1}{A_1}\right)\left(\frac{B_2}{A_2}\right)^2     \nonumber  \\  
&+\frac{\big((k^2_1+k^2_2)^2-4k^2_1k^2_2(\cos^2\varepsilon_1\cos^2\varepsilon_2+
\sin^2\varepsilon_1\sin^2\varepsilon_2) \big)}
 {k_1k_2\cos\varepsilon_1\cos\varepsilon_2\big(k^2_1+k^2_2+2k_1k_2\cos(\varepsilon_1-\varepsilon_2)
\big)\big(k^2_1+k^2_2+2k_1k_2\cos(\varepsilon_1+\varepsilon_2)
\big)}\left(\frac{B_1}{A_1}\right)\left(\frac{B_2}{A_2}\right)        
\Big\}   \label{eqperiodictwoptau}
\end{align}
Here $z_i, i=1, 2, 3, 4$, are given in  \ref{appbKKP}.
$z_i^*$ indicates the complex conjugation of $z_i$.
\end{lemma}
Similar to the two solitons solution of bKK equation, we need
following two Lemmas to find suitable $\frac{B_i}{A_i}, i=1, 2$,
to extract physical $\hat{\tau}^{(2+2)}_{\rm bKK}$ from equation
(\ref{eqperiodictwoptau}) for periodic solution.
\begin{lemma} For $z_i, i=1, 2, 3, 4$,  as given in  \ref{appbKKP},
  the following identities
\begin{equation}
z_2^2=z_1z_3, \quad z_4^2=z_1z_3^*
\end{equation}
hold.
\end{lemma}
\begin{lemma}\label{lemperiodicA1B1A2B2bKK}
Let $z_k=|z_k|e^{i\theta_k}, k=1, 2, 3, 4 $,  be as given in
 \ref{appbKKP}, and
$\frac{B_1}{A_1}=e^{-i\theta_2},\frac{B_2}{A_2}=e^{-i\theta_4}$,
and $g_2=\frac{|z_2|}{|z_1|}$, $ g_3=\frac{|z_3|}{|z_1|}$,
$g_4=\frac{|z_4|}{|z_1|}$, $g_5=\frac{1}{|z_1|}$, then
\begin{align}
&\frac{z_3^*}{z_1^{*}}\left(\frac{B_2}{A_2} \right)^2=\frac{z_3}{z_1^{*}}\left(\frac{B_1}{A_1}\right)^2=g_3,    \\
&\frac{z_1}{z_1^*}\left(\frac{B_1}{A_1}\right)^2\left(\frac{B_2}{A_2}\right)^2=1,\quad
\frac{1}{z_1^{*}}\left(\frac{B_1}{A_1}
\right)\left(\frac{B_2}{A_2} \right)=g_5,  \\ 
&\frac{z_2^*}{z_1^*}\frac{B_2}{A_2}=\frac{z_2}{z_1^{*}}\left(\frac{B_1}{A_1}\right)^2
\left(\frac{B_2}{A_2}\right)=g_2 ,\quad
\frac{z_4^*}{z_1^*}\frac{B_1}{A_1}= \frac{z_4}{z_1^{*}}
\left(\frac{B_1}{A_1} \right)^2\left(\frac{B_2}{A_2} \right)=g_4
\end{align}
hold.
\end{lemma}
We can get the physical $\tau$ function $\hat{\tau}^{(2+2)}_{\rm
bKK}$ by taking $\frac{B_i}{A_i}, i=1, 2$ and relations in Lemma
\ref{lemperiodicA1B1A2B2bKK} back into Lemma
\ref{lemperiodictau2+2bKK}.
\begin{proposition}\label{propperiodictwoparameterbKK}
\begin{align}
&\hat{\tau}^{(2+2)}_{\rm bKK}=  \nonumber \\ 
&\Big\{ \frac{2\cos2(\eta_1+\eta_2)}
{4\big(k^2_1+k^2_2+2k_1k_2\cos(\varepsilon_1-\varepsilon_2)
\big)^2}  + \frac{2g_3 \cos2(\eta_1-\eta_2)
}{4k_1k_2\big(k^2_1+k^2_2+2k_1k_2\cos(\varepsilon_1+\varepsilon_2)
\big)^2}                \nonumber \\  
&+\frac{2g_2\cos2\eta_1}
 {2k_1k_2\cos\varepsilon_2\big(k^2_1+k^2_2+2k_1k_2\cos(\varepsilon_1-\varepsilon_2)
\big)\big(k^2_1+k^2_2+2k_1k_2\cos(\varepsilon_1+\varepsilon_2)
\big)}                        \nonumber \\ 
&+\frac{2g_4\cos2\eta_2}
 {2k_1k_2\cos\varepsilon_1\big(k^2_1+k^2_2+2k_1k_2\cos(\varepsilon_1-\varepsilon_2)
\big)\big(k^2_1+k^2_2+2k_1k_2\cos(\varepsilon_1+\varepsilon_2)
\big)}    \nonumber  \\  
&+\frac{g_5\big((k^2_1+k^2_2)^2-4k^2_1k^2_2(\cos^2\varepsilon_1\cos^2\varepsilon_2+
\sin^2\varepsilon_1\sin^2\varepsilon_2) \big)}
 {k_1k_2\cos\varepsilon_1\cos\varepsilon_2\big(k^2_1+k^2_2+2k_1k_2\cos(\varepsilon_1-\varepsilon_2)
\big)\big(k^2_1+k^2_2+2k_1k_2\cos(\varepsilon_1+\varepsilon_2)
\big)} \Big\}
\end{align}
The periodic solution with two parameters $k_1$ and $k_2$ is
$u=\left(\partial_x^2\log \hat{\tau}^{(2+2)}_{\rm bKK}\right)$.
Furthermore, $\varepsilon_1=\varepsilon_2=\frac{2\pi}{10}$ results
in  two overtaking waves moving in negative direction ;
$\varepsilon_1=\varepsilon_2=\frac{4\pi}{10}$ results in two
overtaking waves moving in positive direction;
$\varepsilon_1=\frac{2\pi}{10},\varepsilon_2=\frac{4\pi}{10}$
results in two head-on colliding waves.
\end{proposition}
We have plotted soliton solutions of bKK in figure
\ref{fig:bKKonePara}, and there periodic solutions with two
spectral parameters in Fig. \ref{fig:bKKtwoPara}.

\section{Periodic and soliton solutions of bSH equation}\label{sectbSH}
The $\tau$ function of the bSH equation is still in the form of a
Grammian although the bSH equation does not belong to the CKP
hierarchy, which is obtained in \cite{vm2} through the B\"acklund
transformation. Similar to the bKK equation, its $\tau$ function
is in the form of Grammian, we can find  $\tau$ function
$\tau^{(1+1)}_{\rm bSH}$ and $\tau^{(2+2)}_{\rm bSH}$of bSH from
Grammian $\tau$ function. Let the initial value be $u = 0$ in
equations (\ref{bSHlax1}) and (\ref{bSHlax2}), then
$\fun{\phi}{0}{i}=\phi(\lambda_i;x,t)$ are solutions of
\begin{align} \label{zerobSHlax1}
& \partial_x^4\phi(\lambda_i;x,t)=\lambda_i \phi(\lambda_i;x,t),
\quad \frac{\partial \phi(\lambda_i;x,t)}{\partial
t}=\left(\partial_x^{3}\phi(\lambda_i;x,t)\right).
\end{align}
\begin{proposition}[see Ref.\ \cite{vm2}]\label{propzerotaubSH}
The $\tau$ function of {\rm bSH} equation  generated by B\"acklund
transformation from initial value $u=0$ is
\begin{align}
&\fun{\tau}{n+n}{\rm bSH} = IW_{n,n}(\fun{\phi}{0}{n},
\fun{\phi}{0}{n-1}, \cdots, \fun{\phi}{0}{1};
      \fun{\phi}{0}{1}, \fun{\phi}{0}{2},\cdots , \fun{\phi}{0}{n} ) \notag \\
&=\left|
\begin{array}{lccccc}
\int \fun{\phi}{0}{n}\cdot\fun{\phi}{0}{1} &
\int\fun{\phi}{0}{n}\cdot\fun{\phi}{0}{2} &
\int\fun{\phi}{0}{n}\cdot\fun{\phi}{0}{3} &\cdots
&\int \fun{\phi}{0}{n}\cdot\fun{\phi}{0}{n-1} &\int\fun{\phi}{0}{n}\cdot\fun{\phi}{0}{n}\\  
\int \fun{\phi}{0}{n-1}\cdot\fun{\phi}{0}{1} & \int
\fun{\phi}{0}{n-1}\cdot\fun{\phi}{0}{2} &\int
\fun{\phi}{0}{n-1}\cdot\fun{\phi}{0}{3} &\cdots
& \int\fun{\phi}{0}{n-1}\cdot\fun{\phi}{0}{n-1} &\int \fun{\phi}{0}{n-1}\cdot\fun{\phi}{0}{n}\\  
\vdots&\vdots&\vdots&\cdots&\vdots\\        
 \int \fun{\phi}{0}{2}\cdot\fun{\phi}{0}{1} &
  \int\fun{\phi}{0}{2}\cdot\fun{\phi}{0}{2}&\int
\fun{\phi}{0}{2}\cdot\fun{\phi}{0}{3} &\cdots
 &\int\fun{\phi}{0}{2}\cdot\fun{\phi}{0}{n-1} &\int
\fun{\phi}{0}{2}\cdot\fun{\phi}{0}{n}\\ 
\int\fun{\phi}{0}{1}\cdot\fun{\phi}{0}{1} & \int
\fun{\phi}{0}{1}\cdot\fun{\phi}{0}{2} &\int
\fun{\phi}{0}{1}\cdot\fun{\phi}{0}{3} &\cdots
 &\int\fun{\phi}{0}{1}\cdot\fun{\phi}{0}{n-1} &\int
\fun{\phi}{0}{1}\cdot\fun{\phi}{0}{n}
\end{array} \right|
\label{taubSHn+n}
\end{align}
and the solution $u$ of {\rm bSH}  from initial value $zero$ is
\begin{equation}
u=\left(\partial_x^2 \log \fun{\tau}{n+n}{\rm bSH}\right)
\label{zeroubSH}
\end{equation}
Here $\fun{\phi}{0}{i}=\phi(\lambda_i;x,t)$ are solutions of
 equation (\ref{zerobSHlax1}).
\end{proposition}
In fact, $\fun{\tau}{n+n}{\rm bSH}$ can be generated by gauge
transformation $\left.T_{n+k}\right|_{n=k}$. The Lax pair of bSH
is\vspace{-0.2cm}
\begin{align*}
&L_{\rm bSH}=\partial_x^4 +4 u \partial_x^2 +4u_x \partial_x +
2u_{xx} +4 u^2+v, \quad M_{\rm bSH}=\partial^3_x +3u\partial_x
+\frac{3}{2}u_x,
\end{align*}
and satisfy $L^*_{\rm bSH}=L_{\rm bSH}, M_{\rm bSH}^*=-M_{\rm
bSH}$. Similar to the CKP hierarchy, let
$T=T_{1+1}=T_I(\fun{\psi}{1}{1})T_D(\fun{\phi}{0}{1})$, and do
gauge transformation  $L_{\rm bSH}^{(2)}=TL_{\rm bSH}T^{-1}$. So
$(L^{(2)}_{\rm bSH})^*=L^{(2)}_{\rm bSH}$ requires
$T_D(\fun{\psi}{1}{1})T_I(\fun{\phi}{0}{1})=
T_I(\fun{\psi}{1}{1})T_D(\fun{\phi}{0}{1})$ as we have seen in CKP
hierarchy. The remaining procedure  is the same as the gauge
transformation of the CKP hierarchy as well as the bKK equation.
Of course, the generating functions
$(\fun{\phi}{0}{i},\fun{\psi}{0}{i})=\left(\phi(\lambda_i; x,t),
\psi(\lambda_i;x,t)\right)$ satisfy equation (\ref{bSHlax1}) and equation
(\ref{bSHlax2}) if the initial values are $u\not=0,v\not=0$, or
equation (\ref{zerobSHlax1}) if the initial values are $u=0,v=0$.
\begin{remark}\label{remleftgoingsolitonbSH}
We should note that $\left. L_{\rm bSH}\right|_{v=0}=\partial_x^4
+4 u
\partial_x^2 +4u_x \partial_x + 2u_{xx} +4
u^2=(\partial_x^2+2u)^2=L^2_{\rm KdV}$. The Lax pair of the KdV
equation is\vspace{-0.2cm}
\begin{align*}
&L_{\rm KdV}=\partial_x^2+2u,\quad M_{\rm KdV}=\partial^3_x
+3u\partial_x +\frac{3}{2}u_x.
\end{align*}
$T_D(\phi^{(0)}_1)$ generates  a single  soliton solution  of the
KdV from zero initial value.  Here
$\phi^{(0)}_1=\phi(\lambda_1;x,t)$ satisfy $L_{\rm
KdV}\phi(\lambda_1;x,t)=\lambda_1\phi(\lambda_1;x,t)$ and
$\frac{\partial \phi(\lambda_1;x,t)}{\partial t}=M_{\rm
KdV}\phi(\lambda_1;x,t) $ simultaneously. The left-going
multi-soliton can be produced by using repeated iteration of
$T_D$.
\end{remark}

\par
In order to get real and smooth solutions, such as soliton and
periodic solution, we should construct {\em physical} $\tau$
function $\hat{\tau}_{\rm bSH}$ from $\fun{\tau}{n+n}{\rm bSH}$
which is complex and related to 4-th roots of $e^{i\varepsilon}$.
The case of $n=1$ and $n=2$ will be discussed in detail. Let us
start to discuss the single soliton with two directional
propagation. To do this, similar to the above two sections, we
should assume the solution of equation (\ref{zerobSHlax1}) as
\begin{align}
&\phi(\lambda_1;x,t)=A_1 e^{p_1 x+ p_1^3 t}+B_1 e^{q_1 x+ q_1^3
t},\quad p_1=k_1 e^{i \varepsilon_1}, q_1=-k_1 e^{-i
\varepsilon_1}, k_1^4=|\lambda_1|,k_1\in \mathbb{R},
\label{bSHzerophi1a}
 \intertext{or }
 &\phi(\lambda_1;x,t)=A_1 e^{p_1 x+ p_1^3 t}+B_1
e^{q_1 x+ q_1^3 t},\quad p_1=k_1 e^{i \varepsilon_1}, q_1= k_1
e^{-i \varepsilon_1},k_1^4=|\lambda_1|,k_1\in \mathbb{R}.
\label{bSHzerophi1b}
\end{align}
For $\phi(\lambda_i;x,t), i=1, 2$, the difference between here and
above two sections is the  $k_i^4=|\lambda_i|, i=1, 2$,  instead
of $k_i^5=|\lambda_i|, i=1, 2$. From Proposition
\ref{propzerotaubSH} we can extract {\em physical} $\tau$ function
$\hat{\tau}^{(1+1)}_{\rm bSH}$ from $\left.\tu{n+n}_{\rm
bSH}\right|_{n=1}$.
\begin{proposition}\label{proponesolitonbSH}
Let $\xi_1= xk_1\cos \varepsilon_1+tk_1^3\cos 3\varepsilon_1$,
$\frac{B_1}{A_1}=ie^{-\varepsilon_1}$,
 and $\fun{\phi}{0}{1}=\phi(\lambda_1;x,t)$ as defined by
equation (\ref{bSHzerophi1a}), then the physical  $\tau$ function of
{\rm bSH} extracted from $\tau^{(n+n)}_{\rm bSH}|_{n=1}$ is
\begin{align}
&\hat{\tau}^{(1+1)}_{\rm bSH}= e^{2 \xi_1}+ e^{-2
\xi_1}+ \frac{2}{\sin\varepsilon_1}\\
\intertext{and the corresponding single soliton $u
=\left(\partial^2_x \log\hat{\tau}^{(1+1)}_{\rm bSH}\right)$  is }
&u=\frac{4 k_1^2 (\cos\varepsilon_1)^2\left(1+ \frac{\cosh
2\xi_1}{\sin\varepsilon_1} \right) }
 { \left(\cosh2\xi_1+\frac{1}{\sin\varepsilon_1} \right)^2}.  \label{rightgoingonesolitonbSH}
\end{align}
Here  $\varepsilon_1=\frac{\pi}{4}$. The velocity of the soliton
is $v=-k_1^2\frac{\cos 3\varepsilon_1 }{\cos
\varepsilon_1}|_{\varepsilon_1=\frac{\pi}{4}}>0$.
\end{proposition}
\begin{proof}
\begin{equation}
\tu{1+1}_{\rm bSH}=\int
(\fun{\phi}{0}{1})^2=\frac{A^2_1e^{2i(xk_1\sin
\varepsilon_1+tk_1^3\sin 3\varepsilon_1)}}{2p_1}\left(
e^{2\xi_1}+e^{-2\xi_1} + \frac{2}{\sin\varepsilon_1}\right)
\end{equation}
\end{proof}
As we discussed in Remark \ref{remleftgoingsolitonbSH}, the
left-going soliton can also be generated by $T_D$.

\begin{proposition}\label{propleftgoingonesolitonbSH}
Let $\xi_1=( xk_1\cos \varepsilon_1+tk_1^3\cos
3\varepsilon_1)|_{\varepsilon_1=0}$,
 and $\fun{\phi}{0}{1}=\phi(\lambda_1;x,t)|_{\varepsilon_1=0}$ as defined by
equation (\ref{bSHzerophi1a}), then the physcial $\tau$ function of
{\rm bSH} generated by $T_D(\fun{\phi}{0}{1})$ is
\begin{align}
&\hat{\tau}^{(1)}_{\rm bSH}= 1+\frac{A_1}{B_1} e^{2 \xi_1}\\
\intertext{and the corresponding single soliton $u
=\left(\partial^2_x \log\hat{\tau}^{(1)}_{\rm bSH} \right)$ is }
&u=\frac{4 k_1^2\frac{A_1}{B_1}}
 { \big(e^{-\xi_1}+\frac{A_1}{B_1}e^{\xi_1} \big)^2}.  \label{leftgoingonesolitonbSH}
\end{align}
Here $\frac{A_1}{B_1}>0$. The velocity of the soliton is
$v=-k_1^2<0$.
\end{proposition}
\begin{proof}
\begin{equation}
\tu{1}_{\rm
bSH}=\fun{\phi}{0}{1}=B_1\left(e^{-\xi_1}+\frac{A_1}{B_1}e^{\xi_1}
\right)
\end{equation}
It can be clarified by $(u=0,v=0)
\xrightarrow{T_D(\phi^{(0)}_{1})}(u^{(1)}\not=0, v^{(1)}=0)$, and
then  $\tu{1}_{\rm bSH}=\fun{\phi}{0}{1}$.
\end{proof}
On the other hand, if $\fun{\phi}{0}{1}=\phi(\lambda_1;x,t)$ as
defined by equation (\ref{bSHzerophi1b}), then we can get periodic
solution from Proposition \ref{propzerotaubSH}.

\begin{proposition}\label{proponeparameterbSH}
Let $\eta_1= xk_1\sin \varepsilon_1+tk_1^3\sin 3\varepsilon_1$,
$A_1=B_1=1$ in $\fun{\phi}{0}{1}$,  then the physical $\tau$
function of {\rm bSH} equation for periodic solution extracted
from $\tau^{(n+n)}_{\rm bSH}|_{n=1}$ is
\begin{align}
&\hat{\tau}^{(1+1)}_{\rm bSH}= \frac{1}{\cos\varepsilon_1}+ \cos(2\eta_1-\varepsilon_1)\\
\intertext{and the corresponding peroidic solution $u
=\left(\partial^2_x \log\hat{\tau}^{(1+1)}_{\rm bSH}\right)$  is }
&u=\frac{-4k_1^2\sin^2\varepsilon_1\left(\frac{\cos(2\eta_1-2\varepsilon_1)}{\cos\varepsilon_1}
+1 \right) }
 { \big(\frac{1}{\cos\varepsilon_1}+ \cos(2\eta_1-\varepsilon_1))^2}.  \label{oneparameterbSH}
\end{align}
Here $\varepsilon_1=\frac{\pi}{4}$. The velocity of the solution
is $v=-k_1^2\frac{\sin 3\varepsilon_1 }{\sin
\varepsilon_1}|_{\varepsilon_1=\frac{\pi}{4}}<0$.
\end{proposition}
\begin{proof}
\begin{equation}
\tu{1+1}_{\rm bSH}=\int (\fun{\phi}{0}{1})^2=\frac{e^{2(xk_1\cos
\varepsilon_1+tk_1^3\cos 3\varepsilon_1)}}{4k_1}\left(
\frac{1}{\cos\varepsilon_1}+\cos(2\eta_1-\varepsilon_1) \right).
\end{equation}
\end{proof}
There are some relationship between the distributions of 4-th
roots of $e^{i\varepsilon}$ and moving direction of solutions.
\begin{enumerate}
\item\
$(e^{i\varepsilon_1},-e^{-i\varepsilon_1})|_{\varepsilon_1=0}$ the
first distribution of 4-th roots of $e^{i\varepsilon}$
$\longrightarrow$
$(p_1=k_1e^{i\varepsilon_1},q_1=-k_1e^{-i\varepsilon_1})|_{\varepsilon_1=0}$
in equation (\ref{bSHzerophi1a}) $\longrightarrow$ left-going  soliton
in equation (\ref{leftgoingonesolitonbSH}); \item\
$(e^{i\varepsilon_1},-e^{-i\varepsilon_1})|_{\varepsilon_1={\pi\over
4}}$ the second distribution of 4-th roots of $e^{i\varepsilon}$
$\longrightarrow$
$(p_1=k_1e^{i\varepsilon_1},q_1=-k_1e^{-i\varepsilon_1})|_{\varepsilon_1={\pi\over
4}}$ in equation (\ref{bSHzerophi1a}) $\longrightarrow$ right-going
soliton in equation (\ref{rightgoingonesolitonbSH});
 \item\
$(e^{i\varepsilon_1},e^{-i\varepsilon_1})|_{\varepsilon_1={\pi\over
4}}$ the third distribution of 4-th roots of $e^{i\varepsilon}$
$\longrightarrow$
$(p_1=k_1e^{i\varepsilon_1},q_1=k_1e^{-i\varepsilon_1})|_{\varepsilon_1={\pi\over
4}}$ in equation (\ref{bSHzerophi1b}) $\longrightarrow$ left-going
periodic wave in equation (\ref{oneparameterbSH}).
\end{enumerate}

In the above discussion, we know the right-going soliton and
left-going periodic wave of of bSH  have the completely same form
with the bKK equation, except $\varepsilon_1=\pi/4$ instead of
$\varepsilon_1=\pi/10$ and $\varepsilon_1=3\pi/10$.  The reason is
that the $\tau$ function of two equations is  in the same Grammian
of generating functions $\fun{\phi}{0}{i}$, and generating
functions $\fun{\phi}{0}{i}$ for two equations satisfy analogous
linear partial differential equations with constant coefficients,
i.e.  equation (\ref{zerobKKlax1}) for bKK equation, equation
(\ref{zerobSHlax1}) for bSH equation. These relations between bKK
and bSH are still true for their two soliton and two parameters
periodic solutions.

\begin{proposition}\label{propright-goingtwosolitonbSH}
The two right-going  solitons are given
\begin{align}
&u=\left(\partial_x^2\log \hat{\tau}^{(2+2)}_{\rm bSH}\right)   \\
\intertext{in which $\hat{\tau}^{(2+2)}_{\rm bSH}$ is}
&\hat{\tau}^{(2+2)}_{\rm bSH}=\hat{\tau}^{(2+2)}_{\rm
bKK}|_{\varepsilon_1=\varepsilon_2=\pi/4}
\end{align}
and $\hat{\tau}^{(2+2)}_{\rm bKK}$ is given by Proposition
\ref{proptwosolitonbKK}.
\end{proposition}
\begin{proposition}\label{propright-goingperiodictwoparameterbKK}
The right-going periodic wave with two spectral parameters $k_1$
 and $k_2$ is given by
\begin{align}
&u=\left(\partial_x^2\log \hat{\tau}^{(2+2)}_{\rm bSH}\right)   \\
\intertext{in which $\hat{\tau}^{(2+2)}_{\rm bSH}$ is}
&\hat{\tau}^{(2+2)}_{\rm bSH}=\hat{\tau}^{(2+2)}_{\rm
bKK}|_{\varepsilon_1=\varepsilon_2=\pi/4}
\end{align}
and $\hat{\tau}^{(2+2)}_{\rm bKK}$ is given by Proposition
\ref{propperiodictwoparameterbKK}.
\end{proposition}

According to the analysis in Remark \ref{remleftgoingsolitonbSH},
two left-going solitons of bSH equation can be generated by a
chain of gauge transformations $(u=0,v=0)
\xrightarrow{T_D(\phi^{(0)}_{1})}\left(u^{(1)}\not=0,
v^{(1)}=0\right)
\xrightarrow{T_D(\phi^{(1)}_{2})}\left(u^{(2)}\not=0,
v^{(2)}=0\right) $(using the notation of \cite{hlc1}),
$\fun{\phi}{0}{i}=\phi(\lambda_i;x,t)|_{\varepsilon_1=0}, i=1, 2$,
are defined by equation (\ref{bSHzerophi1a}). Their $\tau$ function of
bSH generated by $T_2=T_D(\phi^{(1)}_{2})T_D(\phi^{(0)}_{1})$ is
\begin{align}
\tau^{(2)}_{\rm bSH}= \left| \begin{array}{cc}
 \fun{\phi}{0}{1}&
\fun{\phi}{0}{2}\\
 \fun{\phi}{0}{1,x}&
\fun{\phi}{0}{2,x}
 \end{array} \right|.
\end{align}
From $\tau^{(2)}_{\rm bSH}$ we can obtain  the {\em physical}
$\tau$ function $\hat{\tau}^{(2)}_{\rm bSH}$ and two soliton
solution.
\begin{proposition}\label{twoleft-goingsoliton}
Let $\fun{\phi}{0}{i}=\phi(\lambda_i;x,t)|_{\varepsilon_1=0}$ are
defined by equation (\ref{bSHzerophi1a}), $\xi_i=k_ix+k_i^3t, i=1, 2$.
If $\frac{B_1}{A_1}>0,\frac{B_2}{A_2}<0,k_2>k_1$, then the
physical $\tau$ function  $\hat{\tau}^{(2)}_{\rm bSH}$  is given
by
\begin{eqnarray}
&\hat{\tau}^{(2)}_{\rm
bSH}=&(k_2-k_1)e^{\xi_1+\xi_2}-\frac{B_1}{A_1}\frac{B_2}{A_2}(k_1+k_2)e^{-(\xi_1+\xi_2)}
\nonumber\\
& &-(k_2-k_1)\frac{B_2}{A_2}e^{\xi_1-\xi_2}+\frac{B_1}{A_1}
(k_1+k_2) e^{-(\xi_1-\xi_2)}.
\end{eqnarray}
The two soliton solution is $u=\left(\partial^2_x \log
\hat{\tau}^{(2)}_{\rm bSH}\right)$, which is left-going.
\end{proposition}
The collision of two soliton is generated by gauge transformation
chain
$(u=0,v=0)\xrightarrow{T_D(\phi^{(0)}_{1})}\left(u^{(1)}\not=0,
v^{(1)}=0\right)
\xrightarrow{T_I(\psi^{(2)}_{2})T_D(\phi^{(1)}_{2})}\left(u^{(2)}\not=0,
v^{(2)}\not=0\right)$,
$\fun{\phi}{0}{1}=\phi(\lambda_1;x,t)|_{\varepsilon_1=0}$ is defined
by equation (\ref{bSHzerophi1a}),
$\fun{\psi}{0}{2}=\fun{\phi}{0}{2}$, $\fun{\phi}{0}{2}=
\phi(\lambda_2;x,t)$ is defined by equation (\ref{bSHzerophi1a}).
The corresponding $\tau$ function of bSH is
\begin{align}\label{collisiontwosolitonbSH}
\tau^{(2+1)}_{\rm bSH}= \left| \begin{array}{cc}
 \int\fun{\psi}{0}{2}\fun{\phi}{0}{1}&
\fun{\psi}{0}{2}\fun{\phi}{0}{2}\\
 \fun{\phi}{0}{1}&
\fun{\phi}{0}{2}
 \end{array} \right|\stackrel{\fun{\psi}{0}{2}=\fun{\phi}{0}{2}}{===}\left| \begin{array}{cc}
 \int\fun{\phi}{0}{2}\fun{\phi}{0}{1}&
\fun{\phi}{0}{2}\fun{\phi}{0}{2}\\
 \fun{\phi}{0}{1}&
\fun{\phi}{0}{2}
 \end{array} \right|.
\end{align}
Taking $\fun{\phi}{0}{i}, i=1, 2$, back into equation
(\ref{collisiontwosolitonbSH}), we have its explicit expression as
following Lemma.

\begin{lemma}\label{lemcollisiontwosolitonbSH}
Let $\xi_1=xk_1+tk_1^3,
\xi_2=xk_2\cos\varepsilon_2+tk_2^3\cos3\varepsilon_2,
\eta_2=xk_2\sin\varepsilon_2+tk_2^3\sin3\varepsilon_2,
z_i=c_i+d_i, i=1, 3, 5$, as given in  \ref{appbSH}.
\begin{align}
&\tau^{(2+1)}_{\rm bSH}=e^{2i\eta_2}A_2^2A_1\times \nonumber  \\
&\Big\{ \frac{z_1^*e^{\xi_1+2\xi_2}}{2k_2\big(
k^2_1+k^2_2+2k_1k_2\cos\varepsilon_2 \big)}
-\frac{z_1e^{-\xi_1-2\xi_2}}{2k_2\big(
k^2_1+k^2_2+2k_1k_2\cos\varepsilon_2 \big)}
\left(\frac{B_2}{A_2}\right)^2\left(\frac{B_1}{A_1}\right) \nonumber \\
&+\frac{z_3^*e^{-\xi_1+2\xi_2}}{2k_2\big(
k^2_1+k^2_2-2k_1k_2\cos\varepsilon_2 \big)}\left(
\frac{B_1}{A_1}\right)
 -\frac{z_3e^{\xi_1-2\xi_2}}{2k_2\big(
k^2_1+k^2_2-2k_1k_2\cos\varepsilon_2 \big)}
\left(\frac{B_2}{A_2}\right)^2                    \nonumber        \\
&+\frac{z_5e^{\xi_1}}{ik_2\sin\varepsilon_2\big(
k^2_1+k^2_2+2k_1k_2\cos\varepsilon_2 \big) \big(
k^2_1+k^2_2-2k_1k_2\cos\varepsilon_2 \big) }\left(
\frac{B_2}{A_2}\right)                 \nonumber             \\
&+\frac{z_5^*e^{-\xi_1}}{ik_2\sin\varepsilon_2\big(
k^2_1+k^2_2+2k_1k_2\cos\varepsilon_2 \big)\big(
k^2_1+k^2_2-2k_1k_2\cos\varepsilon_2 \big)}
\left(\frac{B_2}{A_2}\right)  \left(\frac{B_1}{A_1}\right)              
 \Big\}.
\end{align}
\end{lemma}
\begin{lemma}\label{relationcollisiontwosolitonbSH1}
For $z_i=|z_i|e^{\theta_i}(i=1,3,5)$, the following identities
\begin{equation}
z_1z_5=z_3z_5^*, \quad e^{i(\theta_3-\theta_1)}= e^{2i\theta_3}
\end{equation}
are true.
\end{lemma}

\begin{figure}[htbp]
  \centerline{
  \includegraphics[width=0.5\textwidth]{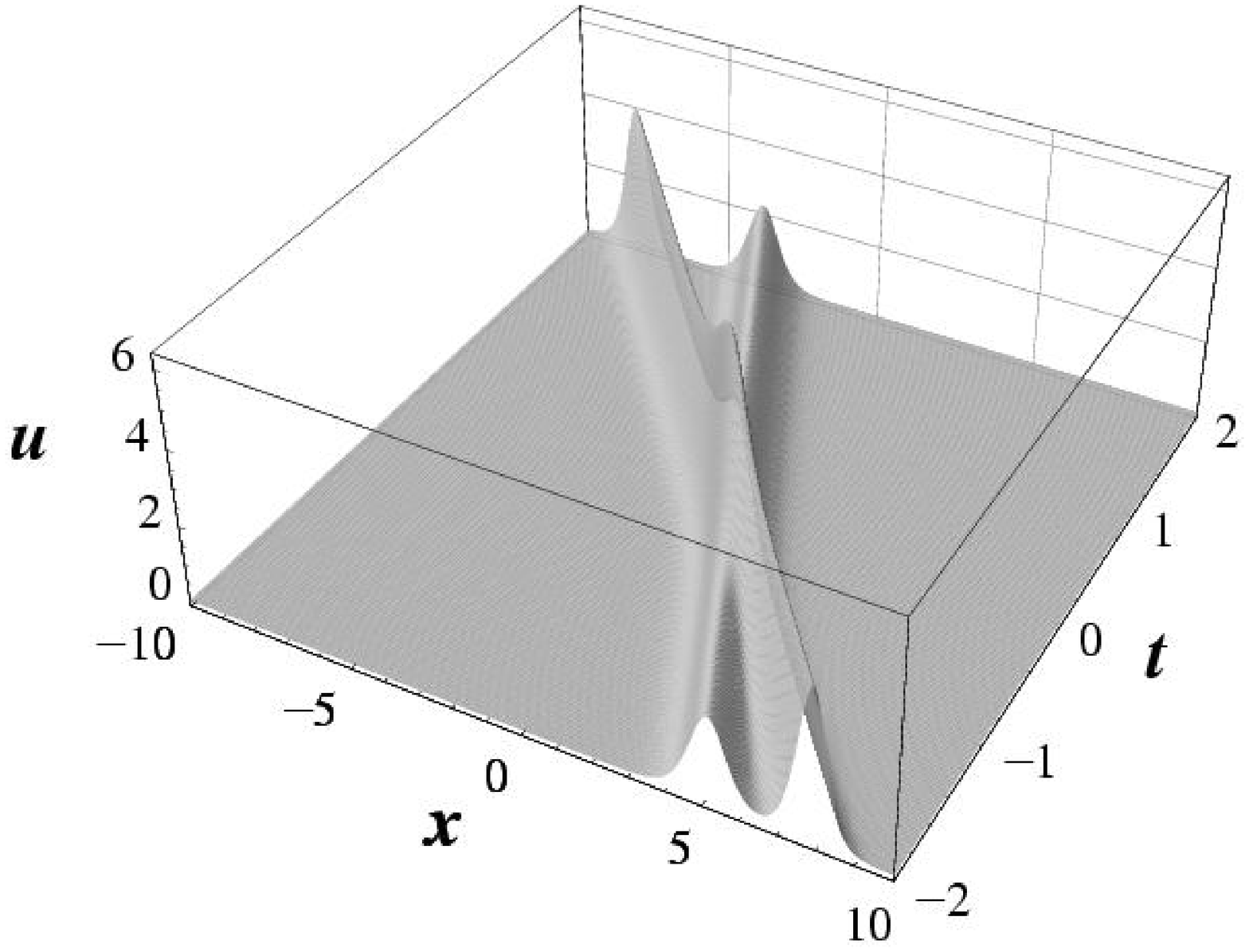}
  \includegraphics[width=0.5\textwidth]{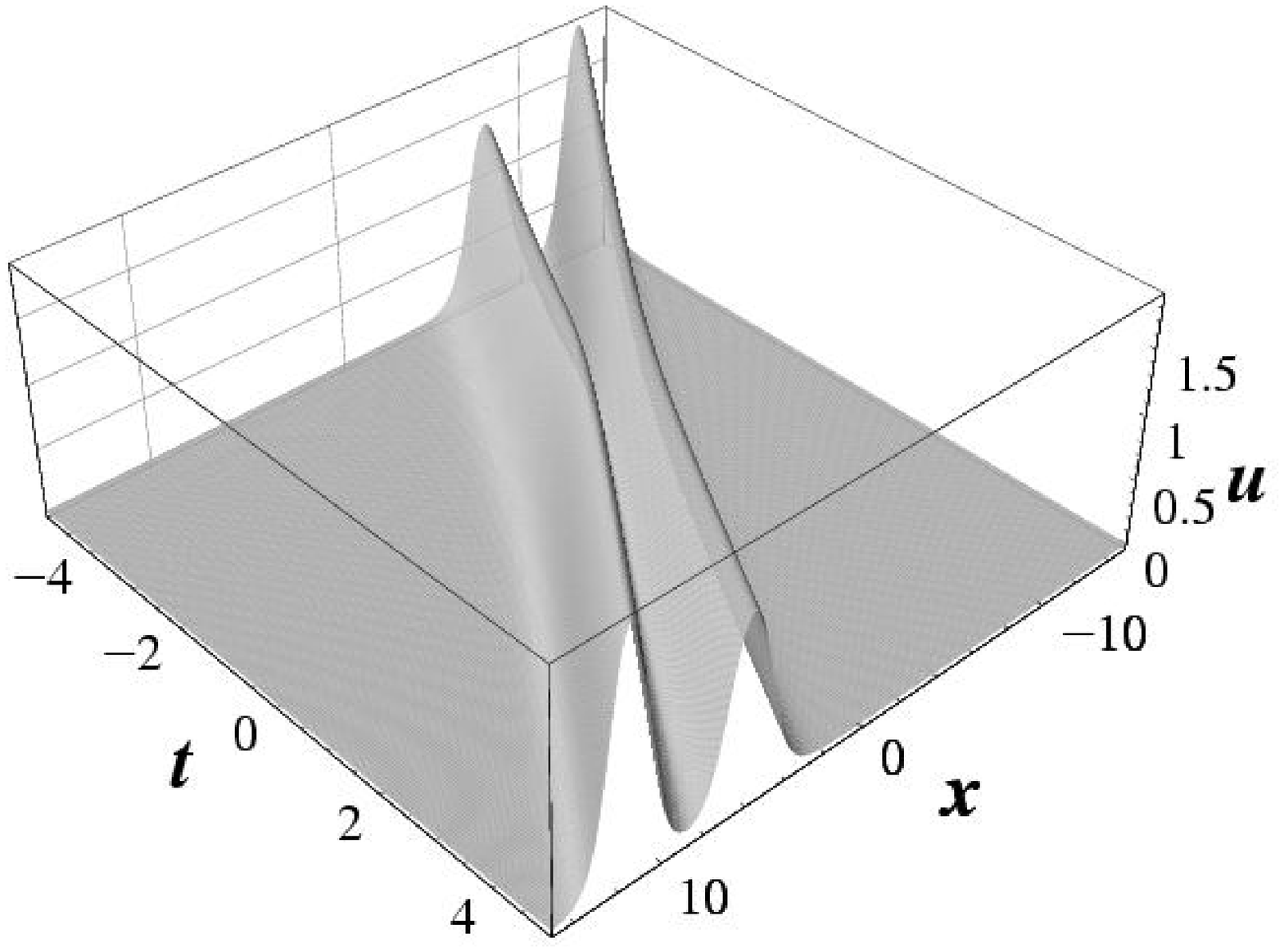}
  }
  \centerline{
  \includegraphics[width=0.5\textwidth]{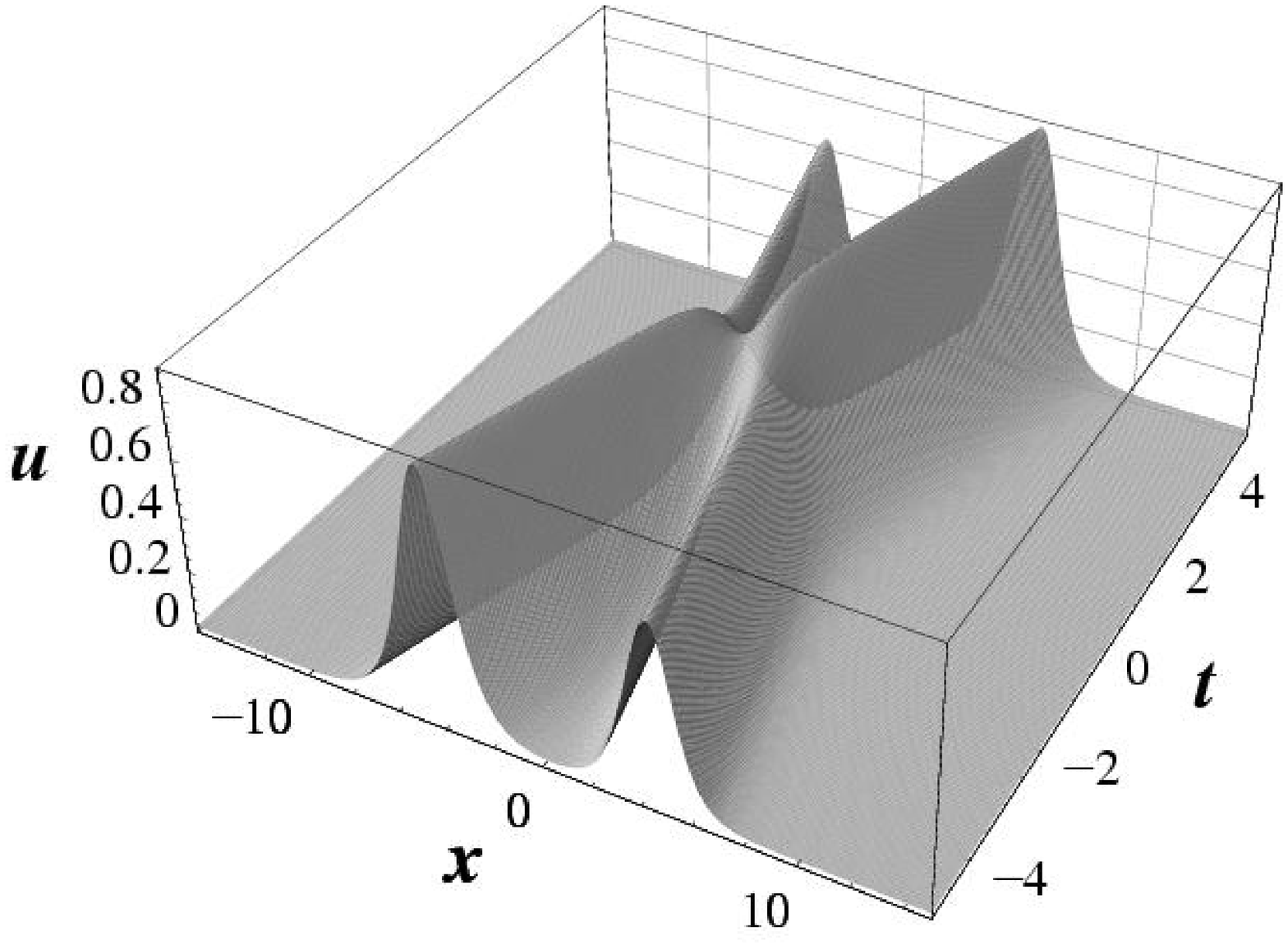}
  }
  \caption{\label{fig:bSHsoliton}
    Two left- and right-going as well as head-on colliding solitons for
    the bSH equation (\ref{bSH}).  Parameters are chosen as:
$A_1=B_1, A_2=2, B_2=-1, k_1=1.5, k_2=2 $ (left);
    $k_1=1.5, k_2=1.3, \varepsilon_1=\varepsilon_2=\frac{\pi}{4}$
    (right);
    $k_1=0.8, k_2=0.9, \varepsilon_2=\frac{\pi}{4}$
    (collision).}
\end{figure}

\begin{figure}[htbp]
\centerline{
  \includegraphics[width=0.5\textwidth]{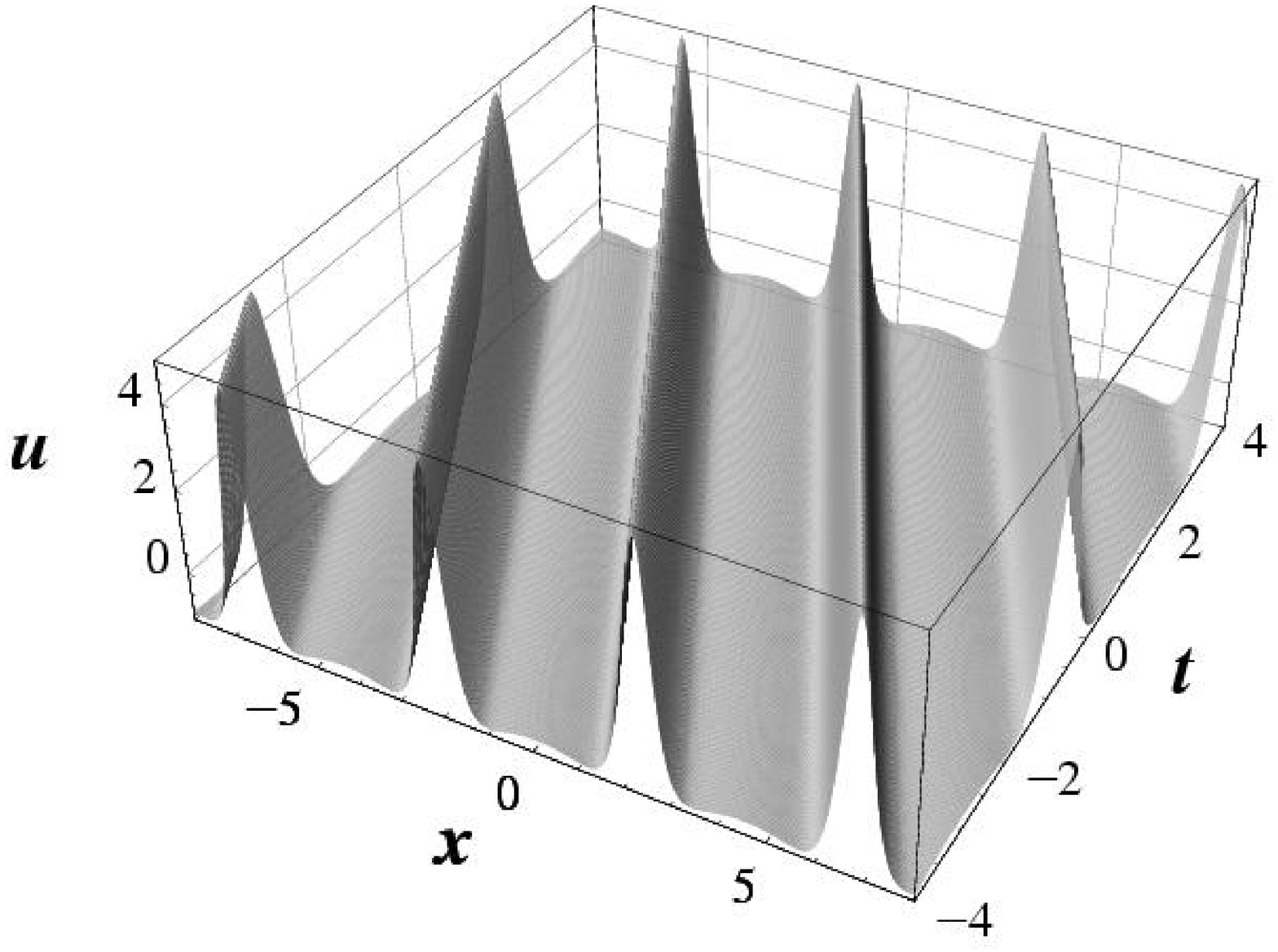}
  \includegraphics[width=0.5\textwidth]{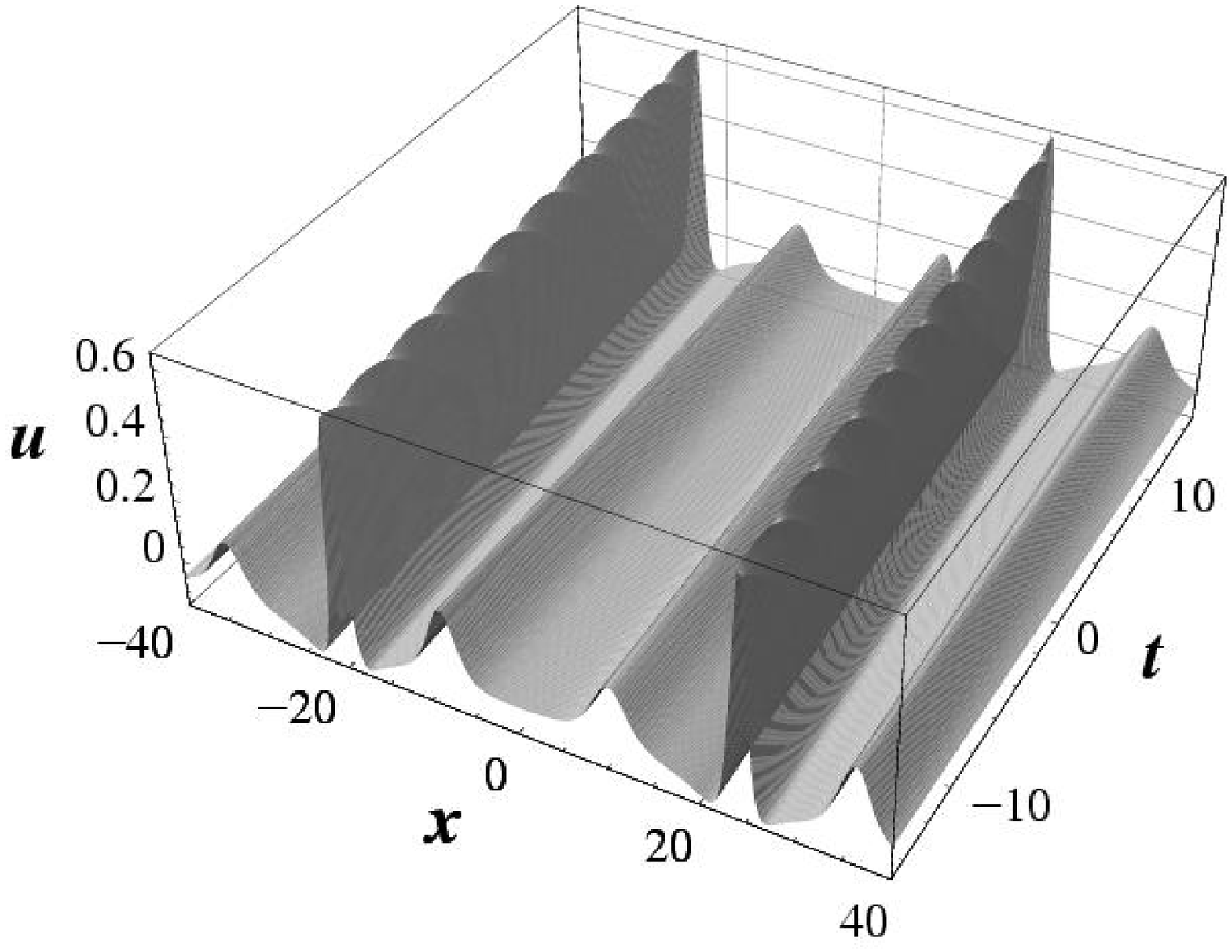}
}
  \caption{\label{fig:bSHperiodic}
    Left-going periodic solutions with one (left) and two (right)
    spectral parameters for the bSH equation (\ref{bSH}).
Parameters: $k_1=1, \varepsilon_1=\frac{\pi}{4}$(left); $k_1=0.2,
k_2=0.3, \varepsilon_1=\varepsilon_2=\frac{\pi}{4} $(right).
    }
\end{figure}

\begin{lemma}\label{relationcollisiontwosolitonbSH2}
Let $z_i, i=1, 3, 5$, be as given by  \ref{appbSH}, if
$\frac{B_1}{A_1}=\frac{z_1^*}{z_3^*}$,
 $\frac{B_2}{A_2}=i\frac{z_1^*}{z_5}$,
 $g_2=g_4=\frac{|z_1|^2}{|z_5|^2}$, $g_6=1$, then
\begin{equation}
g_2=-\frac{z_1}{z_1^*}\left(\frac{B_2}{A_2}
\right)^2\left(\frac{B_1}{A_1} \right), \quad
g_4=-\frac{z_3}{z_1^*}\left(\frac{B_2}{A_2} \right)^2,\
g_6=\frac{z_5^*}{iz_1^*}\left(\frac{B_2}{A_2}
\right)\left(\frac{B_1}{A_1} \right),
\end{equation}
hold.
\end{lemma}
With the help of Lemmata \ref{relationcollisiontwosolitonbSH1} and
\ref{relationcollisiontwosolitonbSH2}, we deduce the {\em physical}
$\tau$ function of colliding two soliton of bSH equation from Lemma
\ref{lemcollisiontwosolitonbSH}.
\begin{proposition}\label{propcollisiontwosolitonbSH}
Let $g_2, g_4$ be given as in Lemma
\ref{relationcollisiontwosolitonbSH2}, then
\begin{align}
&\hat{\tau}^{(2+1)}_{\rm bSH}=\Big\{
\frac{e^{\xi_1+2\xi_2}}{2k_2\big(
k^2_1+k^2_2+2k_1k_2\cos\varepsilon_2 \big)}
+\frac{g_2e^{-\xi_1-2\xi_2}}{2k_2\big(
k^2_1+k^2_2+2k_1k_2\cos\varepsilon_2 \big)}
 \nonumber \\
&\mbox{}\hspace{1cm}+\frac{e^{-\xi_1+2\xi_2}}{2k_2\big(
k^2_1+k^2_2-2k_1k_2\cos\varepsilon_2 \big)}
 +\frac{g_4e^{\xi_1-2\xi_2}}{2k_2\big(
k^2_1+k^2_2-2k_1k_2\cos\varepsilon_2 \big)}   \nonumber        \\
&\mbox{}\hspace{1cm}+\frac{e^{\xi_1}}{k_2\sin\varepsilon_2\big(
k^2_1+k^2_2+2k_1k_2\cos\varepsilon_2 \big) \big(
k^2_1+k^2_2-2k_1k_2\cos\varepsilon_2 \big) }       \nonumber             \\
&\mbox{}\hspace{1cm}+\frac{e^{-\xi_1}}{k_2\sin\varepsilon_2\big(
k^2_1+k^2_2+2k_1k_2\cos\varepsilon_2 \big)\big(
k^2_1+k^2_2-2k_1k_2\cos\varepsilon_2 \big)}
 \Big\}
\end{align}
\end{proposition}
We have plotted the two soliton solutions of bSH equation in figure
\ref{fig:bSHsoliton}, and  periodic solutions with one spectral
parameter and with two spectral parameters of the same equation in
figure \ref{fig:bSHperiodic}.

\section{Lower and Higher order reductions}
In this section, we want to discuss the general character of
soliton equation from lower order to higher order in one same
sub-hierarchy. The purpose is to show the relation between
propagation of soliton on (x,t) plane and the order of Lax pair,
and show the difference between the lower reduction and higher
reduction.  Let Lax pair of soliton equation is $(L,M)$, which
defines $\phi(\lambda;x,t)$ by
\begin{equation}\label{laxpair}
L\phi(\lambda;x,t)=\lambda\phi(\lambda;x,t),  \quad
\frac{\partial \phi(\lambda;x,t)}{\partial t}=M\phi(\lambda;x,t).
\end{equation}
There are some examples of n-reduction of the KP hierarchy.  For
the BKP hierarchy,
\begin{equation}\label{higherordereqBKP}
\left.
\begin{array}{lllll}
\text{Lax pair} \quad &\text{3-reduction}\quad        &\text{5-reduction}\quad    &\text{7-reduction}\quad  &\text{9-reduction}  \\
\text{L}        &B_3                       &B_5                   &B_7                  &B_9 \\
\text{M}        &\widetilde{B}_5           &B_3                   &B_3                  &B_3\\
\text{Equation}  &\text{SK}                &\text{bSK}
&\text{higher\ order} &\text{higher\ order}
\end{array}\right\}.
\end{equation}
Here
\begin{align}
&\widetilde{B}_5=\partial_x^5 +5 u\partial^3_x +5u_x
\partial_x^2+\Big(5u^2+\frac{10}{3}u_{xx}
\Big)\partial_x,\\
&\text{SK\cite{sk,cdg}}:\quad 9u_t+
45u^2u_{x}+u_{xxxxx}+15uu_{xxx}+15u_xu_{xx}=0\label{sk}.
\end{align}
$B_5$ and $B_3$ are given by equations (\ref{bSKlax1}) and
(\ref{bSKlax2}). For the CKP hierarchy,
\begin{equation}\label{higherordereqCKP}
\left.
\begin{array}{lllll}
\text{Lax pair}\quad  &\text{3-reduction}\quad        &\text{5-reduction}\quad    &\text{7-reduction}\quad  &\text{9-reduction}  \\
\text{L}        &B_3                       &B_5                   &B_7                  &B_9 \\
\text{M}        &\widetilde{B}_5           &B_3                   &B_3                  &B_3\\
\text{Equation}  &\text{KK}                        &\text{bKK} &
\text{higher\ order} &\text{higher\ order}
\end{array}\right\}.
\end{equation}
Here
\begin{align}
&\widetilde{B}_5=\partial_x^5 +5 u\partial^3_x +\frac{15}{2}u_x
\partial_x^2+\Big(5u^2+\frac{35}{6}u_{xx}
\Big)\partial_x+ 5uu_x+\frac{5}{3}u_{xxx},\\
& \text{KK\cite{kaup,kupershmidt}}:\quad  9u_t+
45u^2u_{x}+u_{xxxxx}+15uu_{xxx}+\frac{75}{2}u_xu_{xx}=0,
\label{kk}
\end{align}
 $B_5$ and $B_3$  are given in equations
(\ref{bKKlax1}) and (\ref{bKKlax2}). There are several
even-reductions of the KP hierarchy as following,
\begin{equation}\label{evenorderKP}
\left.
\begin{array}{lllll}
\text{Lax pair}\quad  &\text{2-reduction}\quad        &\text{4-reduction}\quad    &\text{6-reduction}\quad  &\text{8-reduction}      \\
\text{L}        &B_2                        &B_4\;                   &B_6\;       &B_8 \\
\text{M}        &B_3                        &B_3\;                   &B_3\;        &B_3\\
\text{Equation} &\text{KdV}                 &\text{bSH}\; &
higher\ order\; & higher\  order
\end{array}
\right\}.
\end{equation}

\par
Now we start to discuss the BKP hierarchy.
\begin{lemma}\label{lemonesolitontauSK}
Let $\widetilde{\xi_1}=xk_1\cos\varepsilon_1+
tk_1^5\cos5\varepsilon_1$, then $\hat{\tau}^{(1+1)}_{\rm SK}$ is
expressed by
\begin{equation}\label{onesolitontauSK}
\hat{\tau}^{(1+1)}_{\rm SK}=\hat{\tau}^{(1+1)}_{\rm
bSK}|_{\xi_1->\widetilde{\xi}_1}
\end{equation}
and the corresponding single soliton is $u=\left(\partial_x^2
\log\hat{\tau}^{(1+1)}_{\rm SK} \right)$.  The  velocity of
soliton  $v_{\_}=\left.-k_1^4\frac{\cos 5\varepsilon_1}{\cos
\varepsilon_1}\right|_{\varepsilon_1={\pi\over 6}}=k_1^4>0$. Here
$\hat{\tau}^{(1+1)}_{\rm bSK}$ is given by Proposition
\ref{proponesolitonbSK}.
\end{lemma}
\begin{proof}
  Because the SK equation and bSK equation belong to the same
  sub-hierarchy BKP, so the results of bSK are also hold by SK equation
  only if we replace $\xi_1$ in bSK by
  $\widetilde{\xi_1}=xk_1\cos\varepsilon_1+ tk_1^5\cos5\varepsilon_1$.
  For SK equation, the generating functions
  $\fun{\phi}{(0)}{i}=\phi(\lambda_i;x,t)$ of gauge transformation
  satisfy
\begin{equation}\label{zerolaxSK}
\partial_x^3\phi(\lambda;x,t)=\lambda \phi(\lambda;x,t), \quad
\frac{\phi(\lambda;x,t)}{\partial t}=\partial^5_x
\phi(\lambda;x,t),
\end{equation}
which are different with equation (\ref{zerobSKlax1}) for bSK equation.
So $k_1^3=|\lambda_1|$. This difference determines replacement in
equation (\ref{onesolitontauSK}).  Of course, similar to the bSK, we
also should assume the solutions of equation (\ref{zerolaxSK}) be the
form of
\begin{align}
&\phi(\lambda_1;x,t)=A_1 e^{p_1 x+ p_1^5 t}+B_1 e^{q_1 x+ q_1^5
t}, p_1=k_1 e^{i \varepsilon_1}, q_1=-k_1 e^{-i \varepsilon_1},
k_1^3=|\lambda_1|,k_1\in \mathbb{R}, \label{SKzerophi1a}
\intertext{or } &\phi(\lambda_1;x,t)=A_1 e^{p_1 x+ p_1^5 t}+B_1
e^{q_1 x+ q_1^5 t},  p_1=k_1 e^{i \varepsilon_1}, q_1= k_1 e^{-i
\varepsilon_1},k_1^3=|\lambda_1|,k_1\in \mathbb{R}.
\label{SKzerophi1b}
\end{align}
Taking the generating functions $\fun{\phi}{(0)}{i}$ in Eq.
(\ref{SKzerophi1a}) back into the Proposition \ref{proptauBKP},
then we can extract $\hat{\tau}^{(1+1)}_{\rm SK}$ from
$\tau^{(1+1)}_{\rm SK}$. The relation $\hat{\tau}^{(1+1)}_{\rm
SK}=\hat{\tau}^{(1+1)}_{\rm bSK}|_{\xi_1->\widetilde{\xi}_1} $  is
given by comparison.
\end{proof}
In particular, there are two distributions of roots of third-order of
$e^{i\varepsilon}$ on circle, which is symmetric with respect to
$y$-axes. However, they are corresponding to same single soliton
solution.
\begin{enumerate}
\item  $\left( e^{i\frac{\pi}{6}}, -e^{-i\frac{\pi}{6}}\right)$
one distribution of 3-order root of $e^{i\varepsilon}$ on unit
circle $\longrightarrow$ $\big(p_1=k_1e^{i\frac{\pi}{6}}$,
$q_1=-k_1e^{-i\frac{\pi}{6}}\big)$ in equation (\ref{SKzerophi1a})
$\longrightarrow$ a single soliton in Lemma
\ref{lemonesolitontauSK}.

\item $\left(e^{i\frac{11\pi}{6}}, -e^{i\frac{-11\pi}{6}}\right)$
one distribution of 3-order root of $e^{i\varepsilon}$ on unit
circle $\longrightarrow$ $\big(p_1=k_1e^{i\frac{11\pi}{6}}$,
$q_1=-k_1e^{-i\frac{11\pi}{6}}\big)$ in
  equation (\ref{SKzerophi1a}) $\longrightarrow$ one soliton as (1).
\end{enumerate}

\begin{lemma}\label{lemonesolitontaunBKP}
The higher order equations of the {\rm BKP} hierarchy are defined
by equation (\ref{higherordereqBKP}).  For the n-reduction equation of
the {\rm BKP} hierarchy {\rm (nBKP)}, $n=2j+1, j=3, 4, 5, \cdots
$, and let $\widetilde{\xi}_{mp}=xk_m\cos
\varepsilon_p+tk^3_m\cos3\varepsilon_p,
k_m^n=k_m^{2j+1}=|\lambda_m|$, then the physical $\tau$ function
of the {\rm nBKP} generated by $T_{1+1}$ is
\begin{equation}
\hat{\tau}^{(1+1)}_{\rm nBKP}=\hat{\tau}^{(1+1)}_{\rm
bSK}|_{\xi_{1}->\widetilde{\xi}_{1p}},
\end{equation}
and the corresponding single soliton of the {\rm nBKP} is
$u=(\partial_x^2 \log\hat{\tau}^{(1+1)}_{\rm nBKP} )$. Here
$\varepsilon_p=\frac{2p-1}{2n}\pi=\frac{2p-1}{4j+2}\pi, p=1,
2,3,\cdots, j$,  $\hat{\tau}^{(1+1)}_{\rm bSK}$ is given by
Proposition \ref{proponesolitonbSK}. So the single soliton can
move along $j$ directions in (x,t) plane, which are given by
$\widetilde{\xi}_{1p}=0$ associated with $j$-value of
$\varepsilon_p$ given before.
\end{lemma}

\begin{proof}
Comparing the nBKP with the bSK equation,  the main change here is
the Lax pair (L,M). The Lax pair of the nBKP defines the
generating functions $\fun{\phi}{(0)}{i}=\phi(\lambda_i;x,t)$ are
slight different as
\begin{equation}
\partial_x^n\phi(\lambda;x,t)=\lambda \phi(\lambda;x,t), \quad
\frac{\phi(\lambda;x,t)}{\partial t}=\partial^3_x
\phi(\lambda;x,t)
\end{equation}
and then we assume
\begin{align}
&\phi(\lambda_1;x,t)=A_1 e^{p_1 x+ p_1^3 t}+B_1 e^{q_1 x+ q_1^3
t}, p_1=k_1 e^{i \varepsilon_p}, q_1=-k_1 e^{-i \varepsilon_p},
k_1^n=|\lambda_1|,k_1\in \mathbb{R}, \label{nBKPzerophi1a}
\intertext{or}
 &\phi(\lambda_1;x,t)=A_1 e^{p_1 x+ p_1^3 t}+B_1
e^{q_1 x+ q_1^3 t}, p_1=k_1 e^{i \varepsilon_p}, q_1= k_1 e^{-i
\varepsilon_p},k_1^n=|\lambda_1|,k_1\in \mathbb{R}.
\label{nBKPzerophi1b}
\end{align}
In order to avoid  the divergence of $u$, we only take
$0<\varepsilon_p<{\pi \over 2}$, and then
$\varepsilon_p=\frac{2p-1}{2n}\pi=\frac{2p-1}{4j+2}\pi, p=1,
2,3,\cdots, j$. This change results to the emergence of
$\widetilde{\xi}_{mp}=xk_m\cos
\varepsilon_p+tk^3_m\cos3\varepsilon_p,
k_m^n=k_m^{2j+1}=|\lambda_m|$. The $\hat{\tau}^{(1+1)}_{\rm nBKP}$
and single soliton solution $u=(\partial_x^2
\log\hat{\tau}^{(1+1)}_{\rm nBKP} )$ can be derived directly from
the Proposition
 \ref{proptauBKP} and the generating functions $\fun{\phi}{(0)}{1}$ in equation
 (\ref{nBKPzerophi1a}) associated with $\lambda_1$ for the gauge transformation.  Further, for a given $p$,
  $\widetilde{\xi}_{1p}=0$
 determines one moving direction of the single soltion on (x,t) plane, then the single
 soliton solution have $j$ directions for propagation because $p=1, 2, \cdots, j$.
\end{proof}
From Lemmata \ref{lemonesolitontauSK}, \ref{lemonesolitontaunBKP} and
the results of the bSK equation, we have
\begin{proposition}\label{propdirections of nBKP}
\begin{enumerate}
\item   The single soliton $u=\left(\partial_x^2
\log\hat{\tau}^{(1+1)}_{\rm nBKP} \right)$ of the {\rm nBKP}
equation, $n=2j+1, j=2, 3, 4, \cdots, $ can move along a direction
defined by $\widetilde{\xi}_{1p}=0$ on (x,t) plane for a given
$p$.

\item $\left( e^{i\varepsilon_p},-e^{-i\varepsilon_p}\right)$ one
distribution of n-th order roots of $e^{i\varepsilon}$ on circle
$\longrightarrow$ $\big(
p_1=k_1e^{i\varepsilon_p}$,$q_1=-k_1e^{-i\varepsilon_p}\big)$ in
equation (\ref{nBKPzerophi1a}) $\longrightarrow$ The single soliton
moves along a line $\widetilde{\xi}_{1p}=0$ on (x, t) plane. Here
$\varepsilon_p \in \Big\{\frac{\pi}{4j+2},\frac{3\pi}{4j+2},
\frac{5\pi}{4j+2}, \cdots,  \frac{(2j-1)\pi}{4j+2}\Big\}$.

\item For a given $n=2j+1$, the single soliton of the {\rm nBKP}
have $j$ directions to propagate on (x,t) plane, which are defined
$\widetilde{\xi}_{1p}=0, p=1, 2, 3, \cdots j.$
\end{enumerate}
\end{proposition}
Note that the result of $j=1$ in above
Proposition is given by Lemma \ref{lemonesolitontauSK}.

\begin{figure}[htbp]
  \centering
  \includegraphics[width=0.8\textwidth]{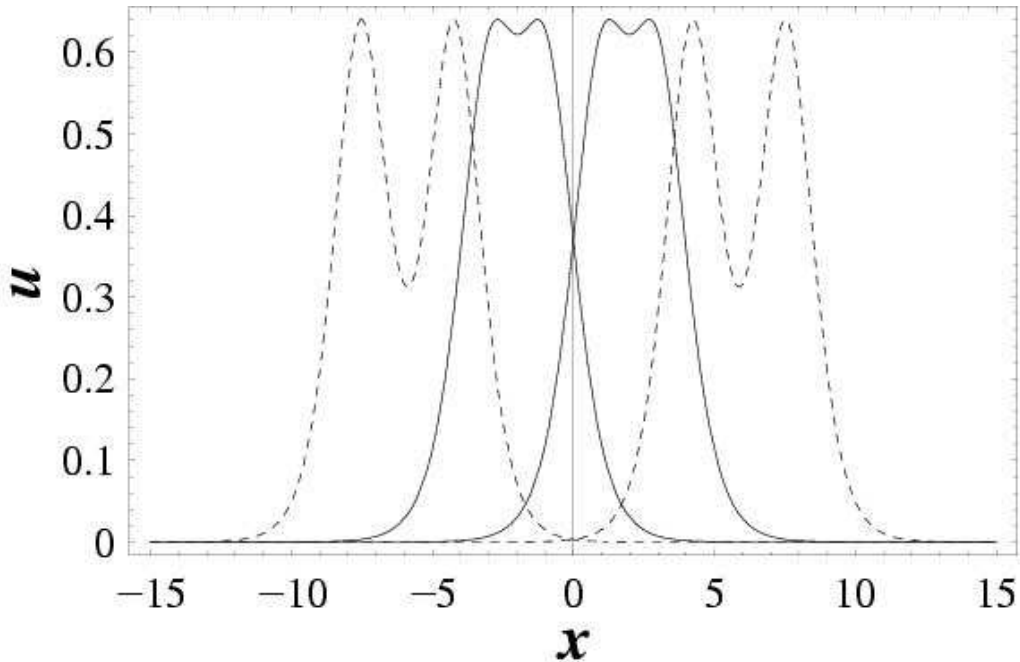}
  \caption{  \label{fig:11ckptwo two-peak}
    Left-going two-peak soliton with dashed line
    ($\varepsilon_1=\pi/22$) is faster, left-going two-peak soliton
    with full line ($\varepsilon_1=3\pi/22$). The left is plotted when
    $t=10$, the right is plotted when $t=-10$}
\end{figure}

Now we turn to the lower and higher reductions of the CKP
hierarchy. Similar to the discussion of the BKP hierarchy in this
section, we can obtain parallel results in the CKP hierarchy, so
we write out the results without proof in the following to save
space.
\begin{lemma}\label{lemonesolitontauKK}
Let $\widetilde{\xi_1}=xk_1\cos\varepsilon_1+
tk_1^5\cos5\varepsilon_1$, then $\hat{\tau}^{(1+1)}_{\rm KK}$ can
be expressed by
\begin{equation}\label{onesolitontauKK}
\hat{\tau}^{(1+1)}_{\rm KK}=\hat{\tau}^{(1+1)}_{\rm
bKK}|_{\xi_1->\widetilde{\xi}_1}
\end{equation}
and the corresponding single soliton is $u=\left(\partial_x^2
\log\hat{\tau}^{(1+1)}_{\rm KK} \right)$.  The  velocity of
soliton is $\hat{v}_{\_}=\left.-k_1^4\frac{\cos
5\varepsilon_1}{\cos \varepsilon_1}\right|_{{\pi \over
6}}=k_1^4>0$.  Here $\hat{\tau}^{(1+1)}_{\rm bKK}$ is given by
Proposition \ref{proponesolitonbKK}.
\end{lemma}
\begin{lemma}\label{lemonesolitontaunCKP}
  The higher order equation of {\rm CKP} defined by equation
  (\ref{higherordereqCKP}).  For n-reduction of {\rm CKP} hierarchy {\rm
    (nCKP)}, $n=2j+1, j=3, 4, 5, \cdots $. Let
  $\widetilde{\xi}_{mp}=xk_m\cos \varepsilon_p+tk^3_m\cos3\varepsilon_p,
  k_m^n=k_m^{2j+1}=|\lambda_m|$, then the $\tau$ function of the {\rm
    nCKP} generated by $T_{1+1}$ is
\begin{equation}
\hat{\tau}^{(1+1)}_{\rm nCKP}=\hat{\tau}^{(1+1)}_{\rm
bKK}|_{\xi_1->\widetilde{\xi}_{1p}},
\end{equation}
and the corresponding single soliton of the { \rm nCKP} equation
is $u=\left(\partial_x^2 \log\hat{\tau}^{(1+1)}_{\rm nCKP}
\right)$. Here $\varepsilon_p=\frac{p}{2n}\pi=\frac{p}{4j+2}\pi,
p=1, 2, 3,\cdots, j$ , and $\hat{\tau}^{(1+1)}_{\rm bKK}$ is given
by Proposition \ref{proponesolitonbKK}. So the single soliton can
move along $j$ directions on (x, t) plane, which are given by
$\widetilde{\xi}_{1p}=0$ associated with $j$-value of
$\varepsilon_p$ given before.
\end{lemma}
Using the Lemmata \ref{lemonesolitontauKK}, \ref{lemonesolitontaunCKP}
and results for the bKK equation, we get
\begin{proposition}\label{propdirections of nCKP}
\begin{enumerate}
\item The single soliton $u=\left(\partial_x^2
\log\hat{\tau}^{(1+1)}_{\rm nCKP} \right)$ of the {\rm nCKP},
$n=2j+1, j=2, 3, 4,\cdots $, can move along a direction defined by
$\widetilde{\xi}_{1p}=0$ on (x, t) plane for a given $p$.

\item $(e^{i\varepsilon_p},-e^{-i\varepsilon_p})$ one distribution
of n-th order roots of $e^{i\varepsilon}$ on circle
$\longrightarrow$
$(p_1=k_1e^{i\varepsilon_p},q_1=-k_1e^{-i\varepsilon_p})$ in
equation (\ref{nBKPzerophi1a}) $\longrightarrow$ the single soliton
moves along a line $\widetilde{\xi}_{1p}=0$ on (x, t) plane. Here
$\varepsilon_p\in
\Big\{\frac{\pi}{4j+2},\frac{3\pi}{4j+2},\frac{5\pi}{4j+2}, \cdots,
\frac{(2j-1)\pi}{4j+2}\Big\}$.

\item For a given $n=2j+1$, the single soliton of the {\rm nCKP}
can move along  $j$ directions on (x,t) plane, which are defined
by $\widetilde{\xi}_{1p}=0, p=1, 2, 3, \cdots, j$.

\item In particular, if $0 < \varepsilon_p <\pi/6$,
  $u=\left(\partial_x^2 \log\hat{\tau}^{(1+1)}_{\rm nCKP} \right)$ is a
  two-peak soliton.
\end{enumerate}
\end{proposition}
In above Proposition, the case of $j=1$ is given by Lemma
\ref{lemonesolitontauKK}. This Proposition shows there exist
several single two-peak solitons for {\rm nCKP} if $ n \geq 11 $ .
\begin{corollary}
There are two single two-peak solitons for 11-reduction of {\rm
CKP} hierarchy, i. e. {\rm 11CKP} equation, $u_{\rm {\tiny
11CKP}}=\left(\partial_x^2\log \hat{\tau}^{(1+1)}_{\rm
bKK}\right)$,
 in which
$\varepsilon_1=\pi/22$ and $\varepsilon_1=3\pi/22$ respectively.
Here $\hat{\tau}^{(1+1)}_{\rm bKK}$ is given in equation
(\ref{efficienttauonesolitonbKK})
\end{corollary}
We have plotted  it out in figure \ref{fig:11ckptwo two-peak} with
$k_1=0.8$.

\begin{figure}[htbp]
\centerline{
  \includegraphics[width=0.5\textwidth]{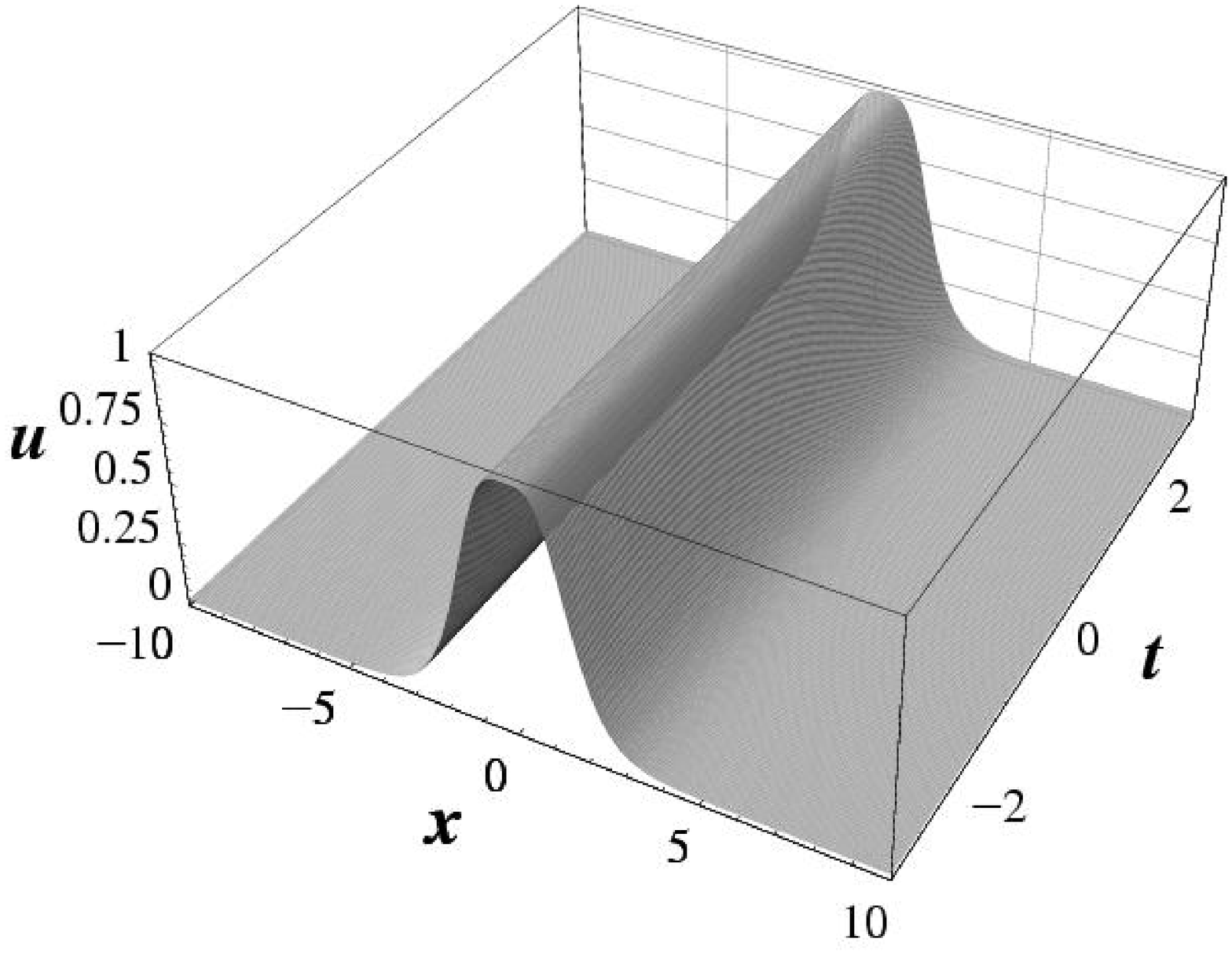}
  \includegraphics[width=0.5\textwidth]{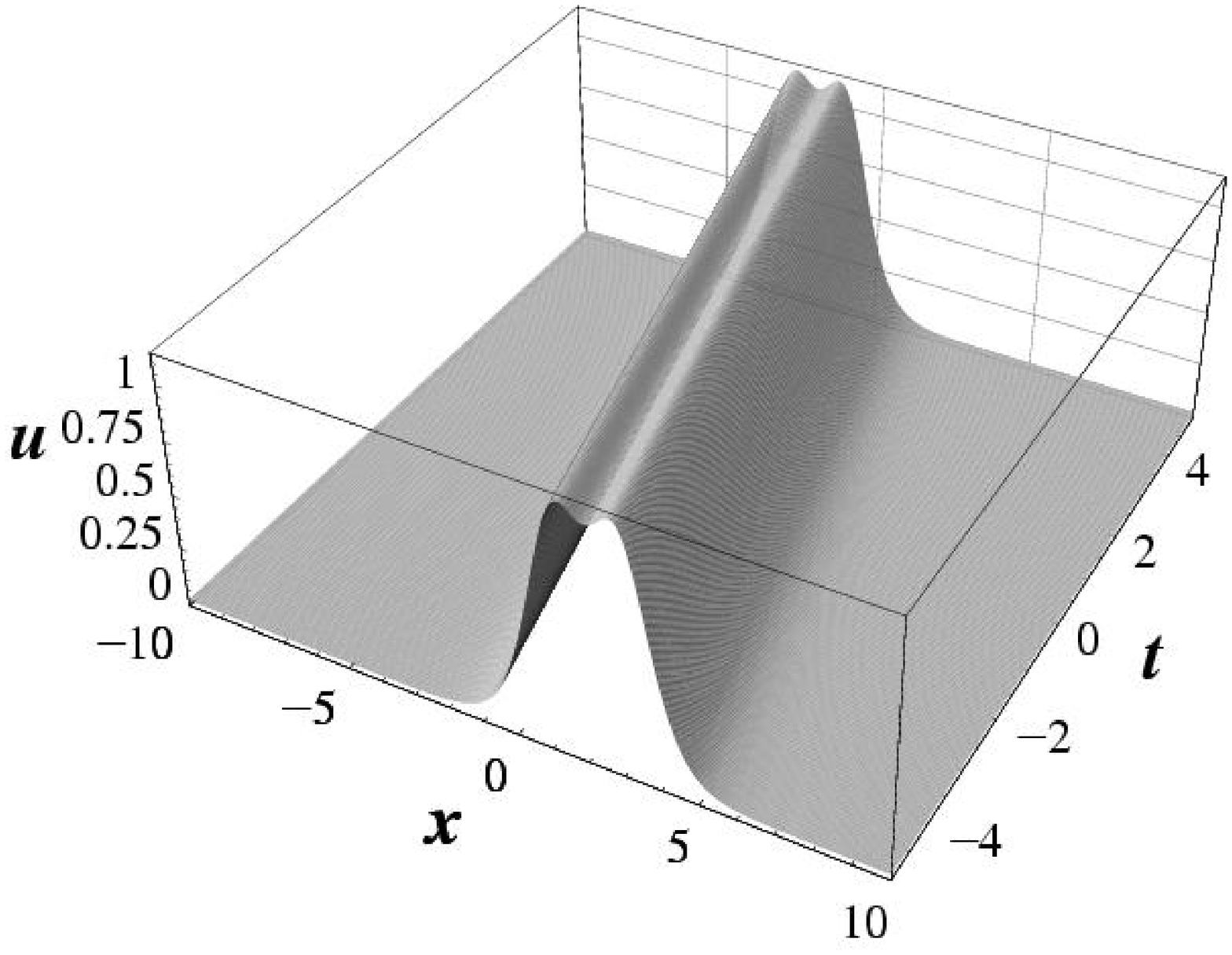}}
  \caption{  \label{fig:9ckpstaionarysoliton}  \label{fig:8ckptwopeaksoliton}
    Left: "stationary" soliton for 9-reduction of CKP, Right: Single
    two-peak soliton for 8-reduction of KP.}
\end{figure}

We have known that $a=1/2$ in Lemma \ref{lembKKpoint} and
Proposition \ref{propdiscusstwopeakbKK} is one crucial point to
exist one-peak soliton or two-peak soliton. It is more interesting
that $a=1/2$ will lead to "stationary" soliton of higher
reductions of the BKP and the CKP hierarchy, which is not moving
on (x,t) plane.  When
$\xi_1|_{\varepsilon_1=\pi/6}=(k_1x\cos\varepsilon_1+tk_1^3\cos
3\varepsilon_1|)_{\varepsilon_1=\pi/6}= (k_1x\cos\varepsilon_1)$,
$\xi_1$ is independent with $t$. So $u$ is independent with $t$ by
taking this $\xi_1$ into Proposition \ref{proponesolitonbSK} and
Proposition \ref{proponesolitonbKK}.
\begin{corollary}
\begin{enumerate}
\item There exists "stationary" single soliton for the 9-reduction
of {\rm BKP} hierarchy,  which is $u_{\rm 9BKP}=\left
(\partial_x^2\log \hat{\tau}^{(1+1)}_{\rm bSK}\right)\left.
\right|_{\varepsilon_1=3\pi/18 }$. Here $\hat{\tau}^{(1+1)}_{\rm
bSK}$ is given by Proposition \ref{proponesolitonbSK};

\item There exists "stationary" single soliton for the 9-reduction
of {\rm CKP} hierarchy, which is $u_{\rm 9CKP}= \left(
\partial_x^2\log \hat{\tau}^{(1+1)}_{\rm
bKK}\right)\left. \right|_{\varepsilon_1=3\pi/18}$. Here
$\hat{\tau}^{(1+1)}_{\rm bKK}$ is given by Proposition
\ref{proponesolitonbKK}.
\end{enumerate}
\end{corollary}
We have plotted out "stationary" soliton for the 9-reduction of
CKP in figure \ref{fig:9ckpstaionarysoliton} when $k_1=1$.
\begin{corollary}
There is single two-peak soliton $u=\left (\partial_x^2 \log
\hat{\tau}^{(1+1)}_{bSH}\right)\left. \right|_{\varepsilon_1=\pi/8
}$ for 8-reduction of the {\rm KP} hierarchy; there is
"stationary" single one-peak soliton $u=\left(\partial_x^2\log
\hat{\tau}^{(1+1)}_{\rm bSH}\right)\left.
\right|_{\varepsilon_1=\pi/6 }$ for the 6-reduction of the {\rm
KP} hierarchy.  Here $\hat{\tau}^{(1+1)}_{\rm bSH})$ is given by
Proposition \ref{proponesolitonbSH}.
\end{corollary}
The two-peak soliton for the 8-reduction of KP hierarchy is
plotted in figure \ref{fig:9ckpstaionarysoliton} when $k_1=1$. For
our best knowledge, this is first time to report the
even-reduction of the KP hierarchy also has two-peak soliton
solution. The possession of two-peak soliton solution is not sole
property of CKP hierarchy.

\section{Conclusions and Discussions}

We have presented a systematic way in which to obtain the solution
of the $n$-reduction ($n=4,5$) from the general $\tau$ function of
the KP hierarchy. Our approach is based on the determinant
representation of gauge transformations $T_{n+k}$\cite{hlc1} and
$\tu{n+k}$ \cite{csy2}. It may be summarized as follows:
\\
$\tu{n+k}\xrightarrow{ \text{
    constraints of GFs and } k=n } \tu{n+k}_{\rm BKP}$
($\tu{n+k}_{\rm CKP}$ ) $\xrightarrow{5-\mbox{reduction}}
\tu{n+k}_{\rm
  Eq}|_{k=n}({\rm Eq}={\rm bSK, bKK})\\ \xrightarrow{ \text{assume the
    form of }\phi_i \text{and find suitable} \frac{B_i}{A_i}}$ efficient
$\tau$ function $\hat{\tau}^{(n+k)}_{\rm Eq}|_{k=n=1,2}$.
We have applied this approach to various equations. The one soliton, two
soliton and periodic solution are constructed for bSK, bKK and bSH. We
show the corresponding relation between the distribution of $5$-th (or
$4$th) roots of $e^{i\varepsilon}$ on the unit circle and several types
of solutions (left-going one soliton, right-going one soliton,
left/right-going periodic solutions).
We also show the reason for the existence of the two-peak soliton.
Furthermore, the lower reduction and higher reduction of BKP, CKP, and
the even-reductions are explored by this method. Our results show that
the soliton of the $n$-reduction (with $n= 2j+1$, $j=1, 2, 3, \ldots$)
of BKP and CKP can move alone $j$ directions, which are defined by
$\widetilde{\xi}_{1p}=0$. Each direction corresponds to one symmetry
distribution of $n$-th roots of $e^{i\varepsilon}$ on the unit circle.
This supplies a very natural explanation why the $5$-reduction BKP (or
CKP) has bi-directional solitons whereas the $3$-reduction of BKP (or
CKP) has only single-directional solitons.
At last, the two-peak soliton is not a monopolizing phenomena of only
the CKP hierarchy. Rather, we find that the higher-order even-reduction
of KP also exhibits two-peak solitons and we elucidate the criterion for
its existence from the Grammian $\tau$ function. At the same time, we
show there is not three and more peak soliton from Grammian $\tau$
function. The "stationary" soliton for higher order reduction of KP
hierarchy is also obtained.

We think that it is possible to construct an $N$-soliton solution of the
bSK, bKK and bSH equations by this approach. Namely, there exist
suitable $\frac{B_i}{A_i}$ ($i=1,2,\cdots,N$) such that we can find a
{\em physical} $\tau$ function $\hat{\tau}^{(N+N)}|_{\rm Eq}$ for these
equations from a complex-valued $\tu{N+N}|_{\rm Eq}$, which is symmetric
because we have assumed generating functions $\phi_i$ in equation
({\ref{zerophi1a}}) and equation ({\ref{zerophi1b}}) with symmetric form.
Here ${\rm Eq}={\rm bSK,bKK,bSH}$. Additionally, it is worthy to discuss
the phase shift in the collision of one-peak soliton and two-peak
soliton. Furthermore, it is possible to construct solutions for bSK, bKK
and bSH from constant initial value $u=constant \not=0$, which is
parallel to present results.

\small{Upon completion of this work, Prof.\ V.\ Sokolov kindly pointed
  out Ref.\ \cite{ds1} where equations (\ref{bSH0}, \ref{sk}, \ref{kk}) and
  their Lax operators as well as the Lax operator $L=\partial +
  u\partial^{-1} u $ for KdV equation have been obtained for the first
  time.}

{\bf Acknowledgement}\\
{\small
  This work was supported partly by project 973 ``Nonlinear Science''
  and the National Natural Science Foundation of China (10301030). We
  thank Prof.\ Musette for his warm and valuable answering of questions
  about Ref.\ \cite{vm1}. We thank Prof.\ Sokolov for information about
  references and useful discussion on the gauge transformation technique
  during his visit at Hefei (P. R. China). Support from the Warwick Chinese
  Fellowship Funds is grateful acknowledged. A short version
  of this paper has been delivered at "2005 International Symposium on Nonlinear
  Dynamics, Celebration of M.S.\ El-Naschie's 60 Anniversary, December 20-21,
 2005, Shanghai, China" }
\newpage

\appendix

\section{bSK equation}\label{appbSK}

\setcounter{equation}{0}

Take $z_k=c_k + i d_k(k=1,3,5,7)$. Then
\begin{eqnarray}
&c_1=&k_1 k_2\cos(\varepsilon_1+\varepsilon_2)\big[
k_1^2\cos2\varepsilon_1+ k_2^2\cos2\varepsilon_2
+2k_1k_2\cos(\varepsilon_1 +\varepsilon_2) \big]
  \nonumber    \\
& &+ k_1 k_2\sin(\varepsilon_1+\varepsilon_2)\big[ k_1^2
\sin2\varepsilon_1+ k_2^2\sin2\varepsilon_2
+2k_1k_2\sin(\varepsilon_1
+\varepsilon_2)\big]   \\
&d_1=&-k_1 k_2\cos(\varepsilon_1+\varepsilon_2)
 \big[
k_1^2 \sin2\varepsilon_1+ k_2^2\sin2\varepsilon_2
+2k_1k_2\sin(\varepsilon_1 +\varepsilon_2)\big]
 \nonumber    \\
& &+ k_1 k_2\sin(\varepsilon_1+\varepsilon_2)\big[
k_1^2\cos2\varepsilon_1+ k_2^2\cos2\varepsilon_2
+2k_1k_2\cos(\varepsilon_1 +\varepsilon_2) \big] \\
&c_3=&k_1 k_2\cos(\varepsilon_1-\varepsilon_2)\big[
k_1^2\cos2\varepsilon_1+ k_2^2\cos2\varepsilon_2
-2k_1k_2\cos(\varepsilon_1 -\varepsilon_2)
 \big] \nonumber    \\
& &+ k_1 k_2\sin(\varepsilon_1-\varepsilon_2)\big[ k_1^2
\sin2\varepsilon_1- k_2^2\sin2\varepsilon_2
-2k_1k_2\sin(\varepsilon_1
-\varepsilon_2)\big]   \\
&d_3=&-k_1 k_2\cos(\varepsilon_1-\varepsilon_2) \big[ k_1^2
\sin2\varepsilon_1- k_2^2\sin2\varepsilon_2
-2k_1k_2\sin(\varepsilon_1
-\varepsilon_2)\big] \nonumber    \\
& &+ k_1 k_2\sin(\varepsilon_1+\varepsilon_2)\big[
k_1^2\cos2\varepsilon_1+ k_2^2\cos2\varepsilon_2
-2k_1k_2\cos(\varepsilon_1 -\varepsilon_2) \big] \\
&c_5=&2k_1 k_2\sin\varepsilon_2 \big[\cos\varepsilon_1\big(
k_1^2\cos2\varepsilon_1- k_2^2 -2k_1k_2\sin\varepsilon_1
\sin\varepsilon_2 \big)
 \big] \nonumber    \\
& &+2k_1k_2\sin\varepsilon_2\big[2k_1\cos\varepsilon_1\sin\varepsilon_1\big(
k_1\sin\varepsilon_1+k_2\sin\varepsilon_2 \big)]   \\
&d_5=&-2k_1 k_2\sin\varepsilon_2 \big[2k_1\cos^2\varepsilon_1\big(
k_1\sin\varepsilon_1+k_2\sin\varepsilon_2 \big)
 \big] \nonumber    \\
& &+ 2k_1k_2\sin\varepsilon_2\big[\sin\varepsilon_1\big(
k_1^2\cos2\varepsilon_1- k_2^2 -2k_1k_2\sin\varepsilon_1
\sin\varepsilon_2 \big)]   \\
&c_7=&2k_1 k_2\sin\varepsilon_1 \big[\cos\varepsilon_2\big( k_1^2-
k_2^2\cos2\varepsilon_2 +2k_1k_2\sin\varepsilon_1
\sin\varepsilon_2 \big)
 \big] \nonumber    \\
& &-2k_1k_2\sin\varepsilon_1\big[2k_2\cos\varepsilon_2\sin\varepsilon_2\big(
k_1\sin\varepsilon_1+k_2\sin\varepsilon_2 \big)]  \\
&d_7=&2k_1 k_2\sin\varepsilon_1 \big[2k_2\cos^2\varepsilon_2\big(
k_1\sin\varepsilon_1+k_2\sin\varepsilon_2 \big)
 \big] \nonumber    \\
& &+ 2k_1k_2\sin\varepsilon_1\big[\sin\varepsilon_2\big( k_1^2-
k_2^2\cos2\varepsilon_2 +2k_1k_2\sin\varepsilon_1
\sin\varepsilon_2 \big)]   
\end{eqnarray}

\section{bKK equation (two solitons)}\label{appbKK2S}

\setcounter{equation}{0}

Take $z_k=c_k + i d_k(k=1,2,3,4)$. Then
\begin{eqnarray}
&c_1=&\cos(\varepsilon_1+\varepsilon_2)\big[ k_1^2+ k_2^2
+2k_1k_2\cos(\varepsilon_1 -\varepsilon_2) \big]^2
 \nonumber    \\
& &-4 k_1 k_2\big[ k_1 \cos2\varepsilon_1+ k_2\cos2\varepsilon_2
+2k_1k_2\cos(\varepsilon_1
+\varepsilon_2)\big]   \\
&d_1=&\sin(\varepsilon_1+\varepsilon_2)\big[ k_1^2+ k_2^2
+2k_1k_2\cos(\varepsilon_1 -\varepsilon_2) \big]^2
 \nonumber    \\
& &-4 k_1 k_2\big[ k_1^2 \sin2\varepsilon_1+
k_2^2\sin2\varepsilon_2 +2k_1k_2\sin(\varepsilon_1
+\varepsilon_2)\big]   \\
&c_2=&\cos\varepsilon_1\big[ k_1^2+ k_2^2
+2k_1k_2\cos(\varepsilon_1 -\varepsilon_2) \big]\big[ k_1^2+ k_2^2
-2k_1k_2\cos(\varepsilon_1 +\varepsilon_2) \big] \nonumber  \\
& &-4k_1k_2\sin\varepsilon_2\big(  k_1^2\sin2\varepsilon_1 +
2k_1k_2\sin\varepsilon_2\cos\varepsilon_1\big)   \\ 
&d_2=&\sin\varepsilon_1\big[ k_1^2+ k_2^2
+2k_1k_2\cos(\varepsilon_1 -\varepsilon_2) \big]\big[ k_1^2+ k_2^2
-2k_1k_2\cos(\varepsilon_1 +\varepsilon_2) \big] \nonumber  \\
& &+4k_1k_2\sin\varepsilon_2\big(  k_1^2\cos2\varepsilon_1 -k_2^2-
2k_1k_2\sin\varepsilon_1\sin\varepsilon_2\big)  \\ 
&c_3=&\cos(\varepsilon_1-\varepsilon_2)\big[ k_1^2+ k_2^2
-2k_1k_2\cos(\varepsilon_1 +\varepsilon_2) \big]^2
 \nonumber    \\
& &+4 k_1 k_2\big[ k_1^2 \cos2\varepsilon_1+
k_2^2\cos2\varepsilon_2 -2k_1k_2\cos(\varepsilon_1
-\varepsilon_2)  \big] \\
&d_3=&\sin(\varepsilon_1-\varepsilon_2)\big[ k_1^2+ k_2^2
-2k_1k_2\cos(\varepsilon_1 +\varepsilon_2) \big]^2
 \nonumber    \\
& &+4 k_1 k_2\big[ k_1^2 \sin2\varepsilon_1-
k_2^2\sin2\varepsilon_2 -2k_1k_2\sin(\varepsilon_1
-\varepsilon_2) \big]  \\
&c_4=&\cos\varepsilon_2\big[ k_1^2+ k_2^2
+2k_1k_2\cos(\varepsilon_1 -\varepsilon_2) \big]\big[ k_1^2+ k_2^2
-2k_1k_2\cos(\varepsilon_1 +\varepsilon_2) \big] \nonumber  \\
& &-4k_1k_2\sin\varepsilon_1\big(  k_2^2\sin2\varepsilon_2 +
2k_1k_2\sin\varepsilon_1\cos\varepsilon_2\big)   \\ 
&d_4=&\sin\varepsilon_2\big[ k_1^2+ k_2^2
+2k_1k_2\cos(\varepsilon_1 -\varepsilon_2) \big]\big[ k_1^2+ k_2^2
-2k_1k_2\cos(\varepsilon_1 +\varepsilon_2) \big] \nonumber  \\
& &+4k_1k_2\sin\varepsilon_1\big(  -k_1^2 +
k_2^2\cos2\varepsilon_2-
2k_1k_2\sin\varepsilon_1\sin\varepsilon_2\big)  
\end{eqnarray}

\section{bKK equation (periodic solutions)}\label{appbKKP}

\setcounter{equation}{0}

Take $z_k=c_k + i d_k(k=1,2,3,4)$. Then
\begin{eqnarray}
&c_1=&\cos(\varepsilon_1+\varepsilon_2)\big[ k_1^2+ k_2^2
+2k_1k_2\cos(\varepsilon_1 -\varepsilon_2) \big]^2
 \nonumber    \\
& &-4 k_1 k_2\big[ k_1^2 \cos2\varepsilon_1+
k_2^2\cos2\varepsilon_2 +2k_1k_2\cos(\varepsilon_1
+\varepsilon_2)\big]   \\
&d_1=&\sin(\varepsilon_1+\varepsilon_2)\big[ k_1^2+ k_2^2
+2k_1k_2\cos(\varepsilon_1 -\varepsilon_2) \big]^2
 \nonumber    \\
& &-4 k_1 k_2\big[ k_1^2 \sin2\varepsilon_1+
k_2^2\sin2\varepsilon_2 +2k_1k_2\sin(\varepsilon_1
+\varepsilon_2)\big]   \\
&c_2=&\cos\varepsilon_1\big[ k_1^2+ k_2^2
+2k_1k_2\cos(\varepsilon_1 -\varepsilon_2) \big]\big[ k_1^2+ k_2^2
+2k_1k_2\cos(\varepsilon_1 +\varepsilon_2) \big] \nonumber  \\
& &-4k_1k_2\cos\varepsilon_2\big(  k_1^2\cos2\varepsilon_1 +k_2^2+
2k_1k_2\cos\varepsilon_1\cos\varepsilon_2\big)   \\ 
&d_2=&\sin\varepsilon_1\big[ k_1^2+ k_2^2
+2k_1k_2\cos(\varepsilon_1 -\varepsilon_2) \big]\big[ k_1^2+ k_2^2
+2k_1k_2\cos(\varepsilon_1 +\varepsilon_2) \big] \nonumber  \\
& &-8k_1^2k_2\cos\varepsilon_2\sin\varepsilon_1\big(
k_1\cos\varepsilon_1 +
k_2\cos\varepsilon_2\big)  \\ 
&c_3=&\cos(\varepsilon_1-\varepsilon_2)\big[ k_1^2+ k_2^2
+2k_1k_2\cos(\varepsilon_1 +\varepsilon_2) \big]^2
 \nonumber    \\
& &-4 k_1 k_2\big[ k_1^2 \cos2\varepsilon_1+
k_2^2\cos2\varepsilon_2 +2k_1k_2\cos(\varepsilon_1
-\varepsilon_2)  \big] \\
&d_3=&\sin(\varepsilon_1-\varepsilon_2)\big[ k_1^2+ k_2^2
+2k_1k_2\cos(\varepsilon_1 +\varepsilon_2) \big]^2
 \nonumber    \\
& &-4 k_1 k_2\big[ k_1^2 \sin2\varepsilon_1-
k_2^2\sin2\varepsilon_2 +2k_1k_2\sin(\varepsilon_1
-\varepsilon_2) \big]  \\
&c_4=&\cos\varepsilon_2\big[ k_1^2+ k_2^2
+2k_1k_2\cos(\varepsilon_1 -\varepsilon_2) \big]\big[ k_1^2+ k_2^2
+2k_1k_2\cos(\varepsilon_1 +\varepsilon_2) \big] \nonumber  \\
& &-4k_1k_2\cos\varepsilon_1\big[ k_1^2+ k_2^2\cos2\varepsilon_2 +
2k_1k_2\cos\varepsilon_1\cos\varepsilon_2\big]   \\ 
&d_4=&\sin\varepsilon_2\big[ k_1^2+ k_2^2
+2k_1k_2\cos(\varepsilon_1 -\varepsilon_2) \big]\big[ k_1^2+ k_2^2
+2k_1k_2\cos(\varepsilon_1 +\varepsilon_2) \big] \nonumber  \\
& &-8k_1k_2^2\cos\varepsilon_1\sin\varepsilon_2\big(
k_1\cos\varepsilon_1+
k_2\cos\varepsilon_2\big)  
\end{eqnarray}

\section{bSH equation}\label{appbSH}
\setcounter{equation}{0}

Take $z_k=c_k + i d_k(k=1,3,5)$. Then
\begin{eqnarray}
&c_1=&2k_2(k_2\cos\varepsilon_2+k_1)-(k_1^2+k_2^2+2k_1k_2\cos\varepsilon_2)\cos\varepsilon_2    \\ 
&d_1=&2k^2_2\sin\varepsilon_2-(k_1^2+k_2^2+2k_1k_2\cos\varepsilon_2)\sin\varepsilon_2    \\ 
&c_3=&2k_2(k_2\cos\varepsilon_2-k_1)-(k_1^2+k_2^2-2k_1k_2\cos\varepsilon_2)\cos\varepsilon_2 \\
&d_3=&2k_2^2\sin\varepsilon_2-(k_1^2+k_2^2-2k_1k_2\cos\varepsilon_2)\sin\varepsilon_2 \\
&c_5=&2k_2^2(k_2^2+k_1^2)\sin^2\varepsilon_2-(k_1^2+k_2^2+2k_1k_2\cos\varepsilon_2)
  (k_1^2+k_2^2-2k_1k_2\cos\varepsilon_2) \\
&d_5=&k_2^2\sin\varepsilon_2\big[
2k_1(k_1^2+k_2^2)-4k_1k_2^2\cos^2\varepsilon_2
\big]
\end{eqnarray}



\end{document}